\documentclass[review]{elsarticle}

\usepackage{lineno,hyperref}
\modulolinenumbers[5]
\usepackage{epsfig}
\usepackage{color}
\usepackage{tipa}

\journal{Progress in Nuclear and Particle Physics}

\def\Journal#1#2#3#4{{#1} {#2} (#4) #3 }

\def\NPA{{\em Nucl. Phys.} A}

\def\PLB{{\em Phys. Lett.} B}

\def\PRL{\em Phys. Rev. Lett.}

\def\PREP{\em Phys. Rep.}

\def\PRD{{\em Phys. Rev.} D}
\def\PRC{{\em Phys. Rev.} C}

\newcommand{\be}{\begin{equation}}
\newcommand{\ee}{\end{equation}}
\newcommand{\bea}{\begin{eqnarray}}
\newcommand{\eea}{\end{eqnarray}}

\begin{document}

\begin{frontmatter}

\title{Novel Techniques for Constraining Neutron-Capture Rates Relevant for \textit{r}-Process Heavy-Element Nucleosynthesis}

\author{A. C. Larsen$^1$, A. Spyrou$^{2,3,4}$, S.N. Liddick$^{2,5}$, M. Guttormsen$^1$}
\address{$^{1}$Department of Physics, University of Oslo, 0316 Oslo, Norway}
\address{$^{2}$National Superconducting Cyclotron Laboratory, Michigan State University, Michigan, USA}
\address{$^{3}$Department of Physics and Astronomy, Michigan State University, Michigan, USA}
\address{$^{4}$Joint Institute for Nuclear Astrophysics, Michigan State University, Michigan, USA}
\address{$^5$Department of Chemistry, Michigan State University, Michigan, USA}

\begin{abstract}
The rapid-neutron capture process ($r$ process) is identified as the producer of about 50\% of elements heavier than iron. This process requires an astrophysical environment with an extremely high neutron flux over a short amount of time ($\sim$ seconds), creating very neutron-rich nuclei that are subsequently transformed to stable nuclei via $\beta^-$ decay. 
In 2017, one site for the $r$ process was confirmed: the advanced LIGO and advanced Virgo detectors observed two neutron stars merging, and immediate follow-up measurements of the electromagnetic transients demonstrated an "afterglow" over a broad range of frequencies fully consistent with the expected signal of an $r$ process taking place. 
Although neutron-star mergers are now known to be $r$-process element factories, contributions from other sites are still possible, and a comprehensive understanding and description of the $r$ process is still lacking. 
One key ingredient to large-scale $r$-process reaction networks  is radiative neutron-capture ($n,\gamma$) rates, for which there exist virtually no data for extremely neutron-rich nuclei involved in the $r$ process. 
Due to the current status of nuclear-reaction theory and our poor understanding of basic nuclear properties such as level densities and average $\gamma$-decay strengths, theoretically estimated ($n,\gamma$) rates may vary by orders of magnitude and represent a major source of uncertainty in any nuclear-reaction network calculation of $r$-process abundances. 
In this review, we discuss new  approaches to provide information on 
neutron-capture cross sections and reaction rates relevant to the $r$ process. 
In particular, we focus on indirect, experimental techniques to measure radiative neutron-capture rates.
While direct measurements are not available at present, but could possibly be realized in the future, the indirect approaches present a first step towards constraining neutron-capture rates of importance to the $r$ process.
\end{abstract}


\end{frontmatter}


\section{Introduction}
\label{sec:intro}
One of the big mysteries that humans have pondered upon, is how and where the elements observed in the Universe were formed. 
The elements are the building blocks of all visible matter, and their distribution is a result of many nucleosynthesis agents acting as ``alchemists'', changing the original Big Bang abundance (consisting of only the lightest elements) into a great variety of nuclides. 

The first attempt to determine the distribution of element abundances of our Solar system was made by Goldschmidt in 1937~\cite{goldschmidt1937}, and has later been substantially improved with precise measurements of CI1 carbonaceous chondrites, terrestrial samples, and analysis of solar spectra ~\cite{lodders2003}. 
In particular, isotopic abundances are mainly derived from terrestrial data, except for hydrogen and the noble gases~\cite{lodders2003}. 
The isotopic abundance distributions are the most revealing fingerprint to the astrophysical processes behind their origin.   

The first direct evidence of heavy element nucleosynthesis in stars came in an observation by Paul Merrill, published in 1952~\cite{merrill1952}. Merrill observed lines of the element Technetium, an element with no stable isotopes, and for which the longest-lived isotope has a half-life of 4 million years. Merrill's surprising observation showed for the first time that stars are the birth place of heavy elements in the Universe, since any Technetium produced during the Big Bang should have decayed away long ago. Following this discovery, in 1957, Burbidge, Burbidge, Fowler and Hoyle~\cite{burbidge1957} and independently Cameron~\cite{cameron1957}
outlined the main nucleosynthesis processes called for to explain the observed abundances of all elements in the Universe. 
For elements heavier than iron (proton number $Z=26$), three main processes were described: 
\begin{itemize}
\item{the rapid neutron-capture process ($r$ process)}
\item{the slow neutron-capture process ($s$ process)}
\item{the proton capture/photodisintegration process ($p$ process)}
\end{itemize}
The two latter processes are not the focus of the present article; excellent reviews of the $s$ process are given in Refs.~\cite{kappeler2009,reifarth2015} and of the $p$ process in Refs.~\cite{arnould2003,rauscher2013}. The present article focuses on the rapid neutron capture process and the impact of the nuclear input on our understanding of $r$-process nucleosynthesis. 
It should be noted that while the three aforementioned processes are most probably the dominant heavy-element production mechanisms, they are not able to reproduce all astrophysical observations, and for this reasons other processes have been proposed, like the $\nu$p process~\cite{wajano2006,pruet2006,frohlich2006, wanajo2011}, the $i$ process~\cite{cowan1977,herwig2011,abate2016,hampel2016,denissenkov2018} and the so-called Light Element Primary Process or LEPP~\cite{arcones2011, montes2007, travaglio2004}. 

\begin{figure}[tb]
\begin{center}
\includegraphics[clip,width=1.\columnwidth]{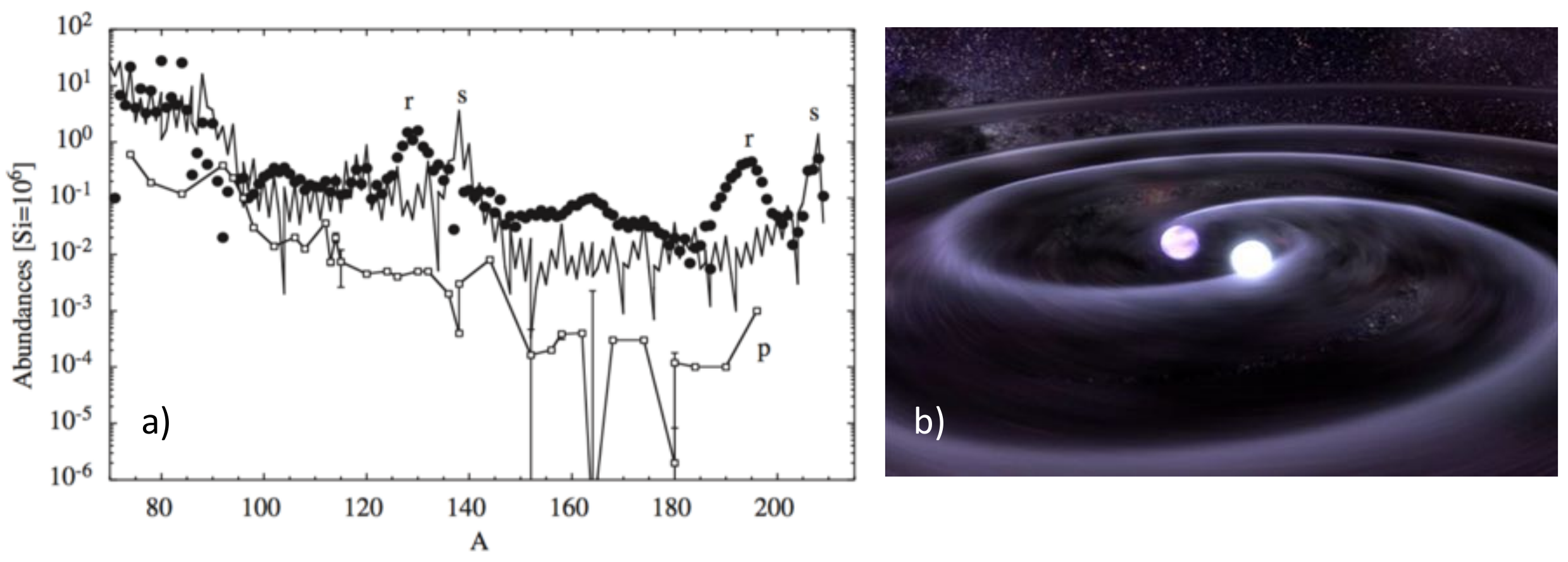}
\caption {(a) Solar-system abundances of heavy-element isotopes broken down to the contributions from the $s$ process (jagged line), the $r$ process (black circles) and the $p$ process (open squares). Figure taken from Ref.~\cite{arnould2007}. (b) Illustration of a neutron-star merger event (credit: Dana Berry and Erica Drezek, NASA/Goddard Space Flight Center).}
\label{fig:abundances}
\end{center}
\end{figure}

Figure~\ref{fig:abundances}, taken from the seminal work of Arnould, Goriely and Takahashi~\cite{arnould2007}, shows the distribution of heavy elements (mass number $A \approx 70-209$) split into  contributions from each of the three main processes.  
Characteristic $s$- and $r$-process peaks are visible around $A \approx 138, 208$ and $A \approx 80, 130, 195$, respectively. These abundance peaks originate from the presence of neutron magic numbers at $N=50$, 82, and 126. Due to the added stability of nuclei with magic neutron numbers, when the reaction flow passes through magic isotones, matter accumulates, and as a result a peak is formed in the abundance distribution. The location of these abundance peaks was the first indication for how far from stability these astrophysical processes might proceed. In the $s$ process, the peaks appear at higher masses, indicating that the $s$ process reaction flow proceeds right around stability. 
In the $r$ process, on the other hand, the abundance peaks appear at lower masses because the $r$ process flows through more neutron-rich nuclei. On top of the main abundance peaks, the $r$-process isotopic distribution exhibits a smaller peak in the rare-earth region around $A\approx 165$. The origin of this structure is not yet well understood, although it is linked to sub-shell closures or other nuclear structure effects~\cite{mumpower2016}.

Amongst the three main processes responsible for heavy-element nucleosynthesis, the 
$r$ process is perhaps the most challenging one to describe, both from 
an astrophysics and a nuclear-physics point of view. 
This process is responsible for the synthesis of about 50\% of the isotopes of elements above iron, and is the only one able to produce actinides~\cite{arnould2007,goriely2015a}. 
While the uncertainties associated with the astrophysical site of the $r$ process are still major, the recent observation of the first neutron-star merger event by gravitational and electromagnetic observatories around the world has at least identified a significant source of $r$-process material in the Universe. 
On the other hand, the nuclear physics properties used in $r$-process calculations are far from constrained. Nuclear masses, $\beta$-decay properties, neutron-capture reactions, and nuclear fission properties are the main quantities needed for a complete description of the $r$-process reaction flow. A comprehensive study of how nuclear physics properties impact $r$-process abundance calculations was presented in the work of Mumpower \textit{et al.}~\cite{mumpower2016}. Here we will focus on one of these properties, namely neutron-capture or $(n,\gamma$) reactions. 

Experimentally, neutron-capture reactions can be studied directly when a neutron beam impinges on a stable or long-lived target nucleus. To-date, the direct measurement of $(n,\gamma$) reactions on short-lived radioactive nuclei is not possible, although some future plans will be discussed in section \ref{sec:direct}. 
For this reason, indirect techniques have been developed that can provide constraints on neutron-capture reaction rates for nuclei far from stability, especially the ones participating in $r$-process calculations. These techniques rely on nuclear structure and nuclear reaction information for the nucleus of interest, eliminating in this way part of the uncertainty in the calculation of $(n,\gamma$) reaction rates. 

The goal of the present review article is to give an overview of the available techniques for constraining neutron-capture reactions involved in the astrophysical $r$ process, as well as experiments and theoretical approaches developed to understand the nuclear-structure aspects relevant for $r$-process nucleosynthesis.

\section{The \textit{r} process: a brief overview}
\label{sec:r-process}

\subsection{$r$-process site and observations}
\label{subsec:site}

In the quest to understand how the heavy elements are created in the Universe we have to look at all observables available to us, and be able to reproduce them with our models. In the case of the $r$ process, until recently, three main observables existed:
\begin{itemize}
\item{Solar-system abundances \textit{e.g.}~\cite{arlandini99}}
\item{Meteoritic data, \textit{e.g.}~\cite{tissot16}}
\item{Observations of $r$-process elements in old stars, \textit{e.g.} \cite{frebel18}}
\end{itemize}

The $r$-process solar-system abundances, such as the ones presented in Fig.~\ref{fig:abundances}, come from the total abundances after subtracting the $s$-and $p$-process contributions, and for this reason they are often called $r$-process residuals. This method introduces significant uncertainties in the residual abundance of some isotopes because of the uncertainty in the other contributions~\cite{arlandini99}. Meteoritic samples, and in particular pre-solar grains that are found in these samples can provide additional information on isotopic ratios originating from before the solar system was formed. Finally, through observations of metal-poor stars we can learn about the composition of stars at an early time, when they have only been enriched by a single or few nucleosynthesis cycles~\cite{frebel18}. It should be noted that astronomical observations can only provide elemental abundances, and the only isotopic information available comes from solar-system and meteoritic samples. 

Using the available observables, it became apparent early on that the $r$ process must proceed through very neutron-rich nuclei. To come to this conclusion one had to look at the abundance distribution of Fig.~\ref{fig:abundances} and make the connection between the different peaks and nuclear magic numbers, as mentioned earlier. In order to get the $r$-process flow far from the valley of stability and into very neutron-rich and short-lived nuclei, the process had to take place in an environment with extreme neutron densities ($\sim 10^{20}$/cm$^3$) and short time scales 
(of the order of seconds). Once the required conditions were known the natural question to ask is  ``which astrophysical environment could host such an extreme event?''. 

While the $r$ process and its general characteristics were introduced more than 60 years ago, a possible host astrophysical site was not unambiguously identified until very recently. Two main candidates were proposed, core collapse supernovae (CCSN) and neutron-star mergers (NSM). Core collapse supernovae were the dominant scenario for many years, but became unfavored when modern simulations showed that a full $r$ process could not be achieved, \textit{e.g.}~\cite{arcones07, fischer10, janka2012,martinez-pinedo2012}. On the other hand, neutron-star mergers seemed more promising due to their natural neutron-rich environment, \textit{e.g.} \cite{goriely2013, just2015}. Initially, the presumed long development time of neutron-star merger systems was difficult to reconcile with observations of $r$-process elements in very old stars~\cite{argast04}. 
However, more recent articles, considering various $r$-process sources within different chemical-evolution models, indicate that neutron-star mergers are fully compatible with the observed abundance patterns in low-metallicity stars, e.g. Refs. ~\cite{matteucci2014,komiya2014,mennekens2014,wehmeyer2015,shen2015,voort2015,ishimaru2015,vangioni2016,cote2018}.
As of today, there is further observational evidence that favors a low-frequency, high-yield scenario for the $r$ process, pointing towards the neutron-star merger picture being correct~\cite{frebel18,hotokezaka2015}. 
Many other possible scenarios have been proposed in the literature as possible hosts for the $r$ process, but will not be discussed here. 

It is important to note that $r$-process abundances can, and most probably do, have contributions from more than one astrophysical site. 
This becomes more apparent when comparing the abundance distributions of metal-poor stars to the solar-system abundances \textit{e.g.} Fig.~11 in~\cite{sneden2008}. 
For elements with atomic number larger than $A \approx$ 56, the abundances from various $r$-process rich stars are in excellent agreement with solar system abundances. 
However, lighter elements do not exhibit the same robustness, presenting significant discrepancies from the solar-system abundance pattern. 
This observation may indicate that multiple astrophysical processes contribute, such as the weak $r$ process~\cite{qian1996, surman2006}, the $i$ process \cite{cowan1977,herwig2011,abate2016,hampel2016,denissenkov2018}, the $\nu$p process  \cite{wajano2006,pruet2006,frohlich2006, wanajo2011}, and the Light Element Primary Process (LEPP)~\cite{arcones2011, montes2007, travaglio2004}.

The $r$-process scene changed completely in 2017 with the first observation of a neutron-star merger event (illustrated in Fig.~\ref{fig:abundances}b) by gravitational and electromagnetic observatories, \textit{e.g.}~\cite{LIGO2017, arcavi2017, cowperthwaite2017, kasliwal2017}. Gravitational waves from GW170817 were detected by the LIGO/VIRGO collaboration revealing the general location of the signal and also the mass of the binary system that produced it~\cite{LIGO2017}.
Numerous telescopes from around the world and in space also observed the same event for several days and weeks~\cite{arcavi2017,cowperthwaite2017, kasliwal2017,goldstein2017,savchenko2017,coulter2017,valenti2017,pian2017,drout2017,tanvir2017}. 
These observations confirmed the predicted ``kilonova'' afterglow~\cite{li1998,metzger2010,kasen17} that is powered by radioactive decays of isotopes of heavy elements. 
The kilonova afterglow associated with GW170817 has been interpreted as
consisting of two components: a blue component that is believed to originate from light $r$-process elements, and that decays away within a few days after the event, and a red component that is interpreted as the result of the radioactive decay of heavy $r$-process elements, in particular lanthanides~\cite{tanaka2013}, and lasts a much longer time than its blue counterpart. Due to the complex atomic structure of lanthanide atoms, the opacity of the stellar environment is much larger, and this results in a wavelength shift towards the red. The effect of the high opacities on the electromagnetic spectrum was predicted in earlier publications~\cite{kasen2013, barnes2013} and was confirmed during the neutron-star merger observation, revealing for the first time, at least one of the sites of $r$-process heavy element production in the universe.

With at least one $r$-process site confirmed, it is now more important than ever to have a good handle on the nuclear physics properties that drive these events, so that we can calculate the final abundances reliably, and also to be able to interpret the plethora of observations from GW170817 and future observations. For this reason, it is critical to understand the sensitivity of $r$-process calculations to nuclear input, and to provide experimental constraints as broadly as possible. 

\subsection{Sensitivity to nuclear input}
\label{subsec:sensitivity}

In general, the astrophysical conditions for the $r$ process can be divided into two broad categories: cold and hot. In a hot $r$ process, the reaction flow proceeds through an equilibrium between neutron-capture reactions and their inverse photodisintegration reactions $(n,\gamma) \Leftrightarrow (\gamma,n)$. Under such conditions, neutron-capture reactions do not affect the flow of matter. The $r$-process path is defined by the neutron-separation energies  $S_n(Z, A) = E_B(Z, A)-E_B(Z, A-1)$, where $E_B(Z,A)$ is the binding energy of a nucleus, and consequently the nuclear masses play a critical role. In this equilibrium scenario, a steady-flow $\beta$ decay and $\beta$-delayed neutron emission occurs, finalizing the abundance pattern back to stability (see \textit{e.g.} the review by Cowan \textit{et al.}~\cite{cowan1991}). This simple picture is not entirely correct, however. At late times, after  the $(n,\gamma) \Leftrightarrow (\gamma,n)$ equilibrium has been broken, neutron-captures start to compete against $\beta$ decay and $\beta$-delayed neutron emission, even photodissociation if the temperature is high enough. Therefore, even in hot $r$-process conditions, where equilibrium is expected to occur, neutron-capture reactions play a critical role. 

On the other hand, in a cold $r$-process scenario, the temperature is not high enough for photodissociation reactions to occur efficiently, and $(n,\gamma) \Leftrightarrow (\gamma,n)$ equilibrium is not reached~\cite{arnould2007}, although re-heating by the radioactive decay may provide such an equilibrium for some of the trajectories at later times~\cite{mendoza-temis2015}.
In this case, neutron-capture reactions play an even more critical role during the full extend of the $r$-process event. 
Here, a steady flow of neutron captures and $\beta$ decays defines the $r$-process reaction path and the final abundance distribution~\cite{arcones2011b,eichler2015}. 

It is clear that even under different and uncertain astrophysical conditions, the general nuclear physics properties that are needed are well defined: nuclear masses, $\beta$-decay half-lives, $\beta$-delayed neutron emission probabilities, and neutron capture reaction rates. On top of these, if the $r$-process flow reaches very heavy nuclei, spontaneous, or neutron-induced fission is possible, and the fission fragments replenish the environment with lighter nuclei, in a circular process known as ``fission recycling''~\cite{goriely2011}. In this case, fission properties become important as well, such as fission barriers and fragment distributions~\cite{goriely2013,eichler2015,goriely2011,beun2008, giuliani2018,goriely2015b}.  

Running $r$-process network calculations requires the use of all aforementioned properties for the $\approx$ 5000 nuclei that participate in the $r$ process from the valley of stability to the neutron drip line. The majority of the involved nuclei are not accessible for experiments in current facilities and $r$-process models rely on theoretical calculations to predict the necessary nuclear properties. It therefore becomes of paramount importance to test the validity of these theoretical calculations, where experiments can reach, and to improve their predictive power if at all possible. 

\begin{figure}[tb]
\begin{center}
\includegraphics[clip,width=0.9\columnwidth]{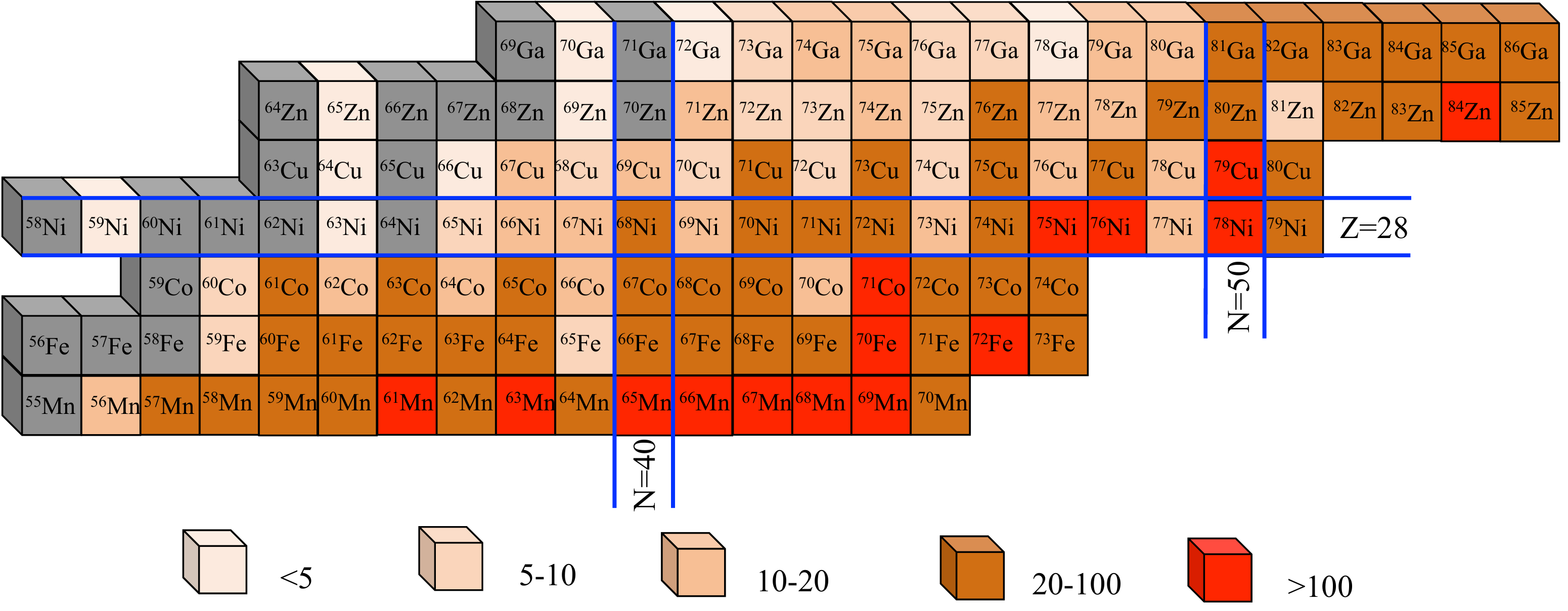}
\caption {Variation in the theoretical prediction of neutron-capture reaction rates around mass 70. The ($n,\gamma$) rates were calculated with the reaction code {\sf TALYS}~\cite{TALYS,koning12} varying the level density and $\gamma$-strength function as listed in Tab.~I of Liddick \textit{et al}.~\cite{liddick2016}.}
\label{fig:ng_variation}
\end{center}
\end{figure}

The present review article focuses on the experimental aspects of one of these nuclear properties, namely neutron capture reactions. The direct measurement of a neutron-capture on a short-lived isotope is extremely challenging due to the fact that none of the reactants can be made as a target. For this reason, to date, no experimental $(n,\gamma$) reaction cross section data exists along the $r$-process path, and astrophysical calculations have to rely on theoretical predictions. A description of the theoretical models used to describe neutron-capture reactions is presented in Sec.~\ref{sec:input}. Here we focus on the impact on $r$-process calculations. Theoretical models that predict $(n,\gamma$) reaction cross sections are well constrained along the valley of stability, and can reproduce experimental data roughly within a factor of 2~\cite{beard2014}. However, moving away from stability, the predictions of these calculations diverge, reaching variations of factors of 100 or more just a few neutrons away from the last stable isotope~\cite{liddick2016, nikas2018}. This can be seen in Fig.~\ref{fig:ng_variation}, which shows part of the chart of nuclei, where the color code represents the variation in the predictions of theoretical calculations. It is clear in Fig.~\ref{fig:ng_variation} that the variation in the theoretical predictions increases as we move away from stable isotopes. 

\begin{figure}[tb]
\begin{center}
\includegraphics[clip,width=0.7\columnwidth]{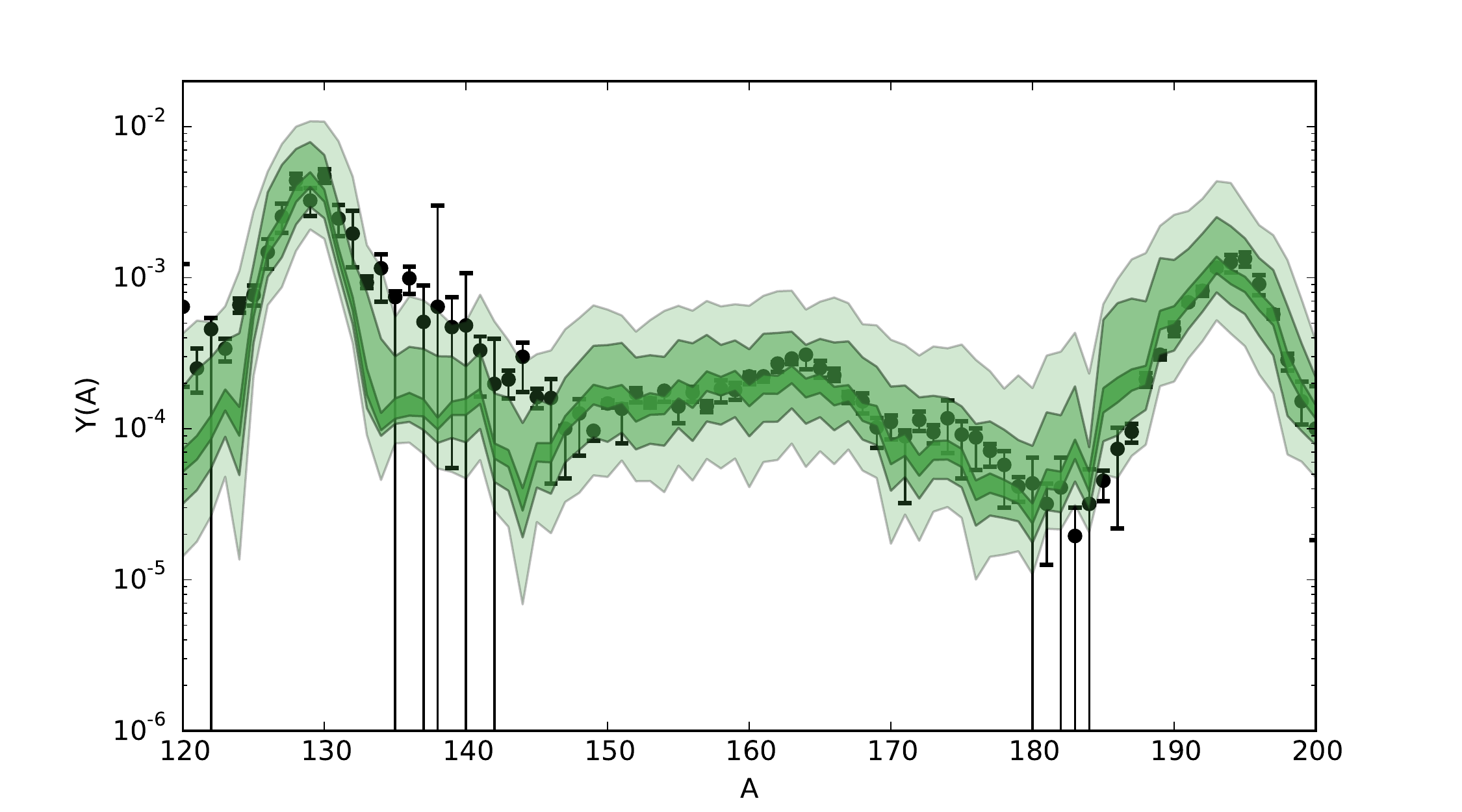}
\caption {Sensitivity of $r$-process abundances to the Monte Carlo variation of neutron-capture rates. The light-color band corresponds to a variation within a factor 100, the medium-color band to a factor of 10 and the dark-color band to a factor of 2. The calculations were done for a neutron-star merger trajectory from~\cite{bauswein2013}. Figure taken from Ref.~\cite{liddick2016}.}
\label{fig:abundance_sensitivity}
\end{center}
\end{figure}

The variation in the theoretical predictions of neutron-capture reaction rates can be used to estimate the impact of these uncertainties on $r$-process calculations. The way to investigate this impact is by performing ``sensitivity studies''. Various techniques exist for identifying the sensitivity of an observable to a particular input. For example through a systematic variation of the input (\textit{e.g.}~\cite{surman2014}), through a Monte Carlo approach (\textit{e.g.}~\cite{mumpower2016}) or through the selection of different input models (\textit{e.g.}~\cite{nikas2018}). A comprehensive study of the sensitivity of the final $r$-process abundance distribution for different astrophysical conditions, and different nuclear physics parameters was investigated in the work of Mumpower \textit{et al.}~\cite{mumpower2016}. In particular for a neutron-star merger scenario, the sensitivity to neutron-capture reactions is shown in Fig.~\ref{fig:abundance_sensitivity}, taken from Ref.~\cite{liddick2016}. In this study, neutron-capture reactions were varied by a factor of 100 (light-color band), by a factor of 10 (medium-color band), and by a factor of 2 (dark-color band). It can be seen in Fig.~\ref{fig:abundance_sensitivity}, that an uncertainty of a factor 100, or even 10, dilutes the abundance distribution produced by the model, and limits our ability to perform meaningful comparisons and to draw conclusions about the applicability of the model or astrophysical conditions.

\begin{figure}[tb]
\begin{center}
\includegraphics[clip,width=1.\columnwidth]{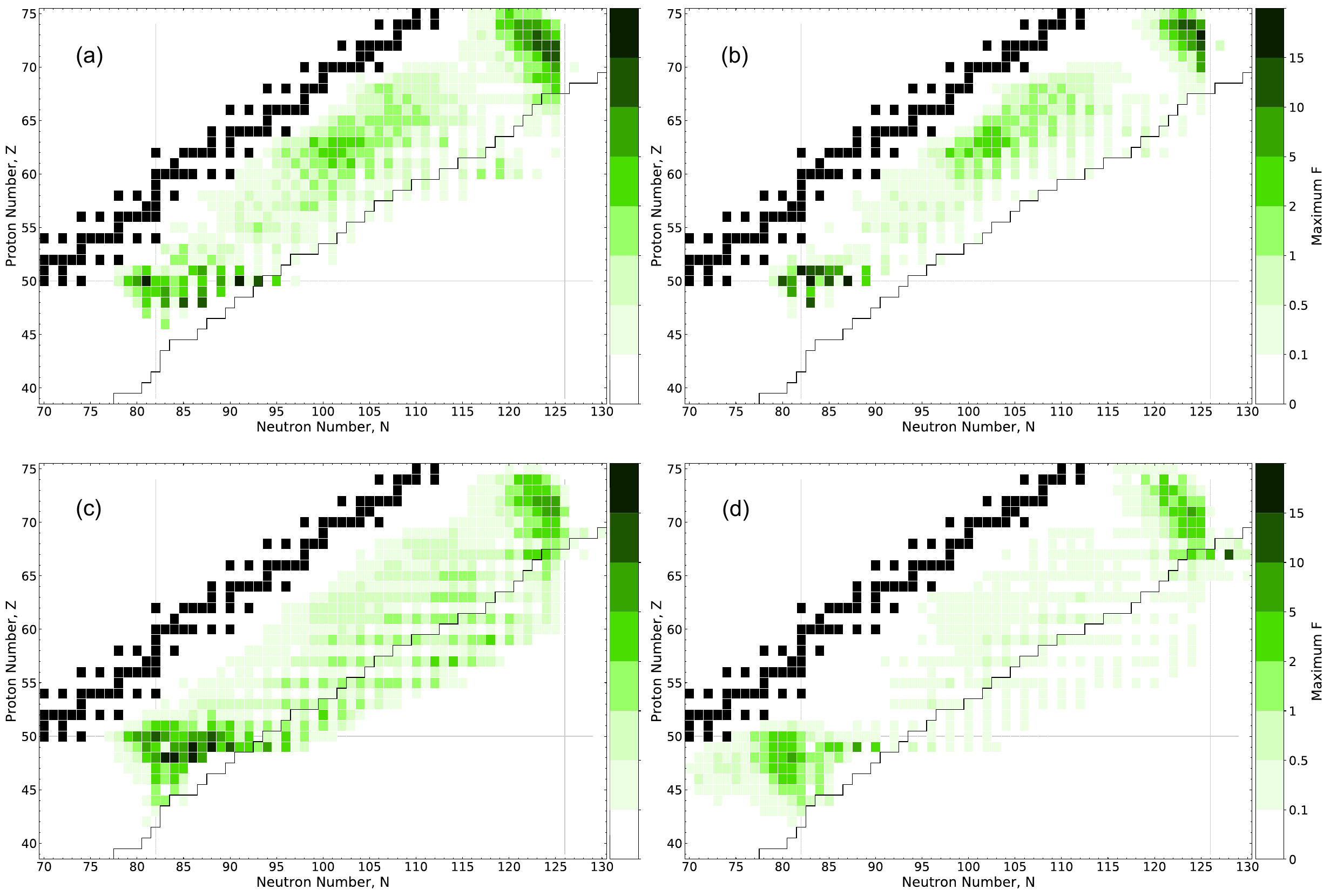}
\caption {Important neutron capture rates in four astrophysical environments: (a) low-entropy hot wind, (b) high-entropy hot wind, (c) cold wind and (d) neutron-star merger. Stable isotopes are depicted as black squares. Estimated neutron-rich accessibility limit is shown by a black line for FRIB with intensity of 10$^{−4}$
particles per second. Figure adapted from Mumpower \textit{et al.}~\cite{mumpower2016}, courtesy of Mumpower.
}
\label{fig:ng_rate_sensitivity}
\end{center}
\end{figure}

Together with showing the impact on the final $r$-process observable, sensitivity studies also serve a second important role, which is to guide experimental studies. Even when future facilities provide access to the majority of $r$-process nuclei, it is not realistic to expect experimental information for all 5000 participating nuclei in a short time. It is therefore critical to identify the nuclei and nuclear properties that have an impact on the final observable, such as the abundance distributions or the kilonova emission. Such sensitivity studies exist for the $r$ process and can provide a list of important isotopes and properties to guide experiments, \textit{e.g.}~\cite{mumpower2016,surman2014,mumpower2012,martin2016}. 
This can be done by systematically and consistently varying each property, \textit{e.g.}~the neutron-capture reaction rate on each isotope, and observing the change in the final abundance distribution, compared to a baseline calculation. The impact can be local, affecting only neighboring nuclei, or it can be global. 
For the case of neutron-capture reactions, on top of the overall impact of the reaction rate variations (as shown in Fig.~\ref{fig:abundance_sensitivity}), the work of Mumpower \textit{et al.}~\cite{mumpower2016} also provided a list of important $(n,\gamma$) reactions. 
This list shows that neutron-capture reactions on nuclei around magic numbers are very important, but also some of the intermediate-mass nuclei between $N=82$ and $N=126$, see Fig.~\ref{fig:ng_rate_sensitivity} (from~\cite{mumpower2016}). 
Unsurprisingly, this study also shows that for hot astrophysical conditions the important neutron-captures are closer to stability, while for cold conditions, neutron-captures matter for more neutron-rich isotopes.  

Neutron-capture reactions play a crucial role in $r$-process nucleosynthesis, under all possible astrophysical conditions. The large theoretical uncertainties come mainly from the fact that no experimental data exist far from stability. For this reason, the development of indirect techniques to provide experimental constraints for these neutron-capture reactions are critical, and this is the focus of the present article.

\section{Input for cross-section and reaction-rate calculations}
\label{sec:input}
To calculate astrophysical Maxwellian-averaged reaction rates, one usually assumes
thermodynamic equilibrium for both the target nucleus and the projectile, 
thus obeying Maxwell-Boltzmann distributions for a given temperature $T$ at the
specific stellar environment. 
Because of the temperature at the astrophysical site, 
the target nucleus might well be in an excited state, which will contribute
to the rate. 
Specifically, the $(n,\gamma)$ reaction rate $N_A \left< \sigma v \right>_{n\gamma}$
is found from integrating 
the cross section over a Maxwell-Boltzmann distribution of energies 
$E$ at a given $T$ (e.g., Ref.~\cite{arnould2007}):
\begin{equation}
N_A \left< \sigma v \right>_{n\gamma} (T) = \left(\frac{8}{\pi\tilde{m}}\right)^{1/2} \frac{N_A}{(k_B T)^{3/2}G_t(T)}\int_0^\infty \sum_\mu \frac{2J_t^\mu +1}{2J_t^0 +1} \sigma_{n\gamma}^\mu (E) E \exp \left[- \frac{E+E^\mu_x}{k_B T} \right] \mathrm{d}E.
\label{eq:rateeq}
\end{equation}
Here,  $N_A$ is Avogadro's number, $\tilde{m}$ is the reduced mass, $\mu$ denotes
an excited state in the target, $J_t^0$ and $J_t^\mu$ are the spin of the ground state and excited states of the target, respectively, 
$E$ is the relative energy of the neutron and target, 
$E_x^\mu$ is the excitation energy for
the state $\mu$, and $k_B$ is Boltzmann's constant. 
Furthermore, the normalized, temperature-dependent partition function is given by $G(T) = \sum_\mu (2J_t^\mu+1)/(2J_t^0 +1) \exp{(-E^\mu_x/k_B T)}$. 
It is seen from Eq.~(\ref{eq:rateeq}) that the reaction rate 
is proportional to the cross section $\sigma_{n\gamma}$. 
Hence, to estimate a correct reaction rate, it is crucial to 
determine this cross section. 

In open-access reaction-rate libraries for $r$-process nucleosynthesis such as JINA REACLIB~\cite{JINA-REACLIB}, BRUSLIB~\cite{BRUSLIB}, and the NON-SMOKER database~\cite{NONSMOKER}, the reaction rates are deduced from cross-section calculations based on the Hauser-Feshbach formalism~\cite{hauser}, which assumes the compound-nucleus picture of Niels Bohr~\cite{bohr1936}. 
As the compound-nucleus concept is the foundation for these cross-section calculations, the main assumptions are outlined here.  

\subsection{The compound nucleus picture: Hauser-Feshbach theory}
For the derivation of the Hauser-Feshbach model, it is assumed that the compound nucleus is created in such a high excitation energy that there is a high number of accessible levels\footnote{"High" is of course a relative term -- one rule of thumb that is often applied is that there should be at least 10 levels within the applied excitation-energy bin~\cite{Rauscher1997}.}.
Further, the corresponding wave functions of the accessible levels are assumed to have a random phase, which means that all interference terms will cancel out when phase averages are performed.
Then, one assigns two separate stages for the  neutron-capture reaction: (\textit{i}) a compound nucleus is formed; (\textit{ii}) the compound nucleus decays by emission of $\gamma$ rays. 
A crucial point here is that the second step is believed to be completely independent of the first step; it is usually said that the compound nucleus``forgets'' the way it was formed and its subsequent decay is fully governed by its statistical properties~\cite{bohr1939,blatt_weisskopf1952,bohr_mottelson1969}.
This condition is believed to be fulfilled if the average level spacing $D$ is sufficiently small, so that the coupling matrix element $V$ can be considered much larger than the level spacing, $|V| \gg D$, and also that the mixing time is short enough for a strong (``complete'') mixing to occur~\cite{Mekjian1973}.
Put in a different way, the compound-nucleus picture is assumed to be valid for slow reactions, where the incident particle remains inside the nucleus and collides with many of the constituent nucleons, and for each collision the incident particle is gradually losing its ``memory'' of the entrance channel~\cite{Ericson1960}. 
Then, the Hauser-Feshbach theory can be applied to describe the radiative neutron-capture cross section. 
If the decay is not completely statistical, \textit{i.e.}, the two steps are not completely independent of each other, a width fluctuation correction factor can be introduced to account for a possible correlation between the entrance and exit channels.  
We refer the reader to Ref.~\cite{hilaire2003} for a discussion of approaches for this width fluctuation correction. 

Adapting the notation in Ref.~\cite{rauscher2000}, for a target nucleus $t$ in a state $\mu$ with spin $J_t^\mu$ and parity $\pi_t^\mu$ hit by an incoming neutron $n$ with spin $J_n$, leading to the creation of a residual nucleus $r$ at a state $\nu$ with spin $J_r^\nu$ and parity $\pi^\nu$ followed by $\gamma$ emission, the radiative neutron-capture cross section is given by
\begin{equation}
\sigma^{\mu\nu}_{n\gamma}(E_{tn}) =  \frac{\pi \hbar^2}{2\tilde{m}_{tn}E_{tn}} \frac{1}{(2J_t^\mu + 1)(2J_n+1)} \sum_{J,\pi} (2J+1) \frac{\mathcal{T}^\mu_n(E,J,\pi,E^\mu_t,J^\mu_t,\pi^\mu_t)\mathcal{T}^\nu_\gamma(E,J,\pi,E_{r}^\nu,J_{r}^\nu,\pi_{r}^\nu)}{\mathcal{T}_{\mathrm{tot}}(E,J,\pi)}.
\label{eq:cross}
\end{equation}
Here, $E_{tn}$ is the center-of-mass energy of the target plus neutron, $\tilde{m}_{tn}$ is the reduced mass, and $E,J,\pi$ are the excitation energy, spin and parity of the compound nucleus. 
Moreover, $\mathcal{T}^\mu_n$ is the transmission coefficient for the neutron and $\mathcal{T}^\nu_\gamma$ for the $\gamma$ ray. 
The total transmission coefficient, $\mathcal{T}_{\mathrm{tot}}$, describes the transmission into all possible bound and unbound states $\nu$ in all energetically accessible exit channels, including the entrance channel. 
Further, as $\mathcal{T}_n \sim \mathcal{T}_{\mathrm{tot}}$, it is easily seen from Eq.~(\ref{eq:cross}) that $\sigma_{n\gamma} \sim \mathcal{T}_\gamma$. 
However, one should keep in mind that if a strong isovector component is present in the imaginary part of the neutron optical-model potential, it could have a drastic impact on ($n,\gamma$) reaction rates for very neutron-rich nuclei~\cite{goriely-delaroche2007}.
To obtain the $(n,\gamma)$ cross section for all possible states $\mu,\nu$, one must take into account that (\textit{i}) the target nucleus might be in an excited state due to thermal excitations caused by the astrophysical plasma~\cite{arnould1972}, and (\textit{ii}) there are many levels in the residual nucleus that contribute to the total transmission coefficient for the $\gamma$ channel.
We focus here on the latter part, where the total $\gamma$-transmission coefficient is given by
\begin{equation}
\mathcal{T}_\gamma(E,J,\pi) =  \sum_{\nu = 0}^{\nu_r} \mathcal{T}^\nu_\gamma(E,J,\pi,E_{r}^\nu,J_{r}^\nu,\pi_{r}^\nu) + \int_{E_r^{\nu_r}}^{E} \sum_{J_r,\pi_r}\mathcal{T}^\nu_\gamma(E,J,\pi,E_{r}^\nu,J_{r}^\nu,\pi_{r}^\nu) \cdot \rho(E_r,J_r,\pi_r)\mathrm{d}E_r.
\end{equation}
The first sum on the right-hand side runs over all experimentally known discrete levels, while the integral and sum run over the product of the nuclear level density $\rho$ and the $\gamma$-ray transmission coefficient $\mathcal{T}_\gamma$, which is directly proportional to the $\gamma$-ray strength function $f(E_\gamma)$ as will be discussed in Sec.~\ref{subsec:gsf}. 
For most nuclei involved in the $r$ process, few or no discrete levels are known except the ground state, and so the models for $f(E_\gamma)$ and $\rho$ become increasingly important.

\begin{figure}[tb]
\begin{center}
\includegraphics[clip,width=0.6\columnwidth]{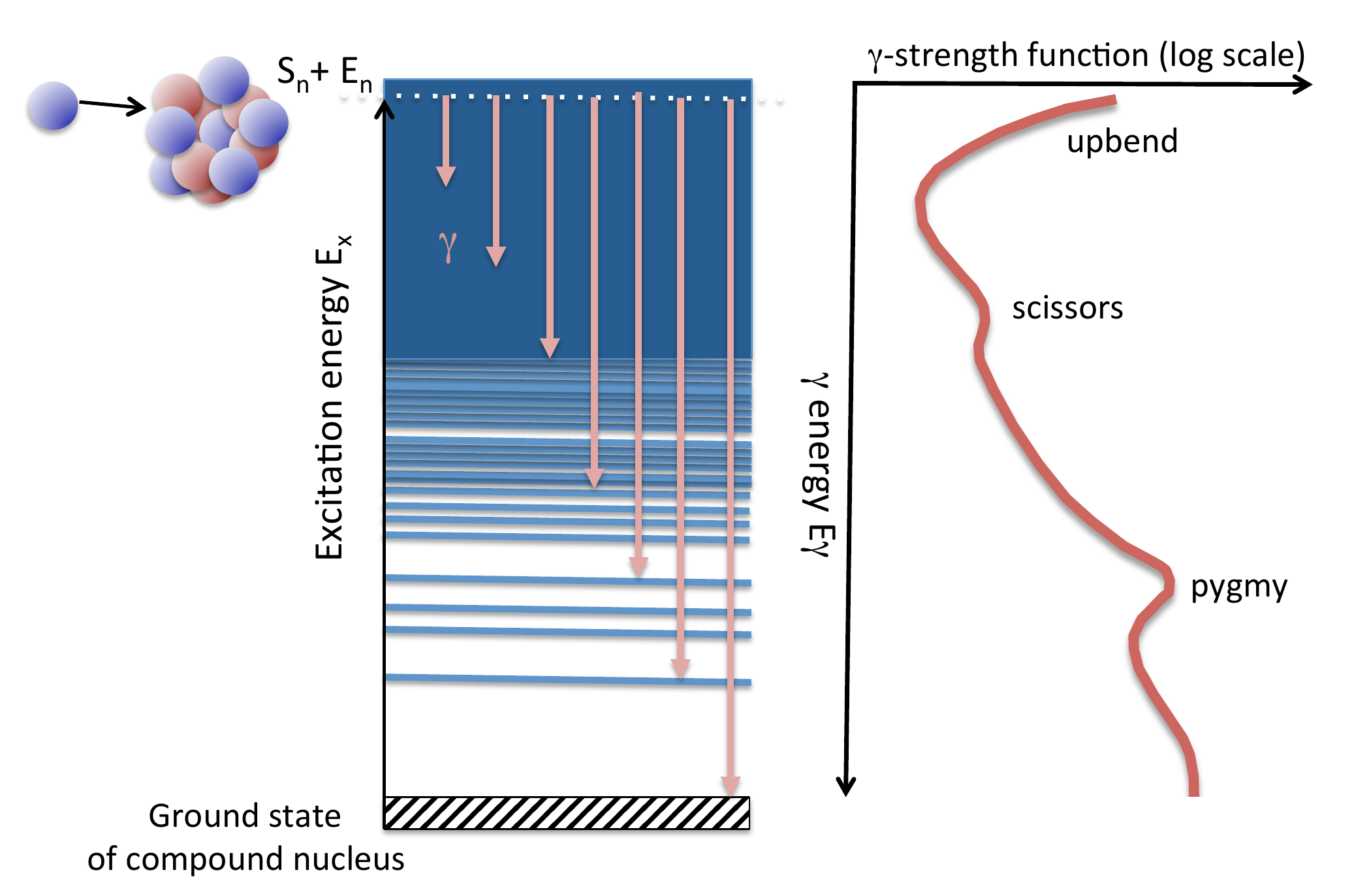}
\caption{(Color online) Schematic representation of radiative neutron capture. A nucleus captures a neutron and the residual (compound) nucleus is excited to a level given by the neutron separation energy and the kinetic energy of the incoming neutron (neglecting recoil). The probability for the neutron to stick inside the compound nucleus, and shed off the excess energy by emitting $\gamma$ rays, is proportional to the accessible levels below $S_n+ E_n$ and the $\gamma$-ray strength function for the transition energy $E_\gamma$. Structures  marked "upbend", "scissors", and "pygmy" represent enhanced $\gamma$-decay strength relative to the tail of the giant, electric dipole resonance.
}
\label{fig:ncap}
\end{center}
\end{figure}
The influence of the level density and $\gamma$ strength function on neutron-capture rates is illustrated in Fig.~\ref{fig:ncap}.
Here, a nucleus consisting of $A$ nucleons captures a neutron, populating a highly excited state in the $A+1$ compound nucleus. 
From perturbation theory~\cite{dirac,fermi}, it follows that the decay rate is proportional to the level density at the final excitation energy, and the square of the matrix element of the initial and final state.
Intuitively, the probability for the compound nucleus to de-excite to the ground state through emission of one or more $\gamma$ rays strongly depends on the number of accessible levels as well as the $\gamma$-ray strength function.
These two crucial quantities will be discussed in some detail in the following. 
For an introduction to neutron optical-model potentials we refer the reader to Hodgson~\cite{Hodgson1984} and Koning and Delaroche~\cite{koning2003}.

\subsection{Nuclear level density}
\label{subsec:nld}
The nuclear level density is a measure of the available quantum levels at a given excitation energy, spin, and parity, and is defined as 
\begin{equation}
\rho(E_x,J,\pi) = \Delta N(E_x,J,\pi)/\Delta E_x,
\end{equation}
where $\Delta N(E_x,J,\pi)$ is the number of levels within the energy bin $\Delta E_x$. 
The level spacing is $D(E_x,J,\pi)$ is simply the inverse of the level density, $D(E_x,J,\pi) = 1/\rho(E_x,J,\pi)$.
In contrast to the \textit{cumulative} number of levels, the level density is dependent on the bin width, but as the level density gets high, this dependence is not as significant as at lower excitation energies where only one or a few levels are contained within the bin.
Moreover, the total level density is given by 
\begin{equation}
\rho(E_x) = \sum_{J,\pi}\rho(E_x,J,\pi),
\end{equation}
and it is common to assume that spin and parity are described by uncorrelated functions, so that 
\begin{equation}
\rho(E_x,J,\pi) = \rho(E_x) g(E_x,J) \mathcal{F}(E_x,\pi),
\end{equation}
where $g$ is the spin distribution and $\mathcal{F}$ the parity distribution.
Following Ericson~\cite{Ericson1960,Ericson1959}, the spin distribution can be derived within the statistical model assuming random coupling of angular momenta, leading to 
\begin{equation}
    g(E_x,J) \simeq \frac{(2J+1)}{2\sqrt{2\pi}\sigma^3} \exp{\frac{-J(J+1)}{2\sigma^2}}.
\end{equation}
Some remarks are in order here. First, one should be aware that the above expression is derived assuming that many particles and holes are excited. 
This is certainly not the case at low excitation energy. Second, for very large values of $J$ the expression is not valid~\cite{Ericson1960}. 

For the parity distribution, it is usually assumed that there is an equal amount of negative and positive parity states; this is approximately true if there is a small admixture of negative-parity states in a region dominated by positive parity states or vice versa~\cite{Ericson1960}. 
Phenomenological models for explicit inclusion of an asymmetric parity distribution have been developed, for example by Al-Quraishi \textit{et al.}~\cite{Al-Quraishi2003}. 
Several authors discuss possible, significant deviations from a symmetric parity distribution, such as Alhassid \textit{et al.}~\cite{alhassid2000} and \"{O}zen \textit{et al.}~\cite{ozen2007}. 
Further, the impact of including parity asymmetry in Hauser-Feshbach calculations has been explored by Mocelj \textit{et al.}~\cite{mocelj2007} and  Loens \textit{et al.}~\cite{loens2008}.
In the latter work, it was concluded that the effect of using parity-dependent level densities was within a factor of $\approx 2$ for Sn isotopes. 
On the experimental side, measurements of $J^\pi=2^+$ and $J^\pi=2^-$ level densities in $^{58}$Ni and $^{90}$Zr by Kalmykov \textit{et al.}~\cite{kalmykov2007} revealed no significant parity asymmetry in the excitation-energy range $E_x \approx 8-14$ MeV.
Furthermore, Agvaanluvsan \textit{et al.}~\cite{agvaanluvsan2003} studied proton-capture reactions on $^{44}$Ca, $^{48}$Ti and $^{56}$Fe target nuclei, and found a rather weak parity dependence on the populated $J=1/2^{\pm}$ and $J=3/2^\pm$ levels in the compound nuclei $^{45}$Sc, $^{49}$V and $^{57}$Co. 
Therefore, it seems like having an unequal amount of positive and negative parity levels is not a major issue; that said, it can be very important for nuclei close to the neutron dripline, where only a few resonance levels might be available (see also Sec.~\ref{sec:direct-preeq}). 

The first theoretical attempt to describe nuclear level densities was done by Bethe in 1936~\cite{bethe1936}. 
In his pioneering work, Bethe described the nucleus as a gas of non-interacting fermions moving freely in equally spaced single-particle orbits. The level density was obtained by the inverse Laplace transformation of the partition function determined from Fermi statistics. Bethe's original results yielded a level density function
\begin{equation}
\rho(E_x) = \frac {\sqrt{\pi}}{12} \frac{{\rm exp} (2\sqrt{aE_x})}{a^{1/4} E_x^{5/4}},
\label{eq:bethe}
\end{equation}
for an excitation energy $E_x$, and where $a$ is the level-density parameter given by 
\begin{equation}
a=\frac{\pi}{6}(g_{\rm p} + g_{\rm n}).
\label{eq:levdenspar}
\end{equation}
The terms $g_{\rm p}$ and $g_{\rm n}$ are the single-particle level density parameters for protons and neutrons, respectively, which are expected to be proportional to the mass number $A$. 
In fact, Bethe's consideration of the nucleus to be a Fermi gas of free protons and neutrons confined to the nuclear volume gives $a = \alpha A$, where $\alpha$ has been found to be about $1/8 - 1/10$ by fitting to experimental data. 

Refined versions of the original Fermi-gas formula attempt to take into account pairing correlations, collective phenomena and shell effects by employing free parameters that are adjusted to fit experimental data on level spacings obtained from neutron and/or proton resonance experiments. Gilbert and Cameron~\cite{Gilbert1965} proposed the following level-density formula in 1965: 
\begin{equation}
\rho(U) = \frac {\sqrt{\pi}}{12} \frac{{\rm exp} (2\sqrt{aU})}{a^{1/4} U^{5/4}} \frac{1}{\sqrt{2\pi}\sigma}.
\label{eq:G&C}
\end{equation}
Here, $U$ is the shifted excitation energy, $U = E_x - \Delta_{\rm p} - \Delta_{\rm n}$, where $\Delta_{\rm p}$ and $\Delta_{\rm n}$ are the pairing energy for protons and neutrons, respectively. The spin cutoff parameter $\sigma$ is given by 
\begin{equation}
\sigma^2 = g\langle m^2 \rangle T,
\label{eq:spincutoff}
\end{equation}
where $g = g_{\rm p} + g_{\rm n}$ relate to the level density parameter as in Eq.~(\ref{eq:levdenspar}), $\langle m^2 \rangle$ is the mean-square magnetic quantum number for single-particle states, and the temperature $T$ is often approximated by 
\begin{equation}
T \approx \sqrt{U/a}.
\label{eq:temp1}
\end{equation}

Another approach for calculating the level density is the constant-temperature (CT) model proposed by Ericson~\cite{Ericson1959}:
\begin{equation}
\rho(E_x) = \frac{1}{T_{\rm CT}}{\exp}\left[(E_x-E_0)/T_{\rm CT}\right],
\label{eq:consttemp}
\end{equation}
where $E_x$ is the excitation energy, and the free parameters $T_{\rm CT}$ and $E_0$ are connected to a constant nuclear temperature (in contrast to Eq.~(\ref{eq:temp1})) and an energy shift, respectively.
The much-used composite formula of Gilbert and Cameron~\cite{Gilbert1965} is basically a combination of the CT model and the Fermi-gas model, with parameters ensuring a smooth connection between the two models. 
Other more or less phenomenological models, such as the Generalized superfluid model of Ignatyuk \textit{et al.}~\cite{ignatyuk1979,ignatyuk1993}, have also been developed; see, for example, Ref.~\cite{RIPL3} for an overview. 
It is also very interesting to note the work of Weidenm\"{u}ller~\cite{weidenmuller1964} and the recent Letter of P\'{a}lffy and Weidenm\"{u}ller~\cite{palffy2012}; the former addresses the shortcomings of phenomenological models especially at high excitation energies, and the need to consider an ``effective'' level density representing levels that are actually populated in a given reaction; the latter presents a new method to calculate level densities within a constant-spacing model that should give reliable results even at very high excitation energies (several hundreds of MeV). 

Although the above-mentioned semi-empirical expressions give reasonable agreement with experimental data on, \textit{e.g.}, neutron resonance spacings, they are not able to describe fine structures in the level density. 
Also, any extrapolation to nuclei far from the valley of stability, where little or no experimental data are known, would be highly uncertain. 
In order to have a predictive power, level densities should ideally be calculated from microscopic models based on first principles and fundamental interactions. 

\begin{figure}[tb]
\begin{center}
\includegraphics[clip,width=1.\columnwidth]{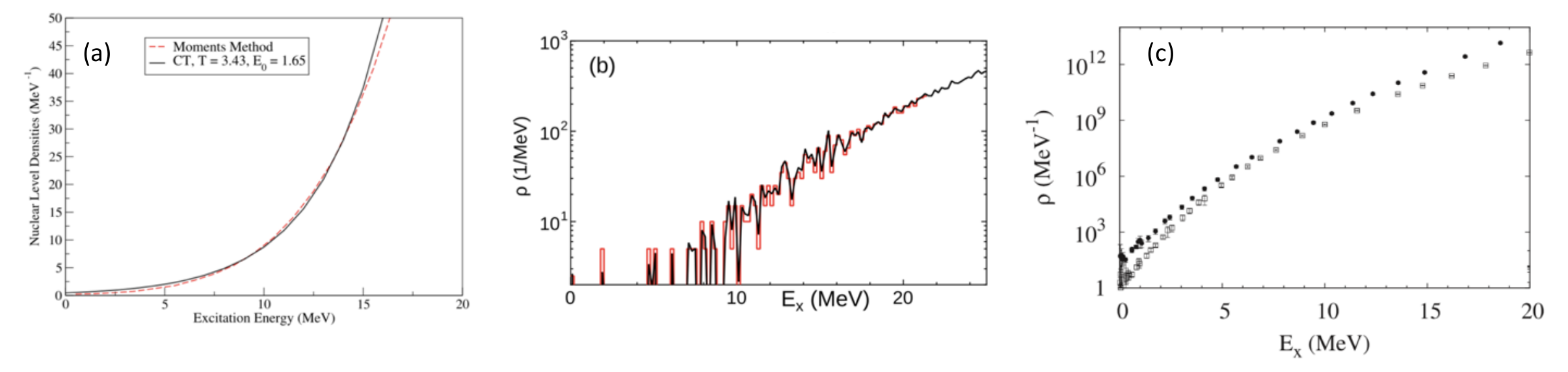}
\caption{(Color online) Microscopic calculations of level densities for (a) $^{24}$Mg, figure from Zelevinsky, Karampagia and Berlaga~\cite{zelevinsky2018}; (b) $^{28}$Si, figure from Shimizu \textit{et al.}~\cite{shimizu2016}; (c) $^{162}$Dy (solid circles) and $^{148}$Sm (open squares), figure from Alhassid \textit{et al.}~\cite{alhassid2016}.
}
\label{fig:shellmodel_nld}
\end{center}
\end{figure}

For a detailed, microscopic description of the nuclear level density, one should solve the exact many-body eigenvalue problem $\hat{H}\left|\Psi\right> = E_x  \left|\Psi\right>$;
however, this has turned out to be a tremendous challenge for mid-mass and heavy nuclei as the dimension of the problem grows rapidly with the number of nucleons. 
For example, within the configuration-interaction shell model, the required model space quickly grows to many orders of magnitude larger than what can be handled with conventional diagonalization methods. 
It is therefore of great importance to introduce methods where level density can be calculated approximately without losing desired microscopic details (see Fig.~\ref{fig:shellmodel_nld}). 

One such method is the shell-model Monte Carlo approach as applied by Alhassid {\it et al}.~\cite{alhassid2008,ozen2015,alhassid2015,alhassid2016}. Here, thermal averages are taken over all possible states of a given nucleus, and two-body correlations are fully taken into account within the model space, see Fig.~\ref{fig:shellmodel_nld}c. 
These calculations are applicable even for rare-earth nuclei.
The drawback is that they are quite computationally costly, and due to the application of canonical-ensemble theory, the excitation energy is not sharp but represents a rather broad distribution for a given temperature, in contrast to experiment. 
Other shell-model approaches have recently appeared in the literature, such as the Moments Method developed by Zelevinsky and coworkers~\cite{horoi2003,horoi2004,horoi2007,senkov2016,zelevinsky2018}, which has so far been applied for lighter nuclei from $^{24}$Mg (Fig.~\ref{fig:shellmodel_nld}a) to $^{64}$Ge. 
Also, the stochastic estimation based on a shifted Krylov-subspace method~\cite{shimizu2016} has recently been put forward. This method seems very promising, although so far it has only been applied to rather light nuclei ($^{28}$Si, see Fig.~\ref{fig:shellmodel_nld}b, and $^{56}$Ni).

Another statistical approach, starting from mean-field theory, is presented by Demetriou and Goriely~\cite{demetriou2001}. 
Here, a global, microscopic prescription of the level density is derived based on the Hartree-Fock-BCS (HFBCS) ground-state  properties (single-particle level scheme and pairing force). 
A combined Hartree-Fock-Bogolyubov and combinatorial model has recently been developed by Goriely, Hilaire and collaborators~\cite{goriely2008}, where the combinatorial predictions provide the non-statistical limit that by definition cannot be described by any statistical approach.
Another advantage of this combined model is that the parity dependence of the level density is obtained in addition to the energy and spin dependence.
Also, a temperature-dependent Hartree-Fock-Bogolyubov-plus-combinatorial method is now available~\cite{hilaire2012}. 

A completely combinatorial level-density model has recently been proposed~\cite{uhrenholt2013}, 
based on the folded-Yukawa single-particle potential and treating pairing and collective states explicitly. 
In particular, the pair gaps for all states are obtained by solving the BCS equations for all individual many-body configurations, demonstrating that there is a substantial pairing effect even at rather high excitation energies (close to the neutron separation energies in even-even rare earth nuclei). 

When it comes to measuring level densities experimentally, several methods have been developed and applied in various excitation-energy regions. 
At low excitation energies it is possible to determine the level density by counting the discrete levels from databases such as the Table of Isotopes~\cite{TOI} and Evaluated Nuclear Structure Data File~\cite{ENSDF}. 
However, this method quickly becomes unreliable when the level density reaches about $\sim 100$ levels per MeV. 

At the neutron (proton) separation energy, the numbers of s- and p-wave neutron (proton) resonances within the energy range of the incoming neutron (proton) reveal the level spacing between the states reached in the capture reaction.
Historically, s-wave neutron resonances have been extensively studied~\cite{RIPL3,mughabghab}, while p-wave neutron resonances are much more scarce. 
Proton s- and p-wave resonances are available for a few cases~\cite{agvaanluvsan2003,Sukhoruchkin2005}.
Neutron (proton) resonances provide parity- and spin-projected level density at and slightly above the neutron (proton) separation energy. 
Obviously, the method is not applicable at other energies, and corrections are needed for missing resonances or contaminating resonances with higher $\ell$ values.

Another appreciable method is the Hauser-Feshbach modelling of evaporation spectra~\cite{vonach1983,wallner1995,Voinov&Grimes,ramirez2015}. 
This method can be applied to the quasi-continuum and provides level density functions, potentially also above the neutron-separation energy. 
However, care has to be taken so that the underlying assumptions of the Hauser-Feshbach theory are met by choosing appropriate reactions, beam energies, ejectile angles and so on. 
Also, {\it a priori}\ knowledge of particle transmission coefficients is needed, and the resulting level density function must be normalized in absolute value to known, discrete levels.

In the so-called Ericson regime (excitation energies $3-4$ MeV above the neutron separation energy for heavy nuclei), the level density can be determined from a fluctuation analysis of total neutron cross sections in the continuum region~\cite{ericsonfluct,Grimes,Salas-Bacci2004}. 
This method relies on specific assumptions concerning how level density can be extracted from cross-section fluctuations. 
In particular, the continuum region must be considered for this analysis, as it is necessary to have the average, total level width $\left< \Gamma \right>$ to be much larger than the average level spacing $\left< D \right>$, which is the case at high excitation energies when many emission channels are open. 

Fluctuations of giant-resonance cross sections have recently been a source of information for spin/parity dependent level densities at high excitation energy and over a rather wide range of excitation energies, typically several MeV~\cite{kalmykov2007,hansen1990,bassauer2016}. 
In contrast to the Ericson-fluctuation analysis, the levels are required to \textit{not} overlap, having $\left< \Gamma \right> < \left< D \right>  < \delta E$ where $\delta E$ is the experimental energy resolution. 
The autocorrelation function, measuring the spectral fluctuations with respect to a local mean value, is proportional to the average level spacing. Thus, by carefully choosing experimental conditions enhancing population of a given spin and parity, an essentially model-independent determination of the level density for that spin and parity can be extracted. 
The method is dependent on the validity of the Wigner~\cite{wigner1967} distribution for the nearest-neighbor level spacing and the Porter-Thomas distribution~\cite{porter-thomas1956} of partial decay widths~\cite{kalmykov2007}.

\subsection{Gamma-ray transmission coefficient and $\gamma$ strength function}

\label{subsec:gsf}
The $\gamma$-ray transmission coefficient $\mathcal{T}_\gamma$ represents, following Blatt and Weisskopf~\cite{blatt_weisskopf1952}, the escape probability for a $\gamma$ ray stuck inside the volume of the nucleus. 
The probability for transmission is in general much smaller than the probability for reflection, \textit{i.e.}, the $\gamma$ ray must try to escape the nucleus many times before it will be emitted.  
The $\gamma$-ray transmission coefficient characterizes the \textit{average} electromagnetic properties of excited states; thus, they are closely connected to radiative decay and photo-absorption processes. 
For a given electromagnetic character $X$ (being electric, $E$, or magnetic, $M$) and with multipolarity $L$, the $\gamma$-ray transmission coefficient ${\mathcal T}_{XL}(E_{\gamma})$ as a function of $E_\gamma$ is proportional to the \textit{$\gamma$-ray strength function} $f_{XL}(E_{\gamma})$ through the relation
\begin{equation}
{\mathcal T}_{XL}(E_{\gamma}) = 2\pi E_{\gamma}^{(2L+1)} f_{XL}(E_{\gamma}).
\label{eq:Ttof}
\end{equation}
Gamma-ray strength functions are also called radiative strength functions and photon strength functions in the literature. 
The concept of strength functions was introduced by Wigner during the development of R-matrix theory for nuclear resonances~\cite{Wigner}. 

The conventional definition of a model-independent $\gamma$-ray strength function was presented by Bartholomew {\it et al}.~\cite{bartholomew1972}:
\begin{equation}
\stackrel{\leftarrow}{f}_{XL}(E_i, J_i, \pi_i, E_\gamma)=\left<\Gamma_{XL}(E_i, J_i, \pi_i, E_\gamma)\right>\rho(E_i,J_i,\pi_i)/E_\gamma^{(2L+1)}.
\label{RSForiginal}
\end{equation}
Here, $\left<\Gamma_{XL}(E_i, J_i, \pi_i, E_\gamma)\right>$ is the average, \textit{partial} radiative width for transitions within an initial excitation-energy bin $E_i$ of levels with spin $J_i$ and parity $\pi_i$, and $\rho(E_i,J_i,\pi_i) = 1/D(E_i,J_i,\pi_i)$ is the level density of those levels. 
Thus, it is seen from Eq.~(\ref{RSForiginal}) that the $\gamma$ strength represents the \textit{distribution} of average, reduced partial $\gamma$-transition widths.
This ``downward'' strength function is related to  $\gamma$ decay, while the ``upward'' strength function is determined by the average photo-absorption cross section $\langle\sigma_{XL}(E_{\gamma})\rangle$ summed over all possible spins of final states~\cite{Axel1968}:
\begin{equation}
\stackrel{\rightarrow}{f}_{XL}(E_f,J_f,\pi_f,E_{\gamma}) = \frac{1}{(2L+1)(\pi\hbar c)^2}\frac{\langle\sigma_{XL}(E_f,J_f,\pi_f,E_{\gamma})\rangle}{E_{\gamma}^{(2L-1)}},
\end{equation}
where $E_f$ is the energy bin reached after photo-absorption, and $J_f,\pi_f$ are the spin and parity of the excited levels, respectively. 
As discussed by Bartholomew \textit{et al.}~\cite{bartholomew1972}, within the extreme statistical model and also the damped harmonic-oscillator model, the strength function is independent of $J$ and $\pi$. 
This assumption is indeed applied in all open-access nuclear-reaction codes such as, e.g, {\sf EMPIRE}~\cite{empire} and {\sf TALYS}~\cite{TALYS,koning12}, providing astrophysical reaction rates for the $r$ process. 
The independence of spin and parity is valid if the wave functions of the highly excited levels within $E_i$ or $E_f$ contain a large number of configurations (high degree of mixing).
Further, it is not obvious that the upward strength equals the downward strength, except for, again, the case of the extreme statistical model. 
At this point, it is however common practice to apply the principle of detailed balance~\cite{blatt_weisskopf1952} and invoke the generalized form of the Brink hypothesis~\cite{brink}, stating that the excitation-energy dependence of the photo-nuclear cross section (and thus $\gamma$ energy in this special case) is \textit{not} dependent on the detailed structure of the initial state. 
In other words, the photo-absorption cross section on an excited state will have the same shape as the photo-absorption on the ground state, and so the upward strength function can be used as a proxy for the downward strength, as applied by Brink~\cite{brink} and Axel~\cite{Axel1962}.
Considering the dependence on \textit{final} states, early work by \textit{e.g.} Bollinger \textit{et al.}~\cite{bollinger1970} indicated no significant sensitivity of  partial radiative widths to the final states for heavy, deformed nuclei. 
That said, for lighter nuclear systems and for very neutron-rich nuclei with low neutron separation energy, there could very well be a strong strength dependence both on the initial and final state and the statistical model might break down. 
  
The simplest model for the strength function, the single-particle model of Blatt and Weisskopf~\cite{blatt_weisskopf1952}, results in $\gamma$-energy independent strength functions. 
This has been long known to be a too simple picture. For instance, the well-known giant electric dipole resonance (GDR) that strongly influences the strength function has been observed throughout the periodic table. 
This resonance is believed to stem from harmonic vibrations where protons and neutrons oscillate off-phase against each other, and is therefore called an \textit{isovector} collective excitation mode. 
Other giant resonances have been discovered as well, such as the giant magnetic dipole resonance (GMDR), which is built of spin-flip transitions between $\ell\pm 1/2$ subshells~\cite{heyde2011}, and the \textit{isoscalar} giant electric quadrupole resonance (GEQR) originating from surface oscillations where the protons and neutrons are distorted in two orthogonal directions. 
For more information on giant resonances in general, see Harakeh and van~der~Woude~\cite{Woude}. 

There is also experimental evidence for other types of resonance-like structures in the $\gamma$ strength function, which are small in magnitude compared to the giant resonances.
Examples of such structures include the $M1$ scissors mode~\cite{heyde2011,Lo_first_scissor_prediction,Leander_first_nilsson_scissor,Magne_scissor_1984,First_e_e,Kneissl,Krticka_letter,schiller2006} and the $E1$ pygmy resonance~(see Refs.~\cite{savran2013,tsoneva2015,savran2018,agvaanluvsan2009} and references therein). 
Moreover, an enhanced $\gamma$-ray strength at low transition energies ($E_\gamma < 3$ MeV) has recently been discovered~\cite{voinov2004}.
We leave the discussion of this very interesting feature to Sec.~\ref{sec:oslo}. 

To describe the $\gamma$-ray strength function, in particular the dominant $E1$ part, several more or less phenomenological models have been developed over the years. 
To this end, the Standard Lorentzian function applied by Brink~\cite{brink} and Axel~\cite{Axel1962},
\begin{equation}
f_{\rm SLO}(E_{\gamma}) = \frac{1}{3\pi^2\hbar^2c^2}\frac{\sigma_{\rm SLO} E_{\gamma} \Gamma_{\rm SLO}^2}{(E_\gamma^2-\omega_{\rm SLO}^2)^2 + E_{\gamma}^2 \Gamma_{\rm SLO}^2},
\label{eq:SLO}
\end{equation}
was originally used for the $E1$ strength, and is still the recommended description of $M1$ and $E2$ contributions~\cite{RIPL3}. Here, the parameters $(\omega_{\rm SLO}, \sigma_{\rm SLO}, \Gamma_{\rm SLO})$ correspond to the centroid, peak cross section, and width of the resonance, respectively. 
Furthermore, the Generalized Lorentzian model of Kopecky and Chrien~\cite{kopecky1987} and Kopecky and Uhl~\cite{ko90} has been widely used in reaction-rate calculations:
\begin{equation}
f_{\rm GLO}(E_{\gamma}) = \frac{1}{3\pi^2\hbar^2c^2}\sigma_{\rm GLO}\Gamma_{\rm GLO}   
 \times \left[ \frac{ E_{\gamma} \Gamma(E_{\gamma},T_f)}{(E_\gamma^2-\omega_{\rm GLO}^2)^2 + E_{\gamma}^2 \Gamma^2(E_{\gamma},T_f)}
+ 0.7 \frac{\Gamma(E_{\gamma}=0,T_f)}{\omega_{\rm GLO}^3}\right]
\label{eq:GLO}
\end{equation}
where the width is given by
\begin{equation}
\Gamma(E_{\gamma},T_f) = \frac{\Gamma_{\rm GLO}}{E_{\rm GLO}^2} (E_{\gamma}^2 + 4\pi^2 T_f^2)
\end{equation}
and $T_f$ is the nuclear temperature of the final states usually calculated using Eq.~(\ref{eq:temp1}). 

In recent years, a phenomenological description of the GDR using a triple Lorentzian parameterization has been applied~\cite{Junghans2008} in order to account for triaxial-shape degrees of freedom. 
Using such an approach, an improved prediction of the $E1$ GDR tail at and below the neutron separation energy can be achieved, leading to a more robust prediction of radiative neutron-capture rates at $s$-process temperatures~\cite{Grosse2014,Junghans2017}.

As in the case of the level density,  a microscopic treatment of the strength function is necessary to obtain information on the underlying nuclear structure and to have predictive power throughout the nuclear chart. 
Many publications have been dedicated to the microscopic description of $\gamma$-ray strength functions;  an overview is given by Paar \textit{et al.}~\cite{paar2007}. 
Goriely and Khan presented in Ref.~\cite{Goriely4} large-scale calculations based on the quasi-particle random-phase approximation (QRPA) model~\cite{Ring&Schuck} to generate excited states on top of the HF+BCS ground state. To account for the damping of the collective motion, the GEDR is empirically broadened by folding the QRPA resonance strength with a Lorentzian function. 

A recent application of the Hartree-Fock-Bogoliubov (HFB) plus QRPA method using the finite-range D1M Gogny force~\cite{goriely2018} combined with shell-model calculations demonstrated that also $M1$ strength can be theoretically obtained for a broad range of nuclei. 
The authors also included effects beyond the one-particle one-hole excitations and the interaction between the single-particle and low-lying collective phonon degrees of freedom through an empirical prescription. 
The shell-model calculations provided extra $M1$ strength below transition energies of $\approx 3-4 $ MeV (this phenomenon will be discussed in Section~\ref{sec:oslo}), which can affect the radiative neutron capture cross section of neutron-rich nuclei as well as proton capture cross section of neutron-deficient nuclei by factors up to a few hundreds for the most exotic cases.
Moreover, using an axially symmetric-deformed HFB-QRPA approach, Martini \textit{et al.}~\cite{martini2016} calculated the $E1$ strength of even-even, open-shell nuclei and were able to reproduce the experimentally-observed splitting of the GDR.  

When it comes to calculations of $E1$ strength within the non-relativistic and relativistic mean-field approach (see Ring~\cite{ring1996} and references therein), much progress has been made in recent years; examples include works of Vretenar \textit{et al.}~\cite{vretenar2001}, Litvinova \textit{et al.}~\cite{litvinova2009a,litvinova2013}, Roca-Maza \textit{et al.}~\cite{roca-maza2012} and Daoutidis and Goriely~\cite{daoutidis2012}.  
These calculations have provided a  good description for the position of the GDR and a theoretical interpretation of the low-lying dipole and quadrupole excitations, and have shed light on the $E1$ pygmy dipole resonance.

Another successful way to treat the collective modes microscopically is the quasi-particle multiphonon (QPM) model introduced by Soloviev~\cite{soloviev1976,soloviev1992}, with further applications and developments by a number of authors, including Andreozzi \textit{et al}.~\cite{NicolaIudice}, Stoyanov and Lo Iudice~\cite{Stoyanov2004,LoIudice2006}, Tsoneva, Lenske and Stoyanov~\cite{tsoneva2004}, and Tsoneva and Lenske~\cite{tsoneva2011}. Within this model, the nuclear eigenvalue problem is solved exactly in a multiphonon space, where the basis states are generated via the Tamm-Dancoff Approximation (TDA)~\cite{Ring&Schuck}. 
In particular, studies of the $E1$ pygmy resonance incorporating energy-density functional theory with the 3-phonon QPM~\cite{tsoneva2015,tsoneva2004} demonstrate the astrophysical impact on calculations of (n,$\gamma$) reaction rates for neutron-rich nuclei, similar to findings within the relativistic quasiparticle time blocking approximation~\cite{litvinova2009b}.  

Regarding the $M1$ strength, shell-model calculations have been employed to study the $M1$-strength distribution of a wide range of nuclei; see e.g. Refs.~\cite{loens2012,schwengner2013,brown2014,schwengner2017,sieja2017,midtbo2018}. Loens \textit{et al.}~\cite{loens2012} also investigated the effect of including the shell-model $M1$ distribution in ($n,\gamma$) reaction cross section and rates for both near-stability and neutron-rich iron nuclei, and demonstrated the importance of considering the full $M1$ distribution for excited states, not only the ground-state transitions, as also pointed out by later works~\cite{brown2014,schwengner2017}. 
Furthermore, the $M1$-strength distribution has recently been studied for heavy nuclei within axially symmetric-deformed HFB-QRPA approach~\cite{goriely2016}. 

The by far largest contribution of experimental information on the $\gamma$-ray strength function is from photoabsorption measurements\footnote{See, \textit{e.g.}, the atlas of ground-state photoneutron and photoabsorption cross sections by S.~S.~Dietrich and B.~L.~Berman~\cite{Dietrich_Berman}, and the Experimental Nuclear Reaction Database~\cite{EXFOR}.}. 
To measure photoabsorption, most often photoneutron cross sections, which provide a good substitute for photoabsorption cross sections, are measured. 
Photoneutron (or photoproton) cross-section measurements are dominated by $E1$ radiation, and are limited to energies above the neutron (proton) separation energy. Also, the absorption cross sections can only be measured on ground states or on very long-lived isomeric states. These measurements are traditionally performed by guiding a beam of photons to impinge on a thick target (typically several grams) of the nucleus that is under study. 
The photons can be of bremsstrahlung type from a betatron or a synchrotron facility, or produced by the in flight annihilation of fast positrons from a linear accelerator giving a quasi-monoenergetic photon beam although still containing some bremsstrahlung components~\cite{Berman_Fultz,Ishkhanov_Varlamov}. 
More recently, the inverse Compton-scattering technique has been utilized to produce  quasi-monoenergetic photon beams (see, \textit{e.g.}, Ref.~\cite{Utsunomiya2003} and references therein).  
The photon beams provided by this technique at the NewSUBARU facility are ideal for studying ($\gamma,n$) with unprecedented precision; see, for example, Refs.~\cite{utsunomiya2013,filipescu2014,nyhus2015}. 
The High Intensity $\gamma$-ray Source (HI$\gamma$S)~\cite{higs2009} is a joint project between the Triangle Universities Nuclear Laboratory (TUNL) and the Duke Free Electron Laser Laboratory (DFELL).
This facility is also capable of providing excellent-quality photon beams through inverse Compton scattering for ($\gamma,n$) measurements, see, \textit{e.g.}, Refs.~\cite{raut2013,sauerwein2014}.

\begin{figure}[tb]
\begin{center}
\includegraphics[clip,width=1.\columnwidth]{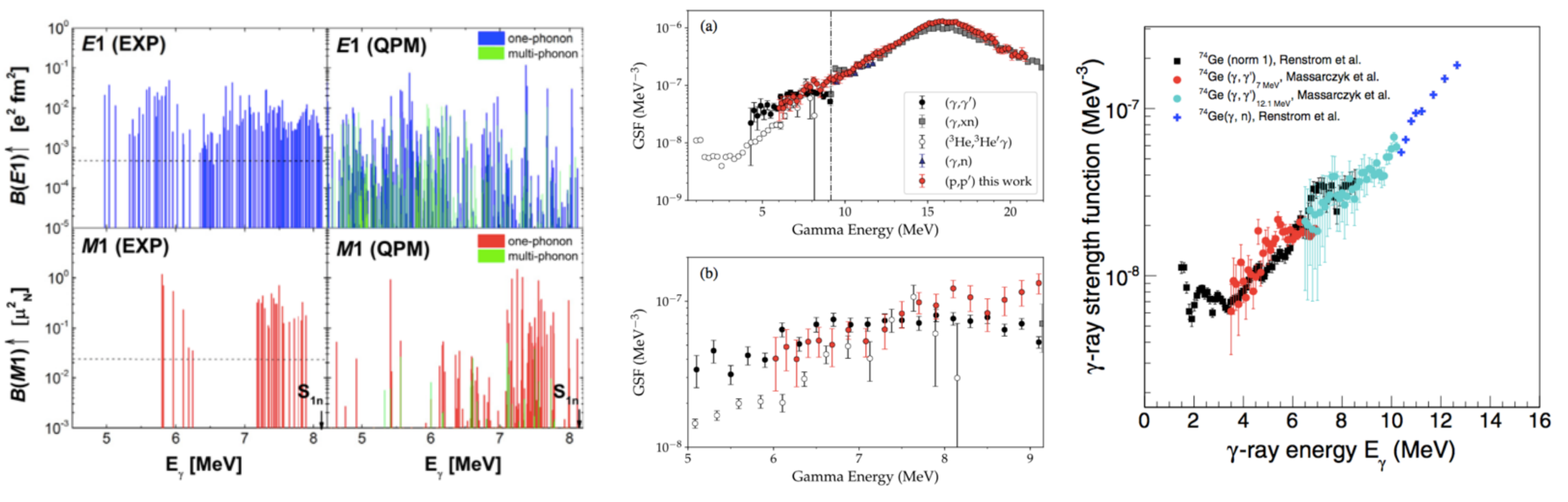}
\caption{(Color online) 
Gamma-ray strength functions from different experiments. 
Left:  $B(E1)$ (top panels) and $B(M1)$ (bottom panels) strength distribution in $^{206}$Pb for resonantly excited states between 4.9 and 8.1 MeV populated with the NRF technique; from Tonchev \textit{et al.}~\cite{tonchev2017}. 
Middle: a comparison of $\gamma$-strength functions for $^{96}$Mo from various reactions and techniques; from Martin \textit{et al.}~\cite{martin2017} demonstrating the use of the ($p,p^\prime$) reaction at very forward angles and high energies. The vertical, dashed line in the upper panel indicates the neutron separation energy of $^{96}$Mo.
Right: $\gamma$-strength functions of $^{74}$Ge using various reactions, showing the consistency of the Oslo method with other measurement techniques for this specific case; from Renstr{\o}m \textit{et al.}~\cite{renstroem2016}.
}
\label{fig:gammastrength_exp}
\end{center}
\end{figure}

To measure the $\gamma$-ray strength function below the particle-emission threshold, photon scattering on isolated levels has been utilized. 
In the so-called Nuclear Resonance Fluorescence (NRF) method, the spins, parities, branching ratios and reduced transition probabilities of the excited states can be extracted in a model-independent way~\cite{Kneissl}. 
Polarization and angular correlation measurements allow the separation of transitions into $E1$, $M1$, and $E2$ transitions, usually with high precision~\cite{Margraf}. 
However, the method is selective with respect to strong transitions, and experimental thresholds might hamper the determination of an \textit{average}  transition strength as represented by the $\gamma$-ray strength function. 
Nevertheless, this method was able to confirm the experimental evidence for a new, low-lying magnetic dipole mode \cite{Kneissl} first discovered in ($e,e^\prime$) experiments~\cite{Bohle} on rare-earth nuclei. Also, a thorough study of the $E1$ pygmy resonance in the $^{40,44,48}$Ca isotopes and in $N=82$ nuclei using photon scattering ($\gamma$,$\gamma'$) reactions  has been presented by Zilges \textit{et al.}~\cite{Zilges}, for Xe isotopes by Massarczyk \textit{et al}.~\cite{massarczyk2014}, and for $N=50$ isotones by Schwengner \textit{et al.}~\cite{schwengner2013prc}. 
The isospin character of the $E1$ pygmy-resonance states has recently been investigated by Crespi \textit{et al.}~\cite{crespi2014} with the inelastic heavy-ion reaction $^{208}$Pb($^{17}$O,$^{17}$O$^\prime \gamma$), extracting the isoscalar component of the $J^\pi = 1^-$ excited states from 4 to 8 MeV.
Furthermore, Massarczyk \textit{et al.}~\cite{massarczyk2015} studied the dipole strength distribution of $^{74}$Ge in photon-scattering experiments using bremsstrahlung produced with electron beams of energies of 7.0 and 12.1 MeV. The results were compatible with other types of experiments, as demonstrated in Fig.~\ref{fig:gammastrength_exp}. 
Using quasi-monochromatic photon beams, Romig \textit{et al.} investigated the low-lying dipole strength of $^{94}$Mo by the use of five beam energies with typical FWHM of 150-200 keV~\cite{romig2013}.
Furthermore, due to the possibility of using polarized beams, $E1$ and $M1$ transitions are easily separated and information on both electromagnetic characters can be obtained~\cite{tonchev2017} (see Fig.~\ref{fig:gammastrength_exp}).
Interestingly, also a pygmy quadrupole resonance has been discovered recently~\cite{pellegri2015,spieker2016}, revealed by a clustering of $J^\pi = 2^+$ states at excitation energies between $\sim 3-5$ MeV. 
The summed $B(E2)\uparrow$ strength yields about 14\% of the isovector giant quadrupole resonance for the $^{124}$Sn case~\cite{pellegri2015}.

Another way of measuring $\gamma$-ray strength functions below the neutron separation energy, is by radiative neutron (or proton) capture reactions into compound states in the final nucleus~\cite{bollinger1970,kopecky1987,ko90,kopecky2017}. 
From such experiments, both average total radiative widths of neutron resonances and individual transition strengths from one or several neutron resonances to one or several lower-lying discrete states can be obtained. 
Such primary $\gamma$-rays are averaged manually to get the $\gamma$-ray strength function, unless ARC neutrons were used, covering a wider range of energy and including many resonances. 
The advantage of measuring individual transition strengths is that since the spin and parity of both the initial and final states are known, $E1$, $M1$, and $E2$ $\gamma$-ray strength functions can be obtained separately. 
The method is however limited in energy in that it provides averages of transitions with energies in the order of $\sim 1-2$ MeV below the neutron separation energy.

Yet another approach in determining the $\gamma$-ray strength experimentally, is the spectrum-fitting method (see Ref.~\cite{Andreas&Thoennessen} and references therein). 
Within this method, a total $\gamma$-cascade spectrum is fitted in terms of trial $\gamma$-ray strength functions and level densities. 
This method has been used extensively for $\gamma$ spectra following, e.g, fusion-evaporation reactions in the search for the temperature response of the giant electric dipole resonance and can cover a wide range of temperatures and spins. 
A special case of the spectrum-fitting method is the two-step cascade (TSC) or (n,$2\gamma$) method, where experimentally, only two-step cascades which connect neutron resonances and discrete low-lying levels with definite parity and spin are recorded. 
In this manner, the method trades flexibility in terms of applicable nuclear reactions~\cite{Krticka_letter,schiller2006}. 
The disadvantage of all spectrum-fitting methods is that the level density remains a large source of systematic uncertainty, unless it is known {\it a priori}.

Wiedeking \textit{et al.}~\cite{wiedeking2012} could extract the shape of the $\gamma$-ray strength function by utilizing the $^{94}$Mo$(d,p\gamma\gamma)^{95}$Mo reaction. 
By tagging on the proton energies, the excitation energy of the residual nucleus could be determined. Furthermore, by gating on known specific transitions at low excitation energies, first generation $\gamma$-rays decaying to known levels with identified spin and parity were selected. Applying the condition that the sum of discrete and primary  $\gamma$-ray energies must be equivalent to the excitation energy provides information on events of unambiguous origin and destination. 
The experimental results from this model-independent $(d,p \gamma \gamma)$ technique verified earlier findings in $^{95}$Mo~\cite{guttormsen2005}, see Fig.~\ref{fig:upbend}.

Proton scattering at very high energies and small forward angles~\cite{tamii2009} is a very promising technique to get information on both the fine and gross structure of the pygmy dipole resonance and the giant dipole resonance~\cite{fearick2018}.
Recently,  Birkhan \textit{et al.}~\cite{birkhan2017} presented the electric dipole polarizability of $^{48}$Ca, using relativistic Coulomb excitation in the $(p,p^\prime)$ reaction at very forward angles. 
The cross sections above 10 MeV show a broad resonance structure identified with $E1$ excitation of the GDR. The resulting dipole response of $^{48}$Ca is found to be remarkably similar to that of $^{40}$Ca, consistent with a small neutron skin.
Furthermore, Martin \textit{et al.}~\cite{martin2017} demonstrated the applicability of the ($p,p^\prime$) reaction to extract information on the $\gamma$-ray strength function both below and above the neutron separation energy (see Fig.~\ref{fig:gammastrength_exp}). 
Also, a comparison is displayed in the right panel of Fig.~\ref{fig:gammastrength_exp}, showing the $^{74}$Ge $\gamma$-ray strength function obtained with three different experimental methods (among them the Oslo method, which will be presented in Sec.~\ref{sec:oslo}); see Ref.~\cite{renstroem2016} and references therein.

Finally, also $\alpha$ scattering at high energies and very forward angles have been applied to study the $\gamma$-ray strength function. 
Endres \textit{et al.}~\cite{endres2012} have compared the $\gamma$ strength in $^{124}$Sn measured with different reactions, namely the NRF $(\gamma, \gamma')$ and the $(\alpha,\alpha' \gamma)$ reactions.
While almost all dipole transitions known from NRF experiments up to about 6.8 MeV were observed in $(\alpha,\alpha' \gamma')$, almost no higher-lying $J^\pi = 1^{\pm}$  states were excited by the $\alpha$ particles. This feature has been interpreted as a splitting of the pygmy dipole resonance into an isoscalar part probed in the $\alpha$ scattering experiment, which is sensitive to the surface oscillations, while the isovector part is only probed in ($\gamma,\gamma^\prime$) experiments. 

\subsection{Breakdown of Hauser-Feshbach: Direct-capture and pre-equilibrium processes}
\label{sec:direct-preeq}
Although the Hauser-Feshbach formalism is the preferred choice of method for reaction-rate calculations, it has long been recognized that it is not applicable for nuclear systems with low level density at the excitation energy reached by neutron-capture, such as  exotic neutron-rich nuclei with very low neutron separation energies. In such cases, the radiative neutron-capture reaction could be dominated by direct electromagnetic transitions to a bound final state instead of going through the step of a compound-state creation. The possibility for a significant  direct-capture component depends on the characteristic time scale of the reaction and decay process~\cite{blatt_weisskopf1952,Ericson1960}. 
In contrast to the compound-nucleus mechanism, where the excitation is a multistep process and the time scale is of the order of 10$^{-14}$--10$^{-20}$ seconds, the direct-reaction process involves a singe-step excitation with a characteristic timescale of the order of 10$^{-21}$--10$^{-22}$ seconds~\cite{krausmann1996}.
Mathews \textit{et al.}~\cite{mathews1983} showed that the direct capture contribution may dominate the total cross section for closed neutron shells (neutron-magic) targets or targets with a low neutron binding energy. 
Moreover, Rauscher \textit{et al.}~\cite{rauscher1998} showed that the direct-capture contribution is highly sensitive to the mass models and nuclear-structure models used. 
Xu \textit{et al.}~\cite{xu2012,xu2014} have studied direct neutron-capture reactions on the basis of the potential model taking into account the $E1$, $E2$, and $M1$ allowed transitions to all possible final states. In Ref.~\cite{xu2012} the authors performed a systematic study for about 6400 nuclei lying between the proton and neutron drip lines, showing that the direct-capture cross section decreases with increasing neutron richness, \textit{i.e.}, with decreasing neutron separation energies. 
Furthermore, in Ref.~\cite{xu2014}, the authors found that for exotic, neutron-rich nuclei, the direct-capture contribution could be two to three orders of magnitude larger than that obtained within the Hauser–Feshbach approach, meaning that the rates traditionally used in $r$-process simulations could be significantly underestimated.

In between the two extremes of the compound-nucleus mechanism and the direct-capture reaction, one encounters pre-equilibrium processes, which are characteristic of high-energy collisions (typically for incident-particle energies of 10-20 MeV). Here, $\gamma$ rays and particles are emitted \textit{after} the first direct interaction and \textit{before}  statistical equilibrium is achieved.
One signature of pre-equilibrium processes is the observed increase of the ($n,\gamma$) cross section for stable nuclei at incident energies typically around 10 MeV. 
However, for exotic neutron-rich nuclei with low neutron separation energies and low level densities, the pre-equilibrium process might influence the neutron channel at much lower energies~\cite{xu2014}, since these nuclei will have difficulties reaching a statistical equilibrium as discussed above. 
To describe the pre-equilibrium cross section, the exciton model~\cite{gadioli1992} has been shown  to provide reasonable results~\cite{koning2004}.

Finally, as discussed recently by Rochman \textit{et al.}~\cite{rochman2017}, for nuclei with very low neutron separation energies, the standard Hauser-Feshbach treatment of neutron-capture cross sections is not valid. 
They present a new method to generate statistical resonances, called the High-Fidelity Resonance method. 
This technique provides similar results as the Hauser-Feshbach approach for nuclei where the level density is high, but deviates significantly for neutron-rich nuclei at relatively low (sub-keV) energies.
Furthermore, in the keV–MeV energy region of astrophysical interest, the High-Fidelity Resonance method produces higher Maxwellian-Averaged cross sections by a factor up to a few hundreds with respect to the corresponding Hauser-Feshbach model.

\section{Approaches for direct measurements of neutron-induced reactions on neutron-rich nuclei}
\label{sec:direct}

As described in Sec.~\ref{sec:r-process}, the $r$ process involves very neutron-rich nuclei with extremely short lifetimes of the order of seconds to down to milliseconds. 
This fact makes it very difficult to perform direct measurements of neutron-induced reactions, as one cannot make a traditional sample of such short-lived nuclei to be placed in a neutron beam, as is done for stable or long-lived nuclei.
For charged-particle induced reactions, one can overcome the problem of short lifetimes by applying inverse kinematics: a beam of the radioactive isotope of interest is created at a radioactive-beam facility and separated from the ``cocktail" of exotic isotopes, before impinging on a hydrogen or helium target (gas-jet targets, gas cells or with hydrogen or deuterium embedded in plastic-type targets such as polyethylene or polystyrene). 
However, this approach is not possible for neutron-induced reactions because that would require a neutron target, which is not achievable as free neutrons are perishables\footnote{The most recent measurement gives a mean lifetime of $\tau_n = 887.7(31)$~s~\cite{yue2013}.} and must continuously be produced by some neutron source with sufficiently high rates to yield a significant neutron flux.  
Hence, new innovative techniques are called for to enable direct measurements of neutron-induced reactions and their corresponding cross section and astrophysical reaction rates.

One of the perhaps most promising initiatives in this respect, is the possibility to perform neutron-induced reactions in inverse kinematics by coupling a radioactive-beam facility with a high-flux neutron source, where the radioactive ions are kept in a storage ring (for reviews on storage rings, see Refs.~\cite{litvinov2013,bosch2013}). 
Recently, for charged-particle induced reactions, the combination of a storage ring and windowless hydrogen micro-droplet target has been successfully applied recently at the Experimental Storage Ring (ESR), GSI~\cite{mei2015}.
The use of a storage ring enables an efficient use of the rare isotopes, and  provides the highest possible luminosity of the rare-isotope beam.
Initial studies of this kind have recently been published by Reifarth \textit{et al.}~\cite{reifarth2014,reifarth2017} and Glorius \textit{et al.}~\cite{glorius2015}, using either a reactor or a spallation target to provide the neutron source. These studies will be outlined and discussed in the following. 

\subsection{Neutron source from a reactor}
As discussed in Ref.~\cite{reifarth2014}, a possible neutron source for rare-isotope inverse-kinematics experiments is the core of a research reactor, where one of the central fuel elements can be replaced with the evacuated beam pipe of the storage ring, \textit{i.e.} the radioactive ions will pass through the reactor core. 
Research reactors \textit{e.g.} of the TRIGA type are rather easily adopted to such a modified geometry. 
Another alternative is to place the beam pipe right next to the reactor core, of course with a loss of neutron flux ($\approx 1$ order of magnitude). 
The neutron energy spectrum would then correspond to a thermal spectrum peaking at $k_B T = 25$ meV, which implies the neutrons are practically at rest compared to the radioactive beam with energy of 0.1 MeV per nucleon or higher. 
\begin{figure}[tb]
\begin{center}
\includegraphics[clip,width=1.\columnwidth]{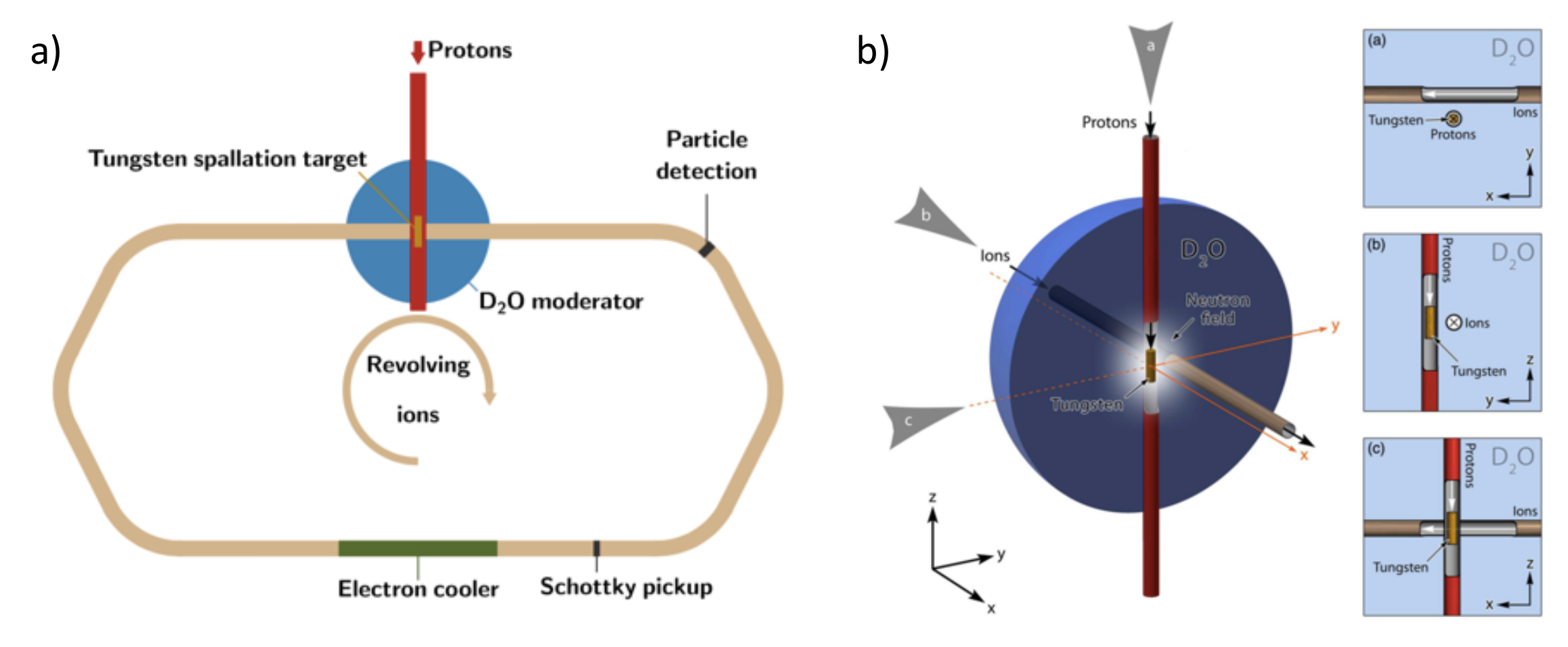}
\caption{(Color online) (a) Sketch of a tungsten spallation target combined with a storage ring. Neutrons are produced by protons impinging on a tungsten spallation target (brown). The proton beam pipe (red) is orientated perpendicular to the ion beam pipe (light brown). 
The neutrons produced in the spallation process are moderated in the surrounding heavy water (blue), penetrate the ion beam pipe and act as a neutron target. 
The ion beam pipe is part of a storage ring outside the moderator. The storage ring may contain additional equipment like an electron cooler (green), Schottky pickups and particle detectors (gray). Figure from Ref.~\cite{reifarth2017}. (b) Sketch of the proposed tungsten spallation target and ion beam pipe. Left: Protons impinge on a tungsten cylinder and produce neutrons. 
The proton beam pipe points along the z-axis and is orientated perpendicular to the ion beam pipe (gray). The beam pipes do not intersect as the ion beam pipe is shifted by several centimeters. Right: 2D projections of the setup (a) along the proton beam pipe, (b) along the ion beam pipe, and (c) perpendicular to both pipes. The lines of sight are indicated by gray arrows in the left sketch. The water moderator is indicated by the light blue background. Figure from Ref.~\cite{reifarth2017}.
}
\label{fig:reifarth}
\end{center}
\end{figure}

One obvious difficulty with ($n,\gamma$) reaction measurements in inverse kinematics, is to detect and separate the  compound nucleus (mass $A+1$) from the radioactive beam (mass $A$).
Since neither the charge nor the momentum of the products are different from the unreacted beam, the neutron capture cannot be detected with particle detectors.
Although the total momentum of the beam-nucleus and compound-nucleus is the same, the velocity, and thus the revolution frequency in the storage ring, is reduced by the factor $A/(A+1)$.
This frequency change, although small, can be measured using Schottky detectors.
The Schottky-noise frequency analysis takes advantage on the fact that each circulating ion in the storage ring induces a mirror charge when passing a pair of capacitive pickup plates~\cite{Bosch1996}. 
These periodic signals reveal the corresponding revolution frequency of each species of stored ions. 
This frequency is a unique function of the mass/charge ratio since for all ions the velocity equals that of the cooler electrons.
The Schottky method has been used successfully at ESR, GSI~\cite{litvinov2011}.

\subsection{Neutron source from spallation}
Instead of using a reactor as a neutron source, it is suggested in Ref.~\cite{reifarth2017} that a spallation neutron source could provide the neutron beam.
There are many advantages to such an approach; from a safety and security regulations point of view, it is far more convenient than a reactor because there is no critical assembly and no use of or production of actinides. 
From a physics point of view, this approach is advantageous because a spallation source will produce much less $\gamma$ rays per neutron as compared to a reactor core. 
A schematic drawing of a spallation source combined with a storage ring is shown in Fig.~\ref{fig:reifarth} (a), while details of the spallation target and the ion beam pipe is outlined in Fig.~\ref{fig:reifarth} (b) (both taken from Ref.~\cite{reifarth2017}).

\section{Indirect measurements}
\label{sec:surrogate}
As direct measurements are currently not possible, constraints on ($n,\gamma$) cross sections must be obtained from indirect methods. 
In this section, we present recently indirect methods that are complementary to each other, and they will all contribute to reduce the theoretically-estimated ($n,\gamma$) reaction rates. 
To reduce systematic errors, it would be preferable to apply more than one method on the same nucleus of interest.

\begin{figure}[tb]
\begin{center}
\includegraphics[clip,width=0.9\columnwidth]{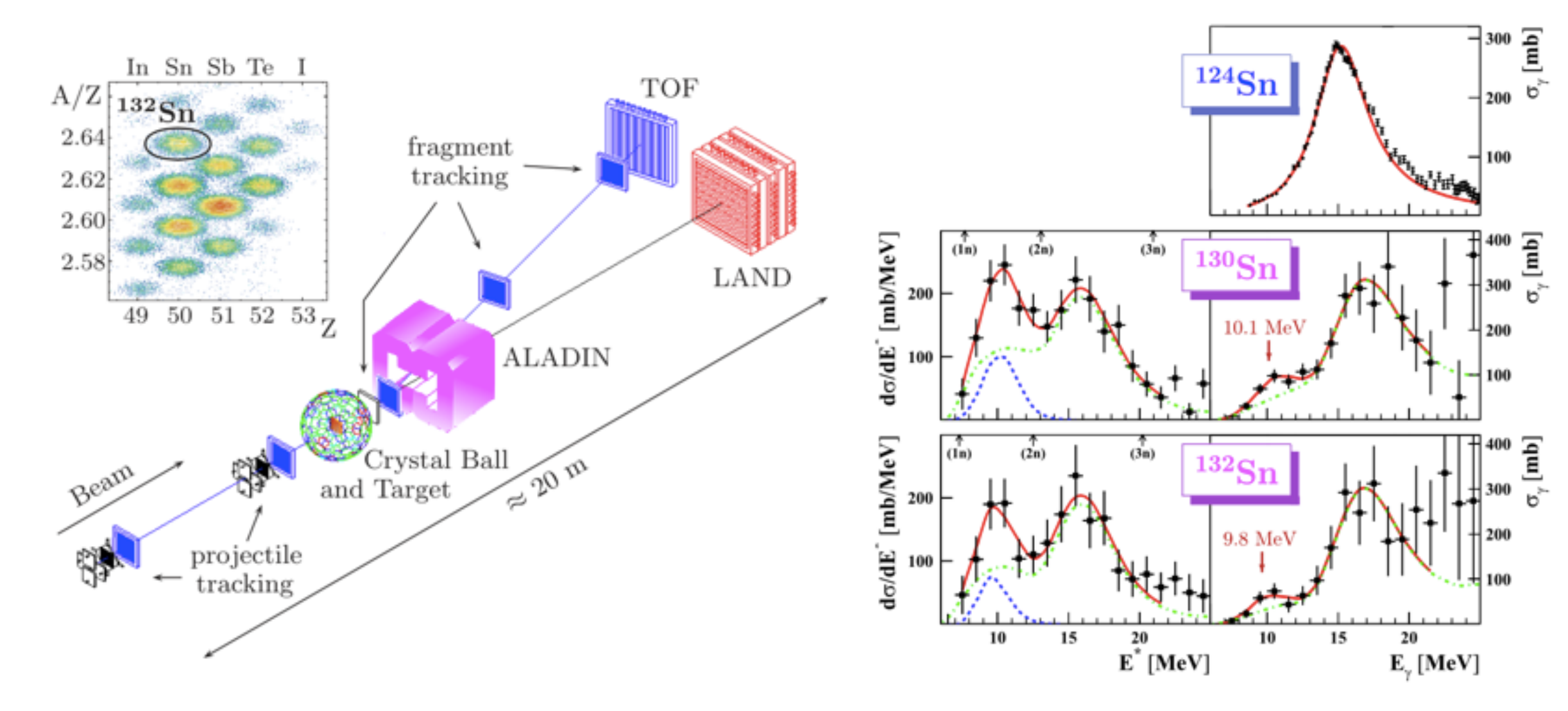}
\caption{(Color online) 
Left: Schematic view of the LAND experimental setup. 
Inset: identification of $^{238}$U fission fragments forming the secondary beam.
Right: left panels show the differential electromagnetic dissociation cross sections (with respect to excitation energy denoted $E^*$) of $^{130,132}$Sn. Arrows indicate the neutron-separation thresholds. 
Corresponding right panels: deduced photo-neutron cross sections. 
The curves represent a fitted Gaussian (blue dashed line) for the pygmy dipole resonance  and a fitted Lorentzian function (green dash-dotted line) for the giant dipole resonance. The sum of the two components is given by the red solid line, after folding with the detector response. In the top right panel, the photo-neutron cross section of stable $^{124}$Sn~\cite{fultz1969} measured in a real-photon absorption experiment is shown for comparison; the solid red line represents a Lorentzian distribution. Figures from Adrich \textit{et al.}~\cite{adrich2005}.
}
\label{fig:adrich}
\end{center}
\end{figure}

\subsection{Coulomb excitation and dissociation}
At GSI, Coulomb dissociation has been applied for measuring the electric dipole strength distribution in exotic nuclei, in particular $^{130,132}$Sn~\cite{adrich2005}. 
Radioactive ions were produced by in-flight fission of a $^{238}$U primary beam hitting a Be target, and various neutron-rich isotopes were identified by using the FRagment Separator (FRS)~\cite{adrich2005}, see inset of Fig.~\ref{fig:adrich}.
The freshly produced heavy-fragment isotopes (secondary beams) were excited through Coulomb interaction with a $^{208}$Pb target, and were de-excited by emission of  neutrons and $\gamma$ rays. 
Neutrons were measured with the large area neutron detector (LAND) while $\gamma$ rays were detected with the $4\pi$ Crystal Ball spectrometer. 
Heavy fragments were identified through energy-loss and time-of-flight measurements and by tracking in the magnetic field of the dipole magnet ALADIN~\cite{adrich2005}.
By measuring the complete kinematics, i.e. the four-momenta of the heavy fragment, neutron(s), and $\gamma$ rays, the invariant mass and thus the initial excitation energy $E^*$ could be extracted.  

The  $E1$-dominated partial cross sections $\textnormal{d}\sigma/\textnormal{d}E^*$ and the deduced total photoneutron cross sections are shown in Fig.~\ref{fig:adrich}. 
The data clearly show a large component in the strength around $E_\gamma \approx 10$ MeV, which is attributed to the $E1$ pygmy resonance above neutron threshold. 
As the measurements only probe the $E1$ $\gamma$-strength above neutron threshold, it could well be that a significant amount of the $E1$ pygmy-resonance strength is also present below threshold, as shown e.g. in Refs.~\cite{Zilges,massarczyk2014}. 
These measurements demonstrate the importance of experimental data of the $\gamma$ strength function for radioactive nuclei; using only phenomenological predictions of the $E1$ strength would completely miss the extra strength close to the 1-neutron separation energy. 
There is no doubt that the presence of the pygmy dipole resonance could impact the $r$-process reaction rates~\cite{tsoneva2015,Goriely4} and thus possibly the final abundance pattern of the $r$-process synthesized elements.  
 
\begin{figure}[bt]
\begin{center}
\includegraphics[clip,width=0.6\columnwidth]{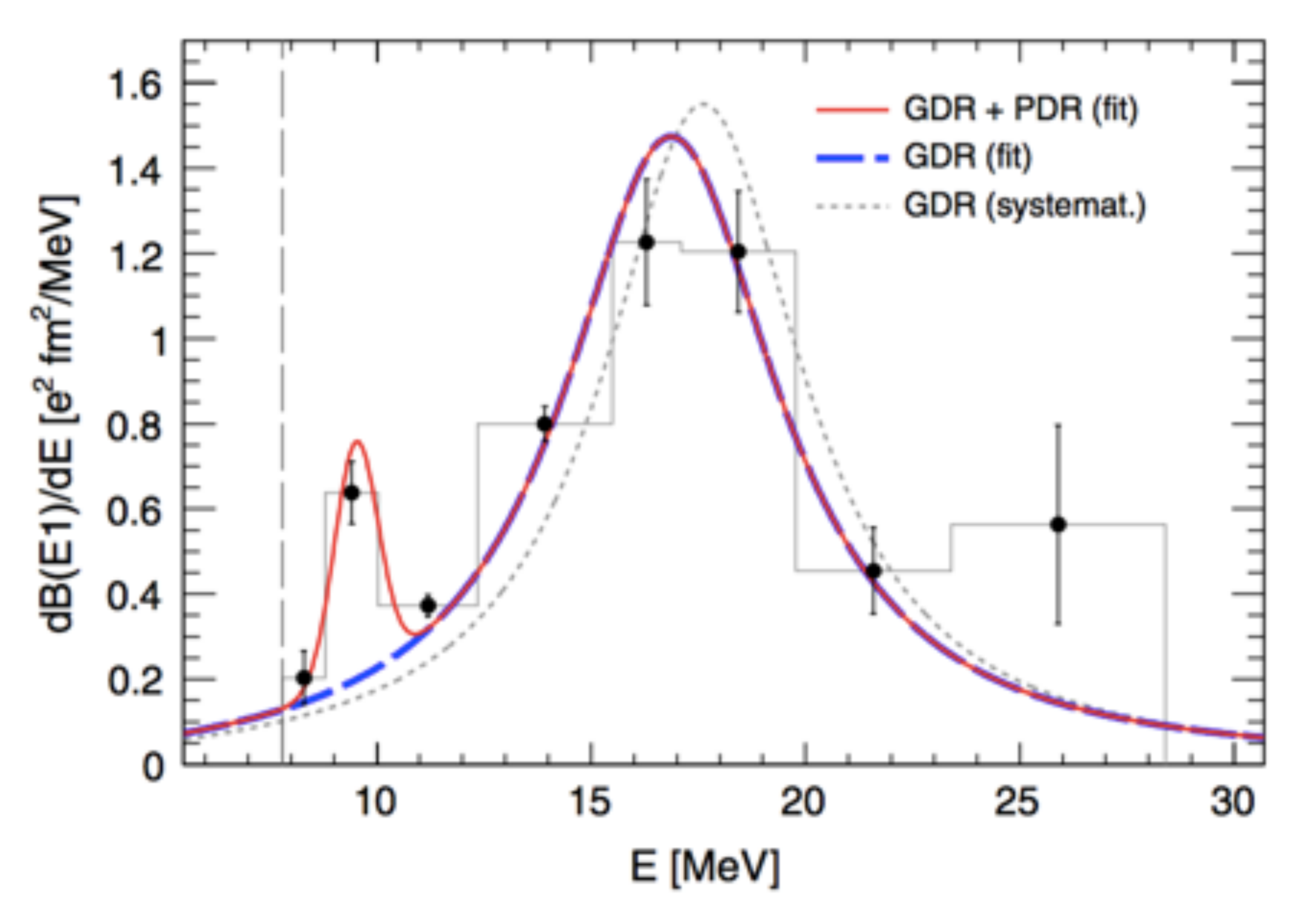}
\caption{(Color online) $E1$ strength distribution  of  $^{68}$Ni (histogram and black data points) with a GDR +  PDR fit function (solid red line). The GDR contribution (dashed blue line) and the GDR from systematics (dotted gray line) are shown for reference. The neutron threshold is indicated by the dashed vertical line at 7.792 MeV. Figure from Ref.~\cite{rossi2013}.
}
\label{fig:rossi}
\end{center}
\end{figure}

A similar structure as for the exotic tin isotopes has been observed in lighter, neutron-rich nuclei as well, such as the case of $^{68}$Ni studied by Wieland \textit{et al.}~\cite{wieland2009} and Rossi \textit{et al.}~\cite{rossi2013}. 
In the GSI experiment of Wieland \textit{et al.}, a $^{68}$Ni beam was produced by fragmentation of a $^{86}$Kr beam at 900 MeV per nucleon focused on a thick (4g/cm$^2$) Be target. 
The $^{68}$Ni ions were selected together with a few similar-mass ions with the FRS, with the $^{68}$Ni ions being the most intense component ($\approx 33\%$).
The $^{68}$Ni nuclei were then impinging on a gold target for Coulomb excitation. 
Gamma rays were measured using the RISING setup with  high-purity Ge detectors and BaF$_2$ detectors, where both types of detectors consistently showing a peak-like structure around 11 MeV. 
This experiment showed for the first time the presence of a pygmy dipole resonance in $^{68}$Ni~\cite{wieland2009}. 

Furthermore, Rossi \textit{et al.} performed a new experiment on $^{68}$Ni at GSI, this time using the R$^3$B-LAND setup and the invariant-mass technique to extract photoneutron cross section using virtual photons through Coulomb excitation. 
The production of $^{68}$Ni ions was again obtained through fragmentation of an $^{86}$Kr beam, this time with an energy of $\approx 650$ MeV per nucleon and with a Be target of thickness 4.2 g/cm$^2$. 
The ions were guided to a secondary target of 519 mg/cm$^2$ natural lead for the Coulomb excitation. 
The obtained $E1$ strength distribution of $^{68}$Ni from Rossi \textit{et al.} is shown in Fig.~\ref{fig:rossi}. As for the tin isotopes, an excess strength is observed around 10 MeV, which is assigned to the pygmy dipole resonance. 
The underlying physics behind the pygmy dipole resonance is still heavily debated in the nuclear-physics community (see Ref.~\cite{savran2013} and references therein, as well as Ref.~\cite{reinhard2013}); it has become clear that theory and experiment must go hand in hand to, in the future, provide a fundamental understanding of this phenomenon.  

As proven by the results described above, Coulomb excitation/dissociation provides invaluable information on the $\gamma$-ray strength function above neutron threshold for radioactive isotopes.
Additional measurements scanning a wider range of masses would be highly desirable, as they complement other measurement techniques and give those techniques guidance for absolute-value normalization.

\subsection{Direct-capture surrogate reactions for nuclei near closed shells}
As mentioned in Sec.~\ref{sec:direct-preeq}, the statistical model can be inadequate for nuclei with very few levels available, such as nuclei at/near shell closure. In such cases, it is necessary to consider direct capture and a state-by-state analysis of available bound and unbound levels in the residual nucleus. 

\begin{figure}[tb]
\begin{center}
\includegraphics[clip,width=0.5\columnwidth]{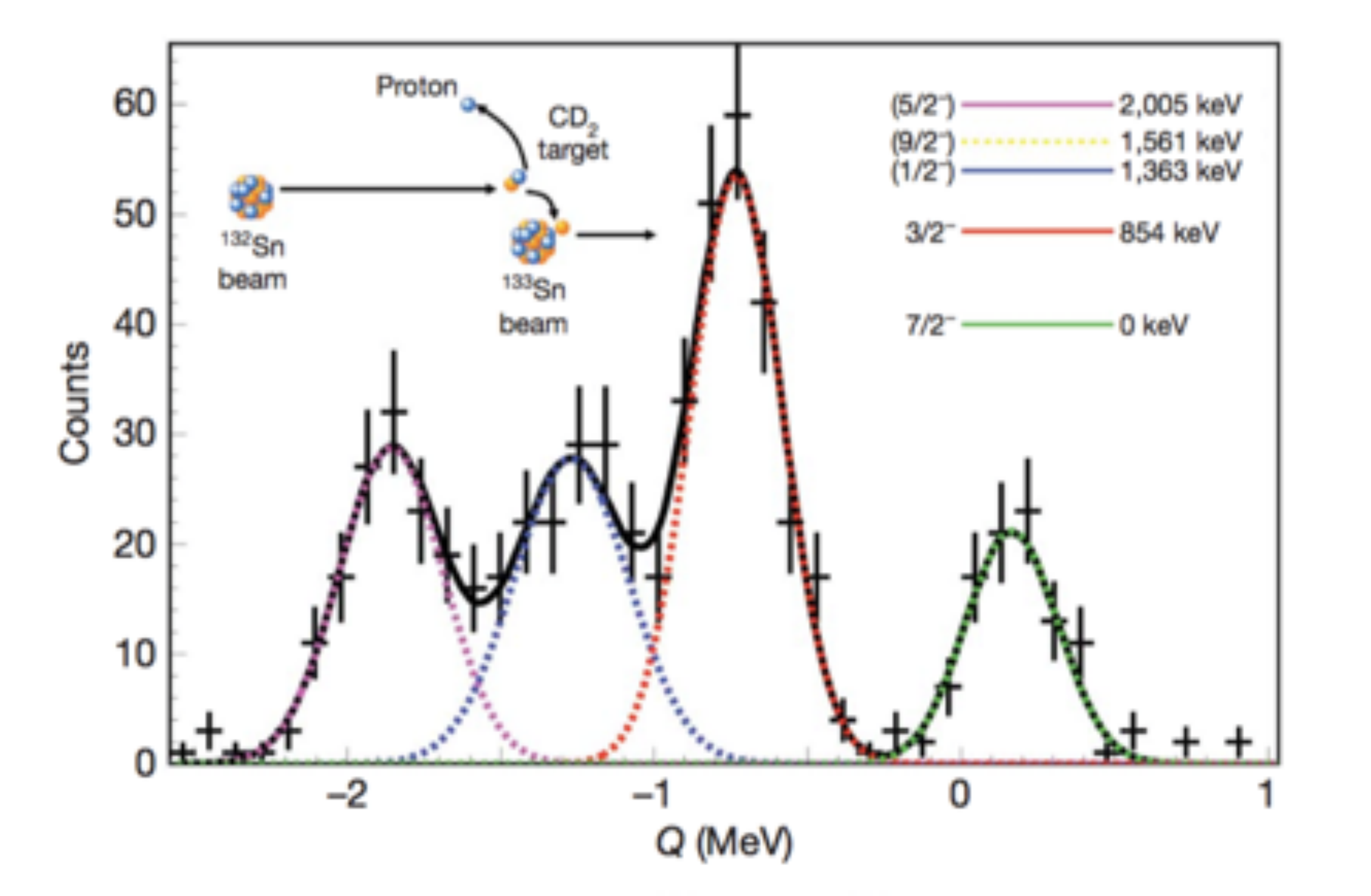}
\caption{(Color online) $Q$-value spectrum for the $^{132}$Sn$(d,p)^{133}$Sn reaction at 54$^{\circ}$ in the centre of mass. The black solid line shows a fit to four peaks: the ground state, the 854-keV state, the 1363-keV state, and the 2005-keV state. The top left inset displays a diagram of the $(d,p)$ reaction in inverse kinematics. The top right inset shows the level scheme of $^{133}$Sn. Figure from Ref.~\cite{jones2010}.
}
\label{fig:jones}
\end{center}
\end{figure}

For this purpose, surrogate reactions probing available levels for the ($n,\gamma$) reaction can be used. A striking example is the measurements of the $^{132}$Sn$(d,p)^{133}$Sn reaction in inverse kinematics by Jones \textit{et al.}~\cite{jones2010,jones2011}. With a low ground state $Q$-value of 0.147 MeV, the $^{132}$Sn$(d,p)$ reaction at energies around the Coulomb barrier populates mainly low-energy, low-angular-momentum, single-particle states~\cite{jones2011}. 
By measuring the angular distribution of the outgoing protons, the $\ell$-transfer of the various states could be determined. The $Q$-value spectrum from Ref.~\cite{jones2010} is shown in Fig.~\ref{fig:jones}.  

Furthermore, the $^{130}$Sn($d,p$)$^{131}$Sn reaction in inverse kinematics was measured by Kozub \textit{et al.}~\cite{kozub2012}, motivated by $r$-process calculations by Beun \textit{et al.}~\cite{beun2008} and Surman \textit{et al.}~\cite{surman2009} suggesting the $^{130}$Sn($n,\gamma$)$^{131}$Sn reaction rate could have a global effect on isotopic abundances during freeze-out. As for the $^{132}$Sn case, direct neutron capture is expected to be significant at late times in the $r$ process near the $N=82$ closed shell.
An apparent single particle spectrum was observed, very similar to the results of
Jones \textit{et al.} in Refs.~\cite{jones2010,jones2011}. 
It was found that the $\ell=1$ single particle states are both bound, 
in favor of a relatively large direct-capture cross section compared to those from models that predict one or both of these states to be unbound~\cite{rauscher1998}. 
From these experimental results, cross sections for $^{130}$Sn($n,\gamma$)$^{131}$Sn direct-semidirect capture have been calculated, reducing the uncertainties by orders of magnitude from previous estimates. However, it is interesting to note that the authors state that contributions from statistical processes were not well understood -- this is still, to our knowledge, the case for ($n,\gamma$) reactions in this mass region and remains an experimental and theoretical task to be investigated.

For lighter nuclei near the $N=28$ shell closure, experiments have been performed by Gaudefroy \textit{et al.}~\cite{gaudefroy2006} on the $^{46}$Ar($d,p$)$^{47}$Ar reaction in inverse kinematics at the SPIRAL facility, GANIL. 
Again, spectroscopic information was obtained from the proton spectra, and the results were used for calculating radiative neutron-capture rates, with possible implications for the the overproduction of the stable $^{48}$Ca isotope as compared to $^{46}$Ca~\cite{gaudefroy2006}.

\subsection{Surrogate reaction method for statistical capture}
\label{subsec:surrogate}
\begin{figure}[tb]
\begin{center}
\includegraphics[clip,width=0.9\columnwidth]{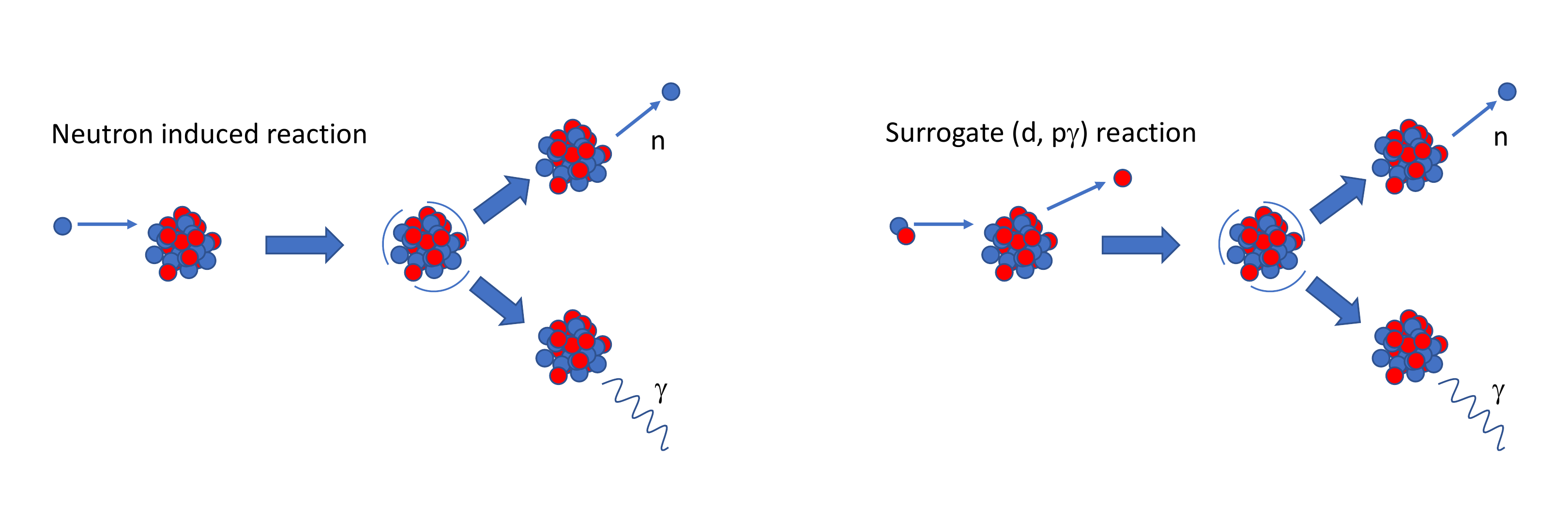}
\caption{(Color online) Schematic view of the surrogate $(n,\gamma)$ method. The
left part shows the desired direct $(n,\gamma)$ reaction. The right part shows the
surrogate reaction used to estimate the neutron capture cross section. Here
the surrogate $(d,p)$  reaction is exploited to mimic the $(n,\gamma)$ reaction.
}
\label{fig:ng}
\end{center}
\end{figure}

The surrogate nuclear reactions method for statistical capture was first established back in the 1970s~\cite{Cramer1970}.
It was designed to indirectly measure nuclear cross sections that otherwise were difficult
or impossible to measure. An illustration of the method is shown in Fig.~\ref{fig:ng}. The
residual nucleus produced by the surrogate reaction is assumed to represent the
 compound nucleus of interest as reached by an ($n,\gamma$) reaction, and the relevant decay probability can thereby be measured.
To extract the neutron-induced cross section, the compound nuclear decay probability
is multiplied by an independently obtained neutron-induced formation cross section,
calculated using an optical-model formalism.
Although applied to different nuclear-astrophysics aspects and for different kinematic energy regions, the surrogate method connects to the Trojan Horse Method~\cite{baur1986,typel2003} in that inclusive non-elastic breakup theory provides a common basis for both methods~\cite{bertulani2018}.

So far, the method has almost entirely been used to estimate neutron-induced fission cross sections. In particular,
the  method turned out to be fruitful for fission studies of actinides
where no stable or long-lived targets were available. For a historical overview,
see Escher \textit{et al.}~\cite{Escher2012}.

In this article, we focus
on the potential of the surrogate method to extract
reaction rates relevant for the $r$-process nucleosynthesis, as during the last decade, the surrogate method has also been applied to the $(n, \gamma)$
neutron capture reaction. 
As the surrogate method proved rather successful for neutron-induced fission cross sections, it was hoped that it would also provide reliable information on radiative neutron-capture cross sections. 
However, the
method did not work as expected for these types of reactions except for high-energy neutrons ($E_n\geq 0.5-1$ MeV, approximately). 
Unfortunately, the
method failed in the few-hundred-keV neutron energy region,
which is relevant for temperatures of the $r$ process.

It became clear that details of the
angular momentum and parity distribution of the surrogate compound
nucleus were essential for estimating the properties of the direct $(n, \gamma)$ reaction channel.
If this distribution of the surrogate compound nucleus $A$
did not match with the low-lying states of the $A+1$ nucleus, the extracted
$(n, \gamma)$ cross section could be $2-3$
times larger than observed in direct measurements. This increase was believed to be caused by the large number of possible decay paths for the surrogate reaction as compared to the $(n, \gamma)$ reaction, mainly due to the difference in spin population.

The nuclear reactions chosen for the surrogate method are usually
light-ion stripping or pick-up transfer reactions with one charged ejectile.
The excitation energy $E_x$ can then
be determined by the reaction kinematics by use of the measured energy of the ejectile.
The corresponding neutron energy in the desired reaction is then given by:
\begin{equation}
E_n=\frac{A+1}{A}(E_x-S_n),
\label{eq:E_n}
\end{equation}
where $A$ is the mass number of the target nucleus and $S_n$ is its neutron separation energy.
Assuming a compound nucleus (CN) formation both for the desired $(n,\gamma)$ and the surrogate reactions,
the neutron-capture cross section can be expressed as a sum of products, each assigned a
specified angular momentum $J$ and parity $\pi$
\begin{equation}
\sigma_{n,\gamma}(E_n)=\sum_{J,\pi}\sigma_{CN}(E_n,J,\pi)P_{\gamma}(E_n,J,\pi),
\label{eq:sigma}
\end{equation}
where $\sigma_{CN}$ is the cross section for forming the compound nucleus with
angular momentum $J$, parity $\pi$ and neutron energy $E_n$. The factor $P_{\gamma}$ is the probability
for decay of this state via emission of one or more $\gamma$ rays. Theoretically,
this quantity is difficult to predict since it directly depends on the
$\gamma$-ray strength function and the angular momentum and parity dependent level densities. As an example,
the level density at $S_n$ may be a factor of two uncertain, even when the
neutron capture level spacing $D_0$ is known~\cite{egidy2005}.

With these difficulties in mind, the \textit{Weisskopf-Ewing limit} has
become the most frequently used approach. Here, the decay probability is assumed
to be angular momentum and parity independent, by simply assuming
\begin{equation}
\sigma_{n,\gamma}(E_n)=\sigma_{CN}(E_n)P_{\gamma}(E_n).
\label{eq:ew}
\end{equation}
By this approximation, one neglects differences that may occur due to mismatch
in the angular momentum and parity distributions of the nucleus $A$ produced
in the surrogate reactions and the $A+1$ nucleus populated in the $n$ channel.
The procedure now is to calculate $\sigma_{CN}(E_n)$ within the Hauser-Feshbach theory~\cite{hauser}
and measure $P_{\gamma}$ determined from the ratio between particle-$\gamma$ coincident and
single-particle spectrum.

\begin{figure}[tb]
\begin{center}
\includegraphics[clip,width=0.6\columnwidth]{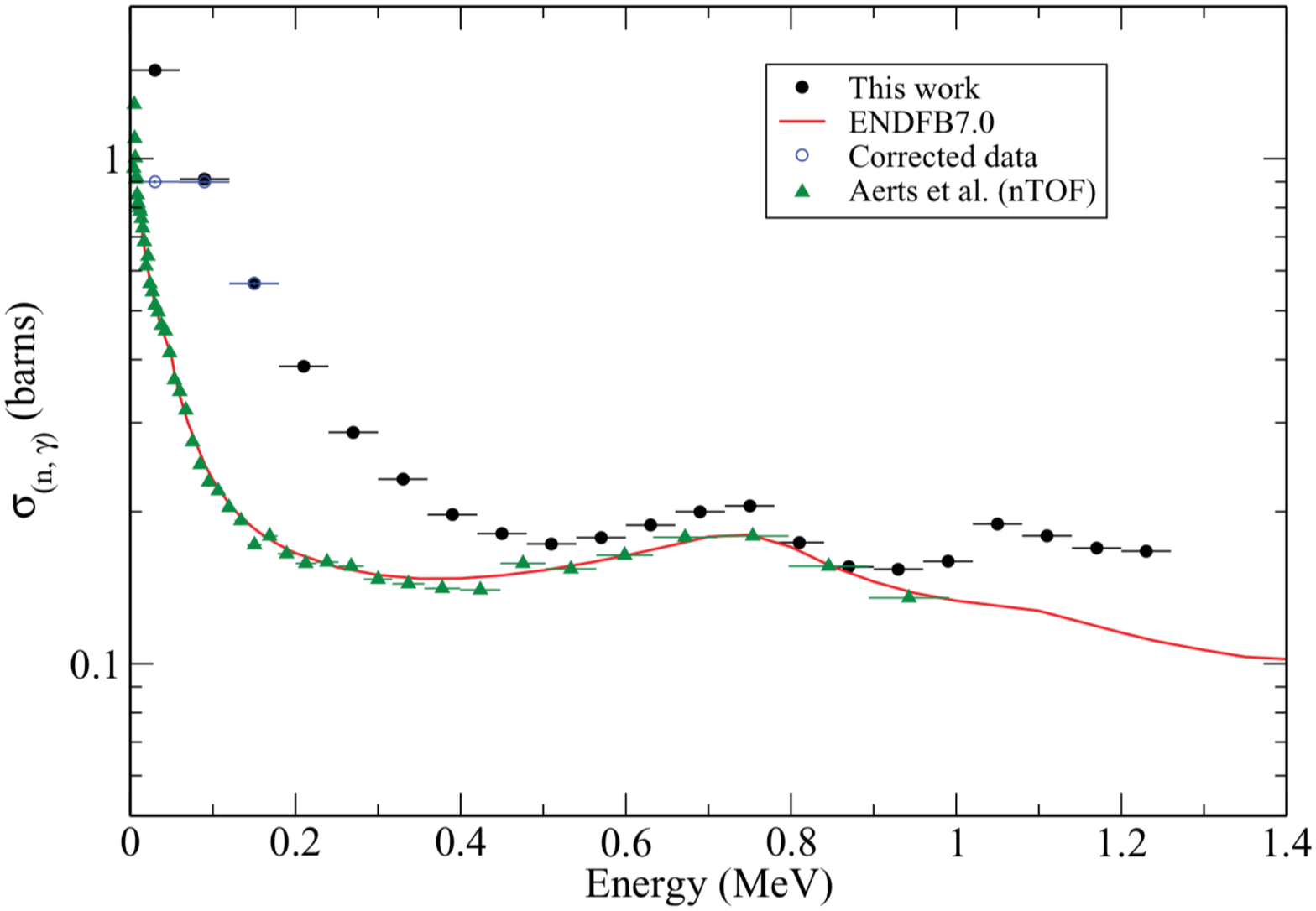}
\caption{(Color online) The indirect $^{232}$Th$(n,\gamma)$ cross
section extracted from the surrogate $^{233}$Th$(d,p)$ reaction (black points).
The results are compared to the ENDFB-7 database (red line)
and the direct neutron-induced data
measured at the n-TOF facility (green triangles).
Figure from Wilson~\textit{et al.}~\cite{Wilson2012}.
}
\label{fig:wilson}
\end{center}
\end{figure}

Figure~\ref{fig:wilson} shows the $^{232}$Th$(n,\gamma)$ cross section~\cite{Wilson2012} extracted by use of
the surrogate $^{233}$Th$(d,p)$ reaction in the Weisskopf-Ewing limit.
The compound nucleus formation cross section was obtained from the {\sf TALYS} optical
model~\cite{TALYS,koning12} calculations with input
parameters adjusted to accurately reproduce the experimental total cross sections.
The surrogate method is seen to work
well in the $E_n \approx 0.5-1$~MeV region where
the data coincide with the direct measured n-TOF data. Recently~\cite{Ducasse2016}, an investigation on the $^{238}$U$(d,p)$ surrogate reaction shows that the $\gamma$-decay probability is several times higher than for the neutron-induced reaction. It is also clear that the deuteron breakup channel complicates significantly the interpretation of the results.

Another approach is the \textit{surrogate ratio method}~\cite{Burke2006,Escher2006,Escher2010}.
In this approach the ratio of two compound reaction cross sections $\alpha$ and $\beta$
is described by invoking the Weisskopf-Ewing approximation
\begin{equation}
R(E_n)=\frac{\sigma^{\alpha}_{n,\gamma}(E_n)}{\sigma^{\beta}_{n,\gamma}(E_n)}
= \frac{\sigma^{\alpha}_{CN}(E_n)P^{\alpha}_{\gamma}
(E_x^{\alpha})}{\sigma^{\beta}_{CN}(E_n)P^{\beta}_{\gamma}(E_x^{\beta})},
\label{eq:srm}
\end{equation}
where the two cross sections $\sigma_{CN}(E_n)$ are calculated from optical compound-nucleus formation models.
It is worth mentioning that the excitation energies $E_x$ of the two systems
(corresponding to the neutron energy $E_n$)
is applied to obtain the ratio for experimentally $\gamma$-decay branchings
$P_{\gamma}$. The branchings are measured by means of $\gamma$-coincidence experiments.
There are various versions of the surrogate ratio method. The \textit{internal} and \textit{external}
surrogate approaches
are based on ratios describing one compound nucleus with different
decay channels or two different compound systems, but with the same exit channels.

\begin{figure}[tb]
\begin{center}
\includegraphics[clip,width=0.5\columnwidth]{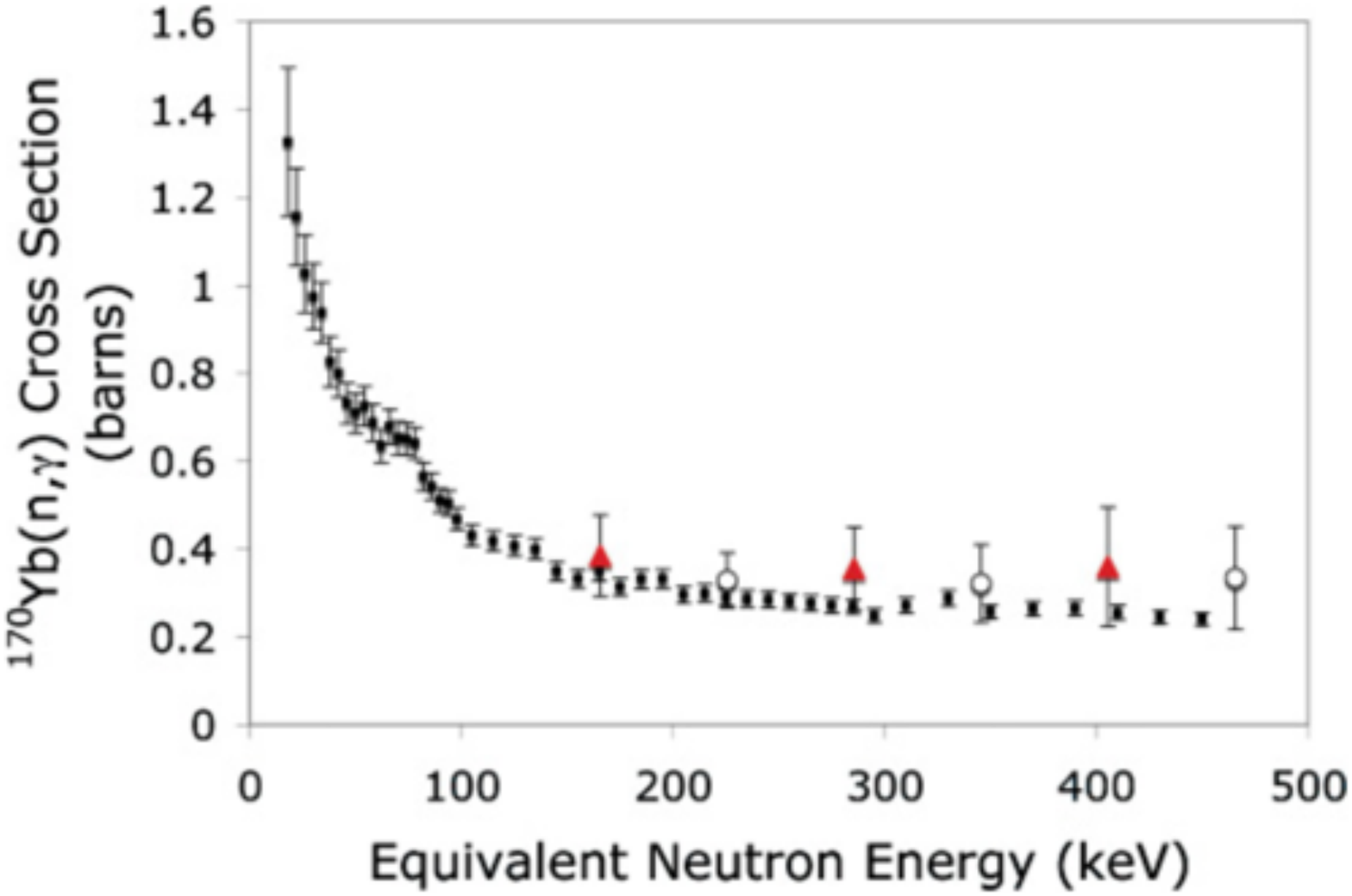}
\caption{(Color online) The $^{170}$Yb$(n, \gamma)$ cross section
extracted from the ratio method using the
$(^3$He, $^3$He$^\prime$)$^{170}$Yb (open circles) and
($^3$He,$\alpha)^{170}$Yb (filled red triangles) as surrogate reactions.
The known reaction used in Eq.~(\ref{eq:srm}), was $^{160}$Dy$(n,\gamma)$.
The data are compared with the previously direct measured $^{170}$Yb$(n,\gamma)$
cross section (filled squares). Figure from Goldblum~\textit{et al.}~\cite{Goldblum2008}.
}
\label{fig:yb}
\end{center}
\end{figure}

Figure \ref{fig:yb} shows results from the surrogate ratio method by
indirectly measuring the $^{170}$Yb$(n, \gamma)$ cross section~\cite{Goldblum2008}.
By using the method in an equivalent neutron energy range of 165 to 465 keV,
the cross sections extracted using the $(^3$He, $^3$He$^\prime$) and ($^3$He,$\alpha)$
surrogate reactions were consistent with the directly measured cross section. The
agreement for $E_n<0.5$ MeV demonstrates
that the surrogate ratio method may work for neutron energies relevant for the $r$ process.
The reason for this achievement is that
the ratios taken in Eq.~(\ref{eq:srm}) tend to cancel the $J,\pi$ dependencies.

\begin{figure}[tb]
\begin{center}
\includegraphics[clip,width=1.0\columnwidth]{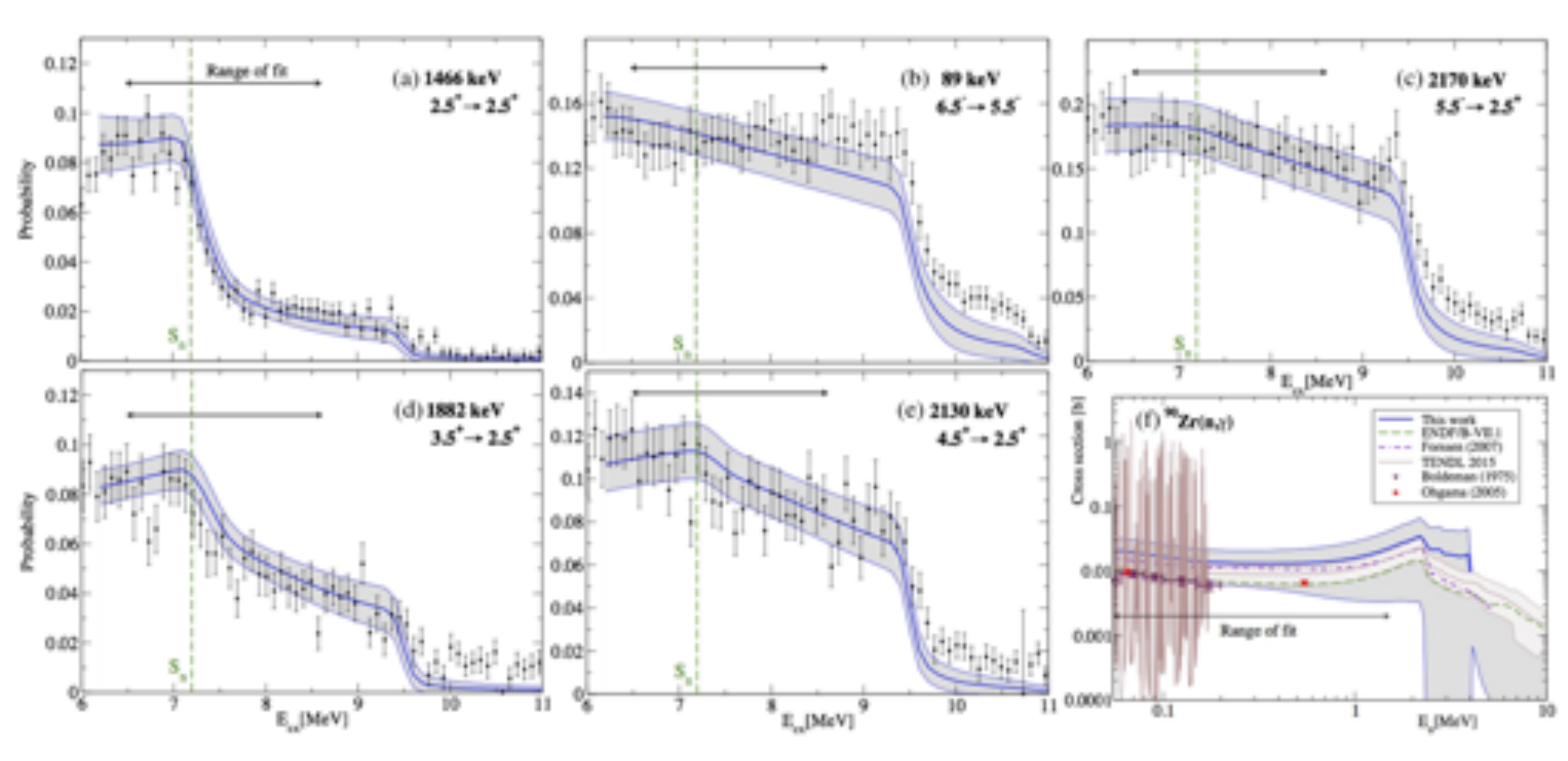}
\caption{(Color online) Probabilities $P_{\gamma}(E_x,J,\pi)$ for observing specific
$\gamma$-ray transitions  in coincidence with the outgoing deuteron
in the surrogate $^{92}$Zr$(p,d)$ reaction (a-e). Panel (f)
shows the extracted $^{90}$Zr$(n, \gamma)$ cross section
compared to direct measurements (red squares and black triangles)
and several evaluations.
Figure from Escher~\textit{et al.}~\cite{Escher2018}.
}
\label{fig:pg}
\end{center}
\end{figure}

Very recently, Escher \textit{et al.}~\cite{Escher2018}
accomplished the surrogate method without exploiting the Weisskopf-Ewing approximation.
By exploring a large parameter space, they demonstrated
that a set of model parameters for the level density and $\gamma$-ray strength function could describe the experimental $\gamma$-decay probabilities $P_{\gamma}(E_n,J,\pi)$. Figure~\ref{fig:pg}
shows that the spin distribution above the neutron separation energy
is well reproduced by these calculations. Putting gates
on transitions between higher-spin states reveal significant
delay in the abrupt drop in $P_{\gamma}$ as function of excitation energy.
This is due to the lack of higher spins in the lower excitation
region of $^{91}$Zr; levels with spin/parity of $(9/2)^+$ and $(11/2)^-$ or higher spins appear first for $E_x>2.1$~MeV.
The parameters from these simulations were then inserted into Eq.~(\ref{eq:sigma})
in order to estimate the neutron capture cross section.
In the work of Ratkiewicz, Cizewski \textit{et al.}~\cite{ratkiewicz2015,ratkiewicz2018}, a similar technique was applied, this time with the $^{95}$Mo($d,p\gamma$)$^{96}$Mo reaction  utilized as a proxy for the desired ($n,\gamma$) reaction, 

One possible drawback of this approach is
that the transfer reaction with its optical potential and nuclear structure
details have to be modelled. The $\gamma$-decay branch above $S_n$ strongly depends on
the $J,\pi$-dependent nuclear level density and the $\gamma$ strength function. In fact,
the $\gamma$-decay probability depends on an integral of the products of these two
functions (see, \textit{e.g.}, Bartholomew \textit{et al.}~\cite{bartholomew1972}). For such calculations, there are several models and parameter sets
that can be chosen from literature, and typically the spread in calculated cross sections may vary within factors of $\approx 5-10$.
To quantitatively determine this spread, Escher \textit{et al.}~\cite{Escher2018} have tested the sensitivity of the calculations using several input options~\cite{RIPL3,Gilbert1965,Schwengner2018,Utsunomiya2008,Mumpower2017}.
The resulting systematic-error band is shown in panel (f) of Fig.~\ref{fig:pg}.

In general, reactions 
that transfer a minimum of angular momentum to the
target nucleus are ideal for the surrogate method. 
Then the experimental results would be easier to interpret
in order to estimate the direct $(n,\gamma)$ cross section.
Therefore, light ion reactions like the $(p,p')$, $(d,p)$ and $(p,d)$ reactions
are all good candidates for surrogate experiments. Using beam energies
below or close to the Coulomb barrier, the $(d,p)$ reaction can be an excellent choice. Enhanced deuteron break-up in the vicinity of the target nucleus would
allow a bare neutron to enter the nucleus, thus effectively mimic a neutron induced reaction.
Furthermore, the surrogate reactions method is a complementary technique to obtain information
on $r$-process nuclei. The experimental set-up for stable targets can easily
by modified to inverse kinematics using radioactive beams on light fixed targets. The next
generation of radioactive beam facilities such as FRIB, Spiral2 and FAIR would support high beam intensities necessary for
such studies. In this way, the surrogate method could be used to make a whole range of
measurements which remain difficult with other methods at the present time.

\section{Brief overview of the standard Oslo method}
\label{sec:oslo}
As the $\beta$-Oslo method is based on the original \textit{Oslo method}~\cite{schiller2000,guttormsen1996,guttormsen1987}, the main features will be outlined in the following. The method allows for simultaneous determination of the functional form of the level density and $\gamma$-ray strength function in one and the same experiment. The method is in principle model independent, relying on a set of assumptions that has been thoroughly tested for the statistical decay region, for which the method is valid. 
However, one should note that the extracted level density and $\gamma$-ray strength results require \textit{absolute normalization}, which will introduce some model dependency, in particular on the applied spin-distribution model. 

\begin{figure}[tb]
\begin{center}
\includegraphics[width=0.8\columnwidth]{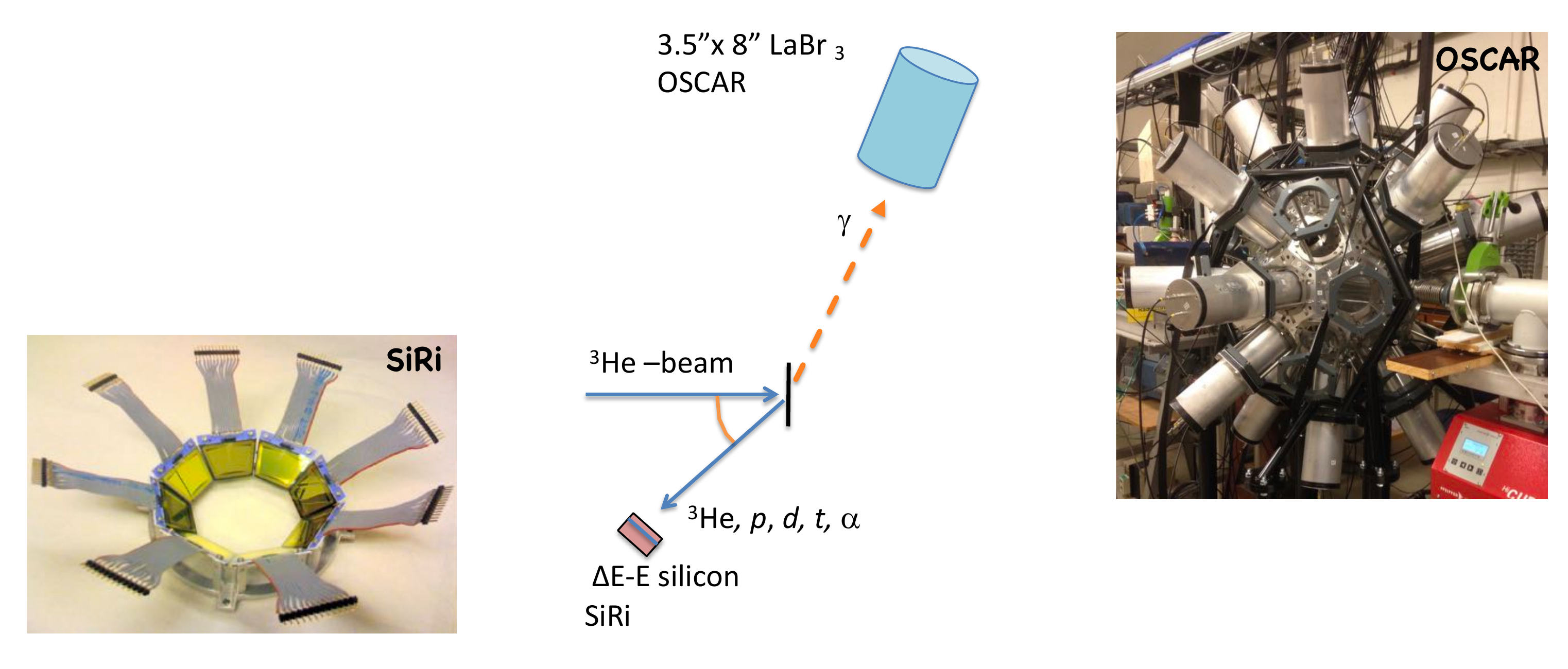}
\caption{(Color online) Typical particle-$\gamma$ coincidence set-up for the standard Oslo method.
The 64 silicon particle telescopes of SiRi are placed in the vacuum chamber at the center of the LaBr$_3$:Ce detector array OSCAR.}
\label{fig:set-up}
\end{center}
\end{figure}

Typically, the standard Oslo method is based on particle-$\gamma$ coincidences using light-ion reactions with one charged ejectile. The most frequently used reactions have been the $(p,p')$, $(d,p)$, $(d,t)$, ($^{3}$He, $^{3}$He') and ($^{3}$He, $^{4}$He) reactions. The method includes the following steps: 
\begin{itemize}
\item{Measure $\gamma$-ray spectra as function of initial excitation energy $E_x$}
\item{Correct the $\gamma$ spectra for the detector response~\cite{guttormsen1996}}
\item{Extract the distribution of the $\gamma$ rays emitted \textit{first} in all
decay cascades (primary $\gamma$ rays) for a given $E_x$~\cite{guttormsen1987}}
\item{Perform a simultaneous fit of all primary $\gamma$-ray spectra for a selected range of excitation regions, and obtain the level density and $\gamma$-ray strength function~\cite{schiller2000}}
\item{Normalize the level density and $\gamma$-ray strength function and evaluate 
systematic errors~\cite{schiller2000,larsen2011}}
\end{itemize}

Figure \ref{fig:set-up} shows the particle-$\gamma$ set-up at the Oslo Cyclotron Laboratory. A silicon particle detection system (SiRi)~\cite{siri}, which
consists of 64 telescopes, is used for the selection of a certain ejectile type and to determine its energy. The
front $\Delta E$ and back $E$ detectors have thicknesses of 130 $\mu$m and 1550 $\mu$m, respectively. SiRi
is usually placed 5 cm from target in backward angles covering $\theta = 126^\circ$ to $140^\circ$
relative to the beam axis. Coincidences with $\gamma$ rays are now performed with the new OSCAR array of 30 $3.5'' \times 8''$ LaBr$_3$:Ce detectors. Previous experiments made use of the CACTUS array~\cite{CACTUS}, consisting of 28 collimated $5'' \times 5''$ NaI:Tl scintillator crystals with a total efficiency of $\approx 15$\%.

\begin{figure}[tb]
\begin{center}
\includegraphics[width=0.9\columnwidth]{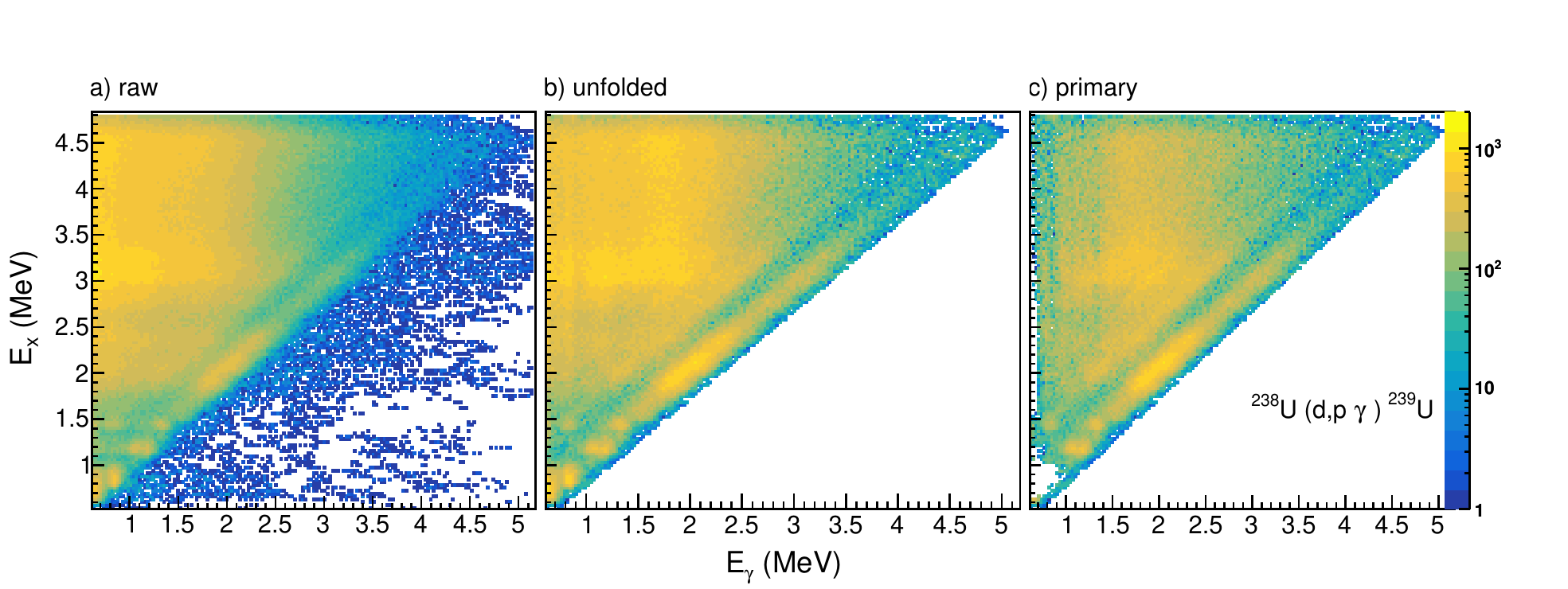}
\caption{(Color online) The raw, unfolded and primary 
matrices for $^{239}$U from the $^{238}$U$(d,p \gamma)$$^{239}$U reaction~\cite{guttormsen2013,guttormsen2014}.}
%
\label{fig:U239matrices}
\end{center}
\end{figure}

The first step in the analysis is to sort the $\gamma$-ray
spectra as function of initial excitation energy $E_x$.
Knowing the details of the reaction kinematics,
$E_x$ is given by the energy of the outgoing charged particle.  The coincidence data are sorted into a raw particle-$\gamma$ matrix $R(E_{\gamma},E_x)$ with proper subtraction of random coincidences. Then, for all $E_x$, the $\gamma$ spectra are unfolded~\cite{guttormsen1996} with the detector response
functions giving the matrix $U(E_{\gamma},E_x)$. The procedure is iterative and stops
when the folding ${\cal F}$ of the unfolded matrix equals the raw matrix within the statistical fluctuations, i.e.~when ${\cal F}(U)\approx R$. The raw and unfolded $\gamma$ matrices are shown in Fig.~\ref{fig:U239matrices} for the $^{238}$U$(d,p \gamma$)$^{239}$U reaction~\cite{guttormsen2013,guttormsen2014}.

\begin{figure}[tb]
\begin{center}
\includegraphics[width=0.8\columnwidth]{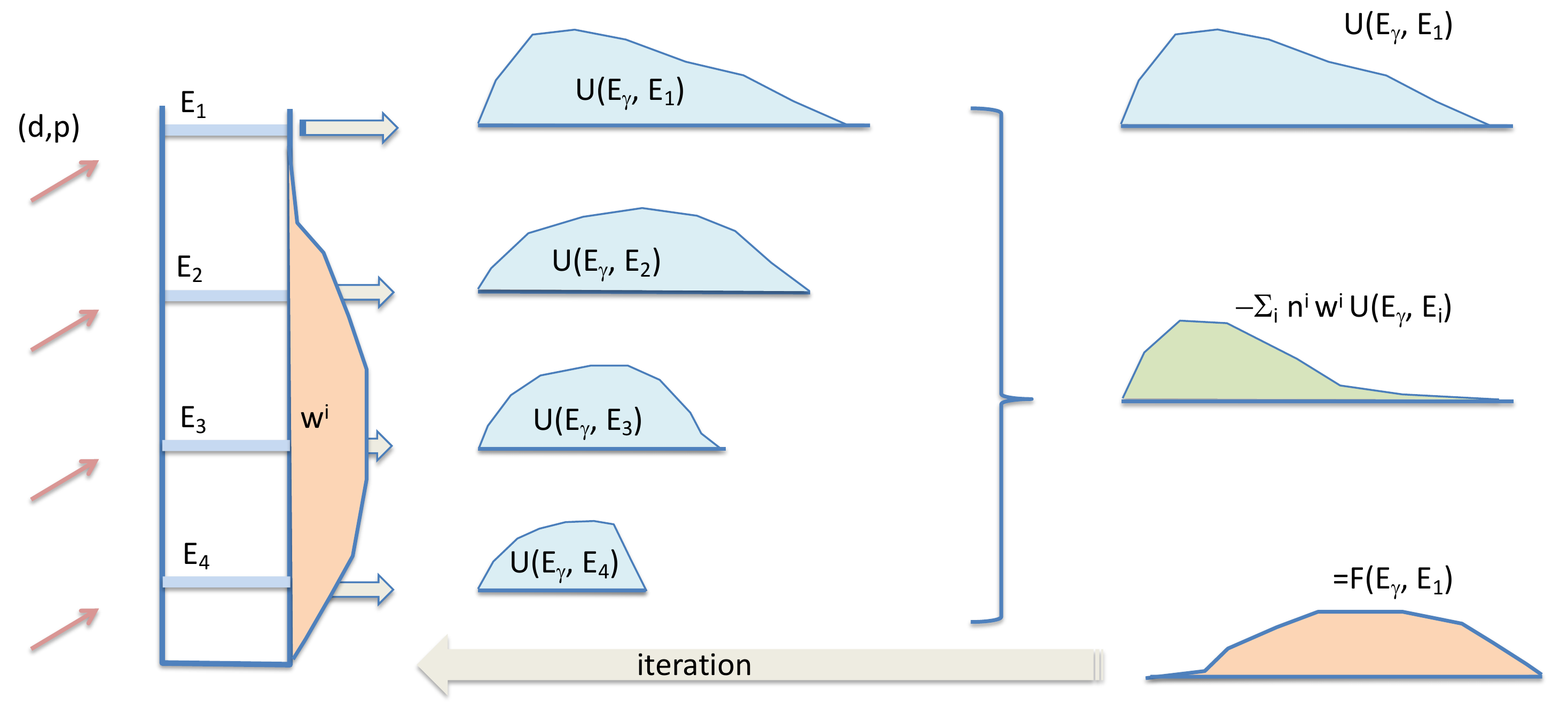}
\caption{(Color online) Illustration of the first-generation method. The nucleus 
is assumed to be populated by the $(d,p)$ reaction shown to  the left. The primary spectrum is obtained by subtracting a weighted sum of the total spectra at lower excitation energies. The difference spectrum $F$ is translated into the next iteration weighting function $w^i$ (see text).}
%
\label{fig:firstgen}
\end{center}
\end{figure}

The first-generation method is based on the assumption that states populated after the first $\gamma$ transition have the same decay properties as states populated directly in the particle reaction at that excitation energy.
Thus, the energy distribution of the first-generation (or
primary) $\gamma$ rays can be extracted from the unfolded total $\gamma$-ray spectra $U$ as illustrated in Fig.~\ref{fig:firstgen}.
The primary spectrum $F$ is obtained by a subtraction of a weighted sum of
$U(E_{\gamma},E_x')$ spectra at or below $E_x$~\cite{guttormsen1987}
\begin{equation}
F(E_{\gamma},E_x)=U(E_{\gamma},E_x) - \sum_{E_x' \leq E_x}n^{E_x}(E_x') w^{E_x}(E_x')U(E_{\gamma},E_x').
\label{eq:primary}
\end{equation}

The factor $n$, which takes care of the specific population cross section,
can be determined by the singles-particle spectrum or by the $U$ matrix~\cite{guttormsen1987}. The weighting function $w^{E_x}(E_x')$, where $\sum_{E_x' \leq E_x} w^{E_x}(E_x')=1$, is unknown. Since $w^{E_x}(E_x')$ corresponds to the branching from levels at $E_x$ to levels at $E_x'$, this quantity corresponds to the normalized primary $F$ spectrum of Eq.~(\ref{eq:primary}). The close connection between $F$ (as a function of $E_{\gamma}$) and $w$ (as a function of $E_x'$) allows for an iteration procedure using Eq.~(\ref{eq:primary}). The first step starts with a trial weighting function $w$ and calculate $F$. Then the primary $F$ spectrum is translated into the next function $w$. In this way the next iteration can proceed and an improved estimate of $F$ is obtained. The iteration procedure is found to converge fast, almost independently on the trial function~\cite{guttormsen1987}. Figure~\ref{fig:U239matrices} shows the
resulting primary $\gamma$ matrix $F(E_{\gamma},E_x)$.

After a successful extraction of the primary $\gamma$-ray spectrum, the procedure allows for a consistency check. The area $A^F$ (number of counts) of the primary spectrum $F$ with $\gamma$-ray multiplicity $M_{\gamma}=1$ should compare to the area $A^U$ of $U$ corresponding to the total $\gamma$ multiplicity of the average cascades from $E_x$. Thus, ideally this total $\gamma$ multiplicity may be expressed by the relations~\cite{guttormsen1987}
\begin{eqnarray}
M_{\gamma}(E_x) &=& \frac{E_x}{\langle E_{\gamma}(E_x) \rangle}\approx \frac{A^U(E_x)}{A^F(E_x)} ,
\label{eq:mult}
\end{eqnarray}
where $\langle E_{\gamma}(E_x)\rangle$ is the average energy of the total $\gamma$ spectrum at $E_x$.

The next step of the Oslo method is to copy $F(E_{\gamma},E_x)$ into a matrix $P(E_{\gamma}, E_x)$ and normalize each $\gamma$-ray spectrum so that
$\sum_{E_{\gamma}}P(E_{\gamma}, E_x)=1$. 
It is then factorized $P$ into two functions by 
\begin{equation}
P(E_{\gamma}, E_x) \propto   \rho(E_x-E_{\gamma}){\cal{T}}(E_{\gamma}) ,\
\label{eqn:rhoT}
\end{equation}
where the decay probability is assumed to be proportional to the
level density at the final energy $\rho(E_x-E_{\gamma})$ according
to Fermi's golden rule~\cite{dirac,fermi}. The decay is also proportional
to the $\gamma$-ray transmission coefficient ${\cal{T}}$, which is assumed
to be independent of excitation energy according to the Brink hypothesis~\cite{brink,guttormsen2016}.
The relation (\ref{eqn:rhoT}) makes it
possible to simultaneously extract the two one-dimensional functions $\rho$
and ${\cal{T}}$ from the two-dimensional landscape $P$. We use the
iteration procedure of Schiller {\em et al.}~\cite{schiller2000}
to determine $\rho$ and ${\cal{T}}$ by a least-$\chi ^2$ fit
using relation (\ref{eqn:rhoT}). For the extraction of $\rho$ and ${\cal{T}}$, it is important to select the higher $E_x$ part of $P$ where the first-generation
procedure works well, i.e. the relation Eq.~(\ref{eq:mult}) is reasonably fulfilled.

There are infinitely many functions  $\rho$ and  $\cal{T}$ that make identical
fits to the experimental $P$ matrix.
These $\tilde{\rho}$ and $\tilde{\cal{T}}$ functions can be generated by the transformations~\cite{schiller2000}
\begin{eqnarray}
\tilde{\rho}(E_x-E_\gamma)&=&A\exp[\alpha(E_x-E_\gamma)]\,\rho(E_x-E_\gamma),
\label{eq:array1}\\
\tilde{{\mathcal{T}}}(E_\gamma)&=&B\exp(\alpha E_\gamma){\mathcal{T}} (E_\gamma).
\label{eq:array2}
\end{eqnarray}
Thus, the normalization of the two functions requires additional information to fix
the parameters $A$, $\alpha$, and $B$ that have to be taken from other experiments.

\begin{figure}[tb]
\begin{center}
\includegraphics[width=0.7\columnwidth]{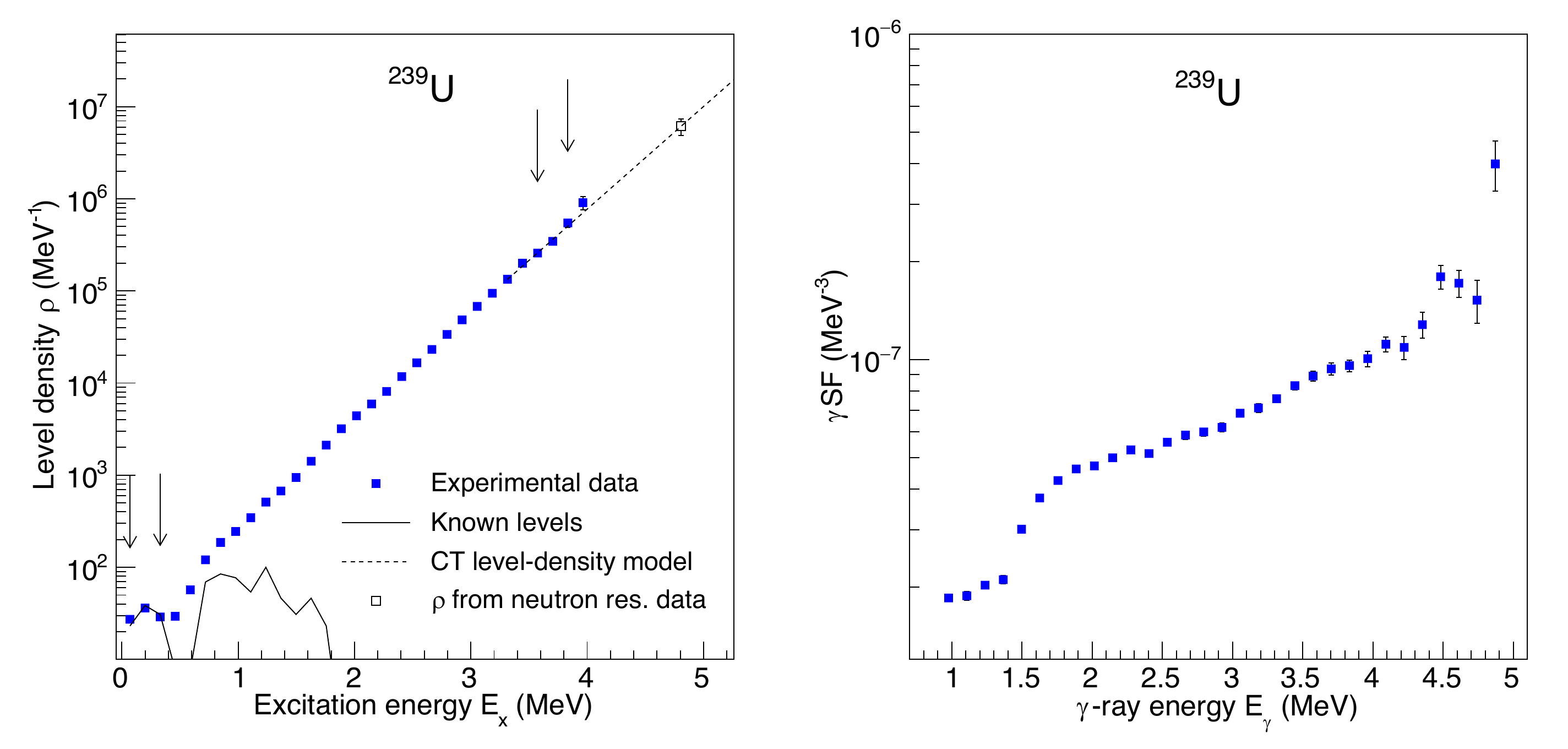}
\caption{(Color online) Level density (left) and $\gamma$-ray strength function (right) for $^{239}$U extracted with the Oslo method, data from Refs.~\cite{guttormsen2013,guttormsen2014}. The bump in the strength for transition energies $E_\gamma \approx 1.5-2.5$ MeV is interpreted as the $M1$ scissors mode (see Ref.~\cite{guttormsen2014} and references therein).}
%
\label{fig:nld_gsf}
\end{center}
\end{figure}

For the level density, two anchor points are used to determine $A$ and $\alpha$.
The procedure is demonstrated for the case of $^{239}$U in the left part of Fig.~\ref{fig:nld_gsf}.
The level density is normalized to known discrete levels at low excitation energy.
At high excitation energy, the neutron resonance spacings $D_0$ are used to estimate the total level density $\rho$
at the neutron separation energy $S_n$. In order to bridge the
energy gap between the data points and the estimated $\rho(S_n)$ value,
the constant-temperature level density formula of Eq.~(\ref{eq:consttemp}) is adopted,
where the temperature $T_{\rm CT}$ and energy shift $E_0$ are free parameters
adjusted to the data. For $^{239}$U in Fig.~\ref{fig:nld_gsf}, $T_{\rm CT}=0.39$~MeV.

In order to translate the spacing $D_0$
to total level density, the spin distribution~\cite{Gilbert1965} is used
\begin{equation}
g(E_x,J) = \frac{2J+1}{2\sigma^2(E_x)}\exp\left[-(J+1/2)^2/2\sigma^2(E_x)\right],
\label{eq:spindist}
\end{equation}
where $E_x$ is the excitation energy and $J$ is the spin. Here, the spin cut-off parameter
$\sigma$ is rather uncertain and may be estimated from various models.
For cases where $D_0$ is unknown, one has to use systematics to estimate $\rho(S_n)$.

The spin cut-off parameter $\sigma$ is traditionally determined by a close-to rigid moment of inertia.
Since $\sigma^2= \Theta T/\hbar^2$~\cite{Ericson1959} and the nuclear temperature $T$
is assumed to be approximately constant for $2\Delta < E_x < S_n$~\cite{Ericson1959,Moretto1975,Moretto2015,zelevinsky2018}, $\sigma^2$
follows the energy dependence of the moment of inertia $\Theta$. $\Theta$ is assumed to be proportional to the number of quasi-particles, which again is proportional to $E_x$. Thus,
\begin{equation}
\sigma^2(E_x)=\sigma_d^2 + \frac{E_x-E_d}{S_n-E_d}\left[\sigma^2(S_n)-\sigma_d^2\right],
\label{eq:sigE}
\end{equation}
which goes through  two anchor points. The first point $\sigma_d^2$
is determined from known discrete levels at low excitation energy $E_x=E_d$.
The second point at $E_x=S_n$ is estimated assuming a rigid moment of inertia~\cite{egidy2005}:
\begin{equation}
\sigma^2(S_n) =  0.0146 A^{5/3} \cdot \frac{1+\sqrt{1+4aU_n}}{2a},
\label{eqn:eb}
\end{equation}
where $A$ is the mass number, and $U_n=S_n-E_1$ is the intrinsic excitation energy.
The single particle level density parameter $a$ and the energy shift parameter $E_1$ is determined according to Ref.~\cite{egidy2005}. In order to obtain a systematic error band, one may multiply
the rigid moment of inertia $\Theta_{\rm rigid}=0.0146 A^{5/3}$ of Eq.~(\ref{eqn:eb})
with a factor $\eta$, which takes the values $\eta =0.8$, 0.9 and 1.0 for the
low, recommended and high values, respectively. Such a systematical error band is consistent with theoretical estimates~\cite{Rauscher1997,RIPL3,uhrenholt2013,alhassid2015,Grimes2016}. 

The usual way to determine the normalization coefficient $B$ of Eq.~(\ref{eq:array2}) is
to constrain the data to the known average total radiative
width $\langle \Gamma_{\gamma} \rangle$ at $S_n$ \cite{schiller2000,voinov2001}, which is defined as 
\begin{eqnarray}
\langle\Gamma_{\gamma } (S_n)\rangle=\frac{1}{2\pi\rho(S_n, I, \pi)} \sum_{I_f}\int_0^{S_n}{\mathrm{d}}E_{\gamma} {\cal {T}}(E_{\gamma})
\times \rho(S_n-E_{\gamma}, I_f),
\label{eq:norm}
\end{eqnarray}
where the summation and integration run over all final levels with spin $I_f$ that are accessible by $E1$ or $M1$ transitions with energy $E_{\gamma}$. This procedure is known to work well
when the individual $\gamma$ widths are centered around a common average value. In order to translate the transmission coefficient to the dipole $\gamma$-ray strength function, including both the $E1$ and $M1$ contributions,
Eq.~(\ref{eq:Ttof}) is used.
The final normalization for the $\gamma$-ray strength function of $^{239}$U is shown in the right part of Fig.~\ref{fig:nld_gsf}. The extra $M1$ $\gamma$ strength of $B= 8.8(14) \mu^2$ located at $E_{\gamma}= 2.41(15)$~MeV is interpreted as the scissors resonance~\cite{guttormsen2014}.
Further description and tests of the Oslo method are given in Refs.~\cite{schiller2000,larsen2011}.

\begin{figure}[tb]
\begin{center}
\includegraphics[width=0.95\columnwidth]{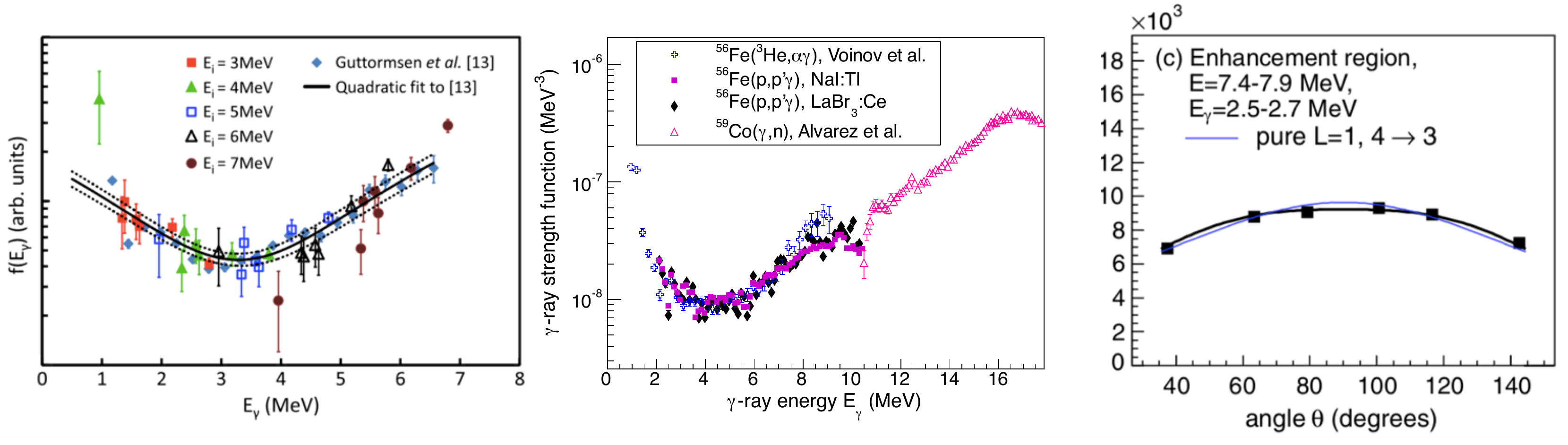}
\caption{(Color online) (a) Comparison between the $\gamma$-ray strength function enhancement (upbend) in $^{95}$Mo from previous experiments of Guttormsen \textit{et al.}~\cite{guttormsen2005} and the new technique of Wiedeking \textit{et al.}~\cite{wiedeking2012}. (b) The upbend in $^{56}$Fe has been verified with different reactions and detectors~\cite{larsen2013}.  (c) The angular distribution of the upbend strength at high excitation energy in $^{56}$Fe is compatible with dipole radiation. Figures from Refs.~\cite{wiedeking2012,larsen2013}.} 
%
\label{fig:upbend}
\end{center}
\end{figure}

One of the most surprising discoveries with the Oslo method is the \textit{upbend},
an unexpected increase in the $\gamma$-ray strength function at low transition energies and high
excitation energies. This peculiar feature was first seen in Fe isotopes~\cite{voinov2004},
and later on found in many nuclei such as $^{43-45}$Sc~\cite{larsen2007,burger2012},
$^{93-98}$Mo~\cite{guttormsen2005}, $^{138,139}$La~\cite{kheswa2015} and recently in the heavy 
$^{151,153}$Sm~\cite{simon2016}. The upbend was independently confirmed in $^{95}$Mo
using a completely different technique to map out the functional form of 
the $\gamma$-ray strength function~\cite{wiedeking2012}, and it was shown to be dominantly 
of dipole nature in $^{56}$Fe~\cite{larsen2013}, see Fig.~\ref{fig:upbend}. At present, it is not clear whether it is of electric~\cite{litvinova2013} or magnetic type~\cite{schwengner2013,brown2014,schwengner2017,sieja2017,midtbo2018}. 
That said, a recent Compton-polarization experiment on the $^{56}$Fe($(p,p'\gamma)$ reaction using GRETINA by M.~D.~Jones \textit{et al.}~\cite{mjones2017} does indicate a small bias towards the upbend being dominantly of magnetic character. 
If the upbend turns out to be present also for very neutron-rich nuclei, 
this could boost the ($n,\gamma$) reaction rates by $\sim 2$ orders of magnitude~\cite{larsen2010}.
To prove or disprove this, one must measure the $\gamma$-ray strength function of neutron-rich nuclei.

In recent publications~\cite{utsunomiya2013,tornyi2014,laplace2016,larsen2016,tveten2016,crespocampo2016},
the level density and $\gamma$-ray strength function measured with the Oslo method
have been successfully used as input to the {\sf TALYS} nuclear reaction code~\cite{TALYS,koning12}
to calculate radiative capture cross-sections. However, the standard Oslo method 
is limited to stable targets giving only information on nuclei close to the valley of stability. In the following,
we describe new approaches to extract the level density and $\gamma$-ray strength function for neutron-rich nuclei.  

\section{New experimental techniques}
\label{sec:exp}
Here we sketch the main principles of two new experimental approaches to measure
level density and $\gamma$-ray strength function for exotic nuclei. Both approaches use the basic
elements of the standard Oslo method, but have different experimental
features to obtain the starting point for the level density and $\gamma$-ray strength function extraction,
namely $\gamma$-ray spectra as function of initial excitation energy $E_i$. 
Also the two new methods encounter different challenges as will be briefly
discussed.
It should be noted that the surrogate technique described in Sec.~\ref{sec:surrogate} can also be applied in inverse kinematics and thus can contribute to the effort of understanding neutron-capture reactions far from stability.

\subsection{The Oslo method in inverse kinematics}
\label{sub:inverse}
The standard Oslo method employs a light-ion beam impinging on a heavy target,
measuring the emitted light-charged particles in coincidence with
$\gamma$ rays from the deexcitation of the populated entry state. The energy and angle of the light-charged particles gives information
on the excitation energy $E_i$ of the residual nucleus (below $S_n$), and therefore the $\gamma$ rays 
are tagged with that specific initial excitation energy. Typically, 
one needs $\approx 40,000$ particle-$\gamma$ coincidences to get reasonable 
error bars on the level density and $\gamma$-ray strength function. 

In recent years, significant effort was made to develop the Oslo method in ``inverse kinematics'', where a heavy beam would impinge on a light target. 
The advantage of this approach is to be able to study nuclei away from the valley of stability using radioactive beams. It can also be useful for some specific stable isotopes, which may be challenging to manufacture as targets. In inverse kinematics, together with the light-charged particles, the recoiling nucleus also typically escapes the target and can be detected if necessary. 
Similar to the standard Oslo method, the combination of beam and target, as well as the type of reaction used, can vary significantly depending on availability, 
and also beam energy, providing a unique complementarity to the various radioactive beam facilities around the world (see Sec.~\ref{sec:facilities}). 

The first proof-of-principle experiment demonstrating the feasibility of
\textit{the Oslo method in inverse kinematics}, was performed at iThemba LABS
in 2015~\cite{siem_wiedeking2015}, with a 300-MeV $^{86}$Kr beam
directed onto C$_2$D$_4$ and C$_8$H$_3$D$_5$ targets. 
Charged particles were measured with two DSSD detectors in 
a ${\Delta}E - E$ configuration, while the $\gamma$ rays were detected with 
the AFRODITE array comprised of eight BGO-shielded CLOVER detectors 
and two large-volume LaBr$_3$:Ce detectors.
The $^{86}$Kr($d,p\gamma$)$^{87}$Kr reaction was selected and analyzed to extract the level density and $\gamma$-ray strength function from the coincidence data. 
The results were very promising and in 
excellent agreement with previous data from other measurements~\cite{ingeberg2016}.

The first radioactive beam experiment applying the Oslo method in inverse kinematics was performed at 
HIE-ISOLDE, CERN, in November 2016~\cite{siem_wiedeking_hieisolde}. A radioactive $^{66}$Ni beam at $\approx 4.5$-MeV/nucleon, 
was combined with a C$_2$D$_4$ target, providing access to two reactions:  $^{66}$Ni($d,p\gamma$)$^{67}$Ni and  $^{66}$Ni($d,t\gamma$)$^{65}$Ni.
Charged particles from the two reactions were collected using C-REX (an upgrade of the T-REX detector array~\cite{TREX}), while the $\gamma$ rays were measured
with MINIBALL~\cite{reiter2002} together with six large-volume LaBr$_3$:Ce detectors. 
The run was very successful and the data are currently being analyzed. The success of this experiment is of major importance for the field because on top of 
providing the first ever measurement of the level density and $\gamma$-ray strength function for the particular isotopes, it opens the path towards more radioactive beam experiments at HIE-ISOLDE and other facilities, 
provided that the beam intensity and reaction cross section result in sufficient yields. 

A new approach to enhance the reaction yield in inverse kinematics experiments is currently under development by the University of Guelph, TRIUMF, Colorado School of Mines and the Technical University of Munich, to be implemented at the TRIUMF laboratory, in Canada (see Sec.~\ref{sec:facilities}). The TIGRESS Silicon Tracker Array (TI-STAR)~\cite{muecher2018} will be a compact detector inside the TIGRESS $\gamma$-ray spectrometer~\cite{scraggs2005,hackman2014}, designed for transfer reactions with heavy radioactive ion beams. The detector will host an extended deuterium gas target at atmospheric pressure to induce one-neutron transfer reactions with the incoming exotic ion beam. Different from (active) gas detectors, TI-STAR will use a silicon tracker with an ultra-thin first silicon layer to track the emitted target-like nuclei. Event data of the 250~$\mu m$ pitch silicon strip detectors $(\Delta E- \Delta E-E)$ will be collected using the SKIROC-2 ASIC~\cite{callier2011}, connected to a custom-made digitizer card. Due to its tracking capabilities, TI-STAR will allow to maintain the good energy resolution of a thin target experiment ($\sim$ 150~keV), while increasing the reaction yield by 1-2 orders of magnitude due to the use of a thick pure deuterium gas target ($\sim$ 2.8~mg/cm$^2$). Experiments with radioactive beams down to $\approx$ 1000 pps are expected to be feasible with TI-STAR, giving access to nuclei much further from the valley of stability. First experiments are planned for 2020 at TRIUMF and in the future at the ARIEL facility, focusing on the mass region $A=130$, around the second $r$-process peak~\cite{muecher2018}. 

While transfer reactions are well suited for beam energies of a few MeV/u, different types of reactions can be used at fragmentation facilities around the world, which typically offer much higher beam energies. One such reaction-type that is currently investigated at the NSCL, at Michigan State University (see Sec.~\ref{sec:facilities}) is the charge-exchange reaction. In this reaction a proton in the beam nucleus is exchanged with a neutron from the target nucleus (or vice versa), in a reaction that mimics $\beta$ decay or electron capture~\cite{ichimura2006, harakeh2001, osterfeld1992, fujita2011}. Charge exchange reactions can be used with various probes other than the traditional $(p,n)$ and $(n,p)$ reactions, such as $(d,^2$He), $(t,^3$He), ($^3$He,$t$), and heavy-ion reactions, that offer experimental advantages such as improved energy resolution, or selectivity to specific types of excitations. In a recent charge-exchange reaction experiment performed at the NSCL, the charge exchange reaction of $^{46}$Ti$(t,^3$He $\gamma)^{46}$Sc was measured~\cite{noji2014, noji2015}. The S800 spectrograph was used to momentum-analyze the recoiling $^3$He nuclei and extract the excitation energy of the populated nucleus $^{46}$Sc. The deexcitation $\gamma$ rays were detected in the GRETINA $\gamma$-ray tracking array~\cite{paschalis2013}. These data were recently investigated as a proof of principle for using charge-exchange reactions within the Oslo method. Preliminary results are encouraging and have spurred interest for applying the Oslo technique using charge-exchange reactions in forward and inverse kinematics at the NSCL and in the future at FRIB (see Sec.~\ref{sec:facilities}).

\subsection{The $\beta$-Oslo method}

In many cases it is hard to achieve an intense enough radioactive beam 
to apply the Oslo method in inverse kinematics. One of the main 
strengths with the recently invented \textit{$\beta$-Oslo method}~\cite{spyrou2014}
is that one can obtain sufficient statistics with an implantation rate down to
1 particle per second. Here, one makes use of the fact that for
neutron-rich nuclei, the $Q$-value for $\beta$ decay is comparable to or even 
higher than the neutron separation energy, and so will populate excited states
in a broad energy range in the daughter nucleus. Further, using a 
\textit{segmented} total-absorption spectrometer such as the 
SuN detector~\cite{SuN} at the National Superconducting Cyclotron Laboratory (NSCL), Michigan State
University (MSU), one gets the sum of all $\gamma$ rays 
in the cascades providing a measure of the initial excitation energy,
while the single segments give the individual $\gamma$ rays. Thus,
one can obtain a set of excitation-energy tagged $\gamma$-ray spectra
and apply the Oslo method to extract level density and $\gamma$-ray strength function for the
daughter nucleus. 

\begin{figure}[bt]
\begin{center}
\includegraphics[width=0.9\columnwidth]{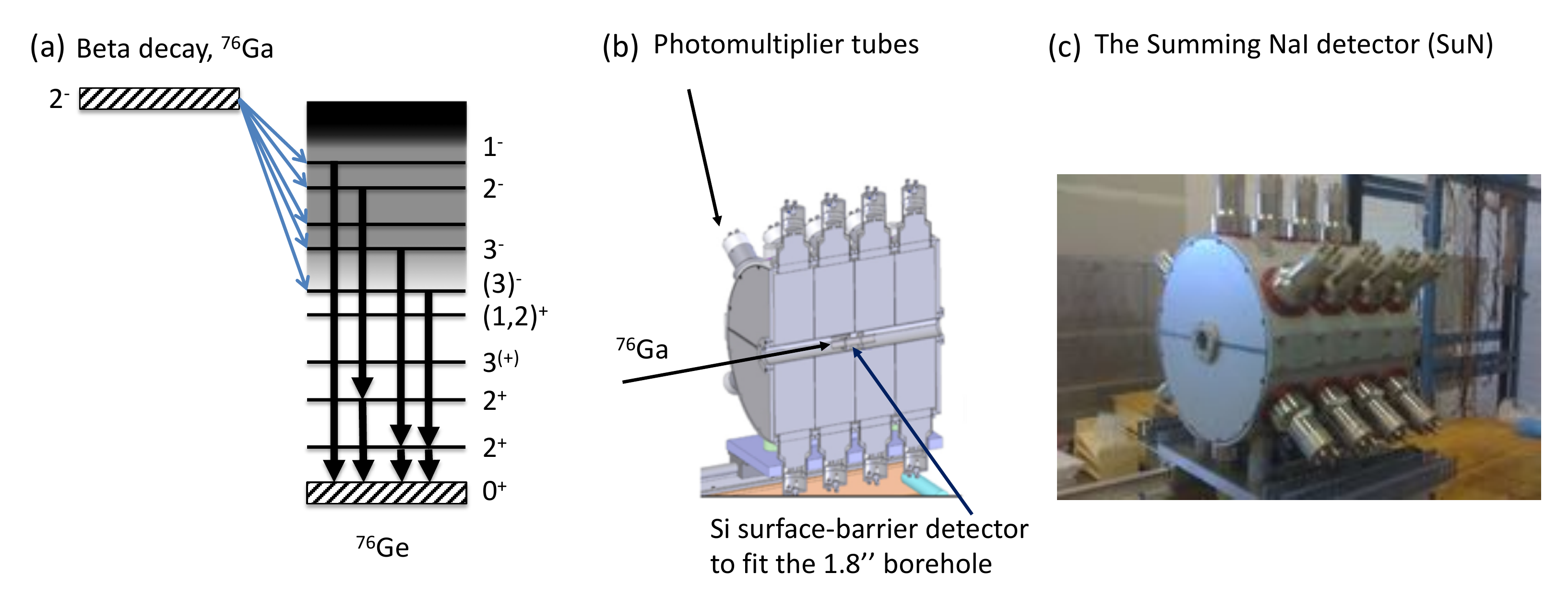}
\caption{(Color online) (a) Feeding of levels in $^{76}$Ge from the $\beta$ decay of the  2$^-$ ground state of $^{76}$Ga. (b) The beam of $^{76}$Ga is implanted in the center of the summing NaI detector SuN.  (c) The photo shows SuN with all PMTs.}
%
\label{fig:betaOslo}
\end{center}
\end{figure}

The $\beta$-Oslo method was first applied on $^{76}$Ga $\beta$-decaying into 
$^{76}$Ge~\cite{spyrou2014}. 
Figure \ref{fig:betaOslo} (a) illustrates the population of
levels in the $^{76}$Ge nucleus by Gamow-Teller  $\beta$ decay ($\Delta L = 0,1$, no parity change) from the 2$^-$ ground state of $^{76}$Ga. 
The excited levels will then decay by emitting $\gamma$ rays through all possible branchings for the initial levels. 
The $^{76}$Ga experiment was performed at the NSCL (MSU) using a 130-MeV/nucleon $^{76}$Ge beam producing
$^{76}$Ga by fragmentation on a thick beryllium target. The $^{76}$Ga
secondary beam was first guided through the
A1900 fragment separator~\cite{morrissey2003}, and then 
thermalized in the  
large-volume gas cell~\cite{cooper2014} and delivered to the 
experimental setup, where the $^{76}$Ga beam was 
implanted on a silicon surface barrier detector mounted inside SuN, detecting only the 
$\beta$ particles due to the low beam energy ($\approx 30$ keV after the gas cell). The implantation
into the center of SuN is illustrated in Fig.~\ref{fig:betaOslo}.
SuN was then used for detecting the subsequent $\gamma$-ray cascades in the 
daughter nucleus $^{76}$Ge. 

\begin{figure}[tb]
\begin{center}
\includegraphics[clip,width=1.\columnwidth]{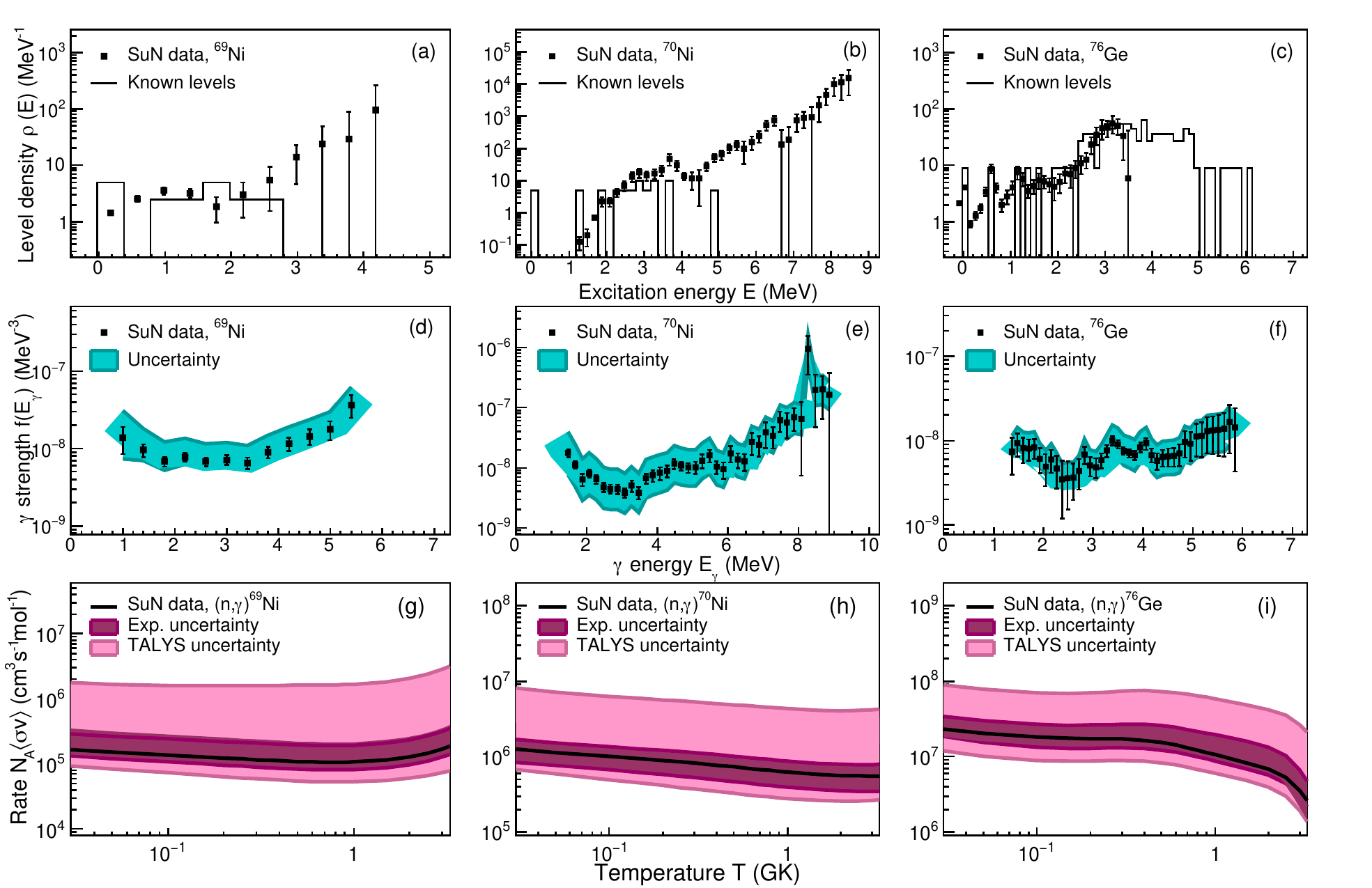}
\caption {(Color online) Data obtained with the $\beta$-Oslo method. Upper panels (a)-(c): Level densities of $^{69}$Ni, $^{70}$Ni and $^{76}$Ge.
Middle panels (d)-(f): 
Gamma strength functions of $^{69}$Ni, $^{70}$Ni and $^{76}$Ge.
Lower panels (g)-(i):
Radiative neutron-capture reaction rates for
$^{68}$Ni($n,\gamma$)$^{69}$Ni,
$^{69}$Ni($n,\gamma$)$^{70}$Ni and
$^{75}$Ge($n,\gamma$)$^{76}$Ge compared to the uncertainty band obtained by varying the input models for the level density and $\gamma$-strength function in {\sf TALYS}. 
The data are adapted from Ref.~\cite{spyrou2017} for $^{69}$Ni, Refs.~\cite{liddick2016,larsen2018} for $^{70}$Ni, and Ref.~\cite{spyrou2014} for $^{76}$Ge. }
\label{fig:rates}
\end{center}
\end{figure}

The measured level density and $\gamma$-ray strength function of $^{76}$Ge are shown in 
Fig.~\ref{fig:rates} (c) and (f), respectively (adapted from Ref.~\cite{spyrou2014}).
Although $^{76}$Ge is the most neutron-rich, stable Ge
isotope in nature, the uncertainty of the 
$^{75}$Ge($n,\gamma$)$^{76}$Ge reaction rate obtained by varying 
all level density, $\gamma$-ray strength function and n-OMP models in {\sf TALYS} is a factor of $\approx 8$. 
As shown in Fig.~\ref{fig:rates} (i), using the SuN data as input in {\sf TALYS}, the uncertainty
was significantly reduced to about a factor of $\approx 2$. 

Further, the $\beta$-Oslo method has recently been applied on the 
neutron-rich $^{70}$Co $\beta$-decaying into $^{70}$Ni~\cite{liddick2016,larsen2018}.
The experiment was performed in 2015 at NSCL, MSU,
where a primary 140-MeV/nucleon $^{86}$Kr beam hit a beryllium target
to produce $^{70}$Co through fragmentation. The fragmentation reaction products 
were separated with the A1900 separator and delivered to the experimental
setup, this time to a DSSD inside SuN, detecting both the fragment and the
$\beta$ particle. 
Spatial and time considerations were applied to properly correlate the $\beta$ particles and the implanted ions. Again, SuN was used to detect the $\gamma$-ray cascades 
from the daughter nucleus, $^{70}$Ni. 
The extracted level-density and $\gamma$-strength data are shown in Fig.~\ref{fig:rates} (b), (e). 
Complementary data from GSI on the $^{68}$Ni $\gamma$-ray strength function~\cite{rossi2013}
\textit{above} the neutron separation energy allowed for a well-determined
absolute normalization
of the full $\gamma$-ray strength function, giving uncertainties down to a factor $2-3$ in the
deduced $^{69}$Ni($n,\gamma$)$^{70}$Ni reaction rate (see Fig.~\ref{fig:rates} (h)). 
This is to be compared with the uncertainty band considered, \textit{e.g.}, in Ref.~\cite{surman2014}
of a factor of 100 (multiplying the JINA REACLIB rate~\cite{JINA-REACLIB} 
with a factor 0.1 and 10). It is clear that the data-constrained rates 
represent a significant improvement. 
During the experimental campaign leading to the $^{70}$Ni results discussed above,  data were also taken on $^{69}$Co populating excited levels in $^{69}$Ni as presented in Ref.~\cite{spyrou2017}.  
The resulting level density and $\gamma-$ray strength are presented in Fig.~\ref{fig:rates} (a) and (d), respectively, while the experimentally constrained rate is shown in Fig.~\ref{fig:rates} (g), again representing a significant reduction of the uncertainty. As shown in Ref.~\cite{liddick2016} and demonstrated in Fig.~\ref{fig:abundance_sensitivity},
the rates need to be determined within at least a factor of 10 
to be able to meaningfully compare fine structures in the isotopic abundances
with the nucleosynthesis simulation.

\section{Current and future facilities and equipment }
\label{sec:facilities}

\begin{figure}[tb]
\begin{center}
\includegraphics[clip,width=0.6\columnwidth]{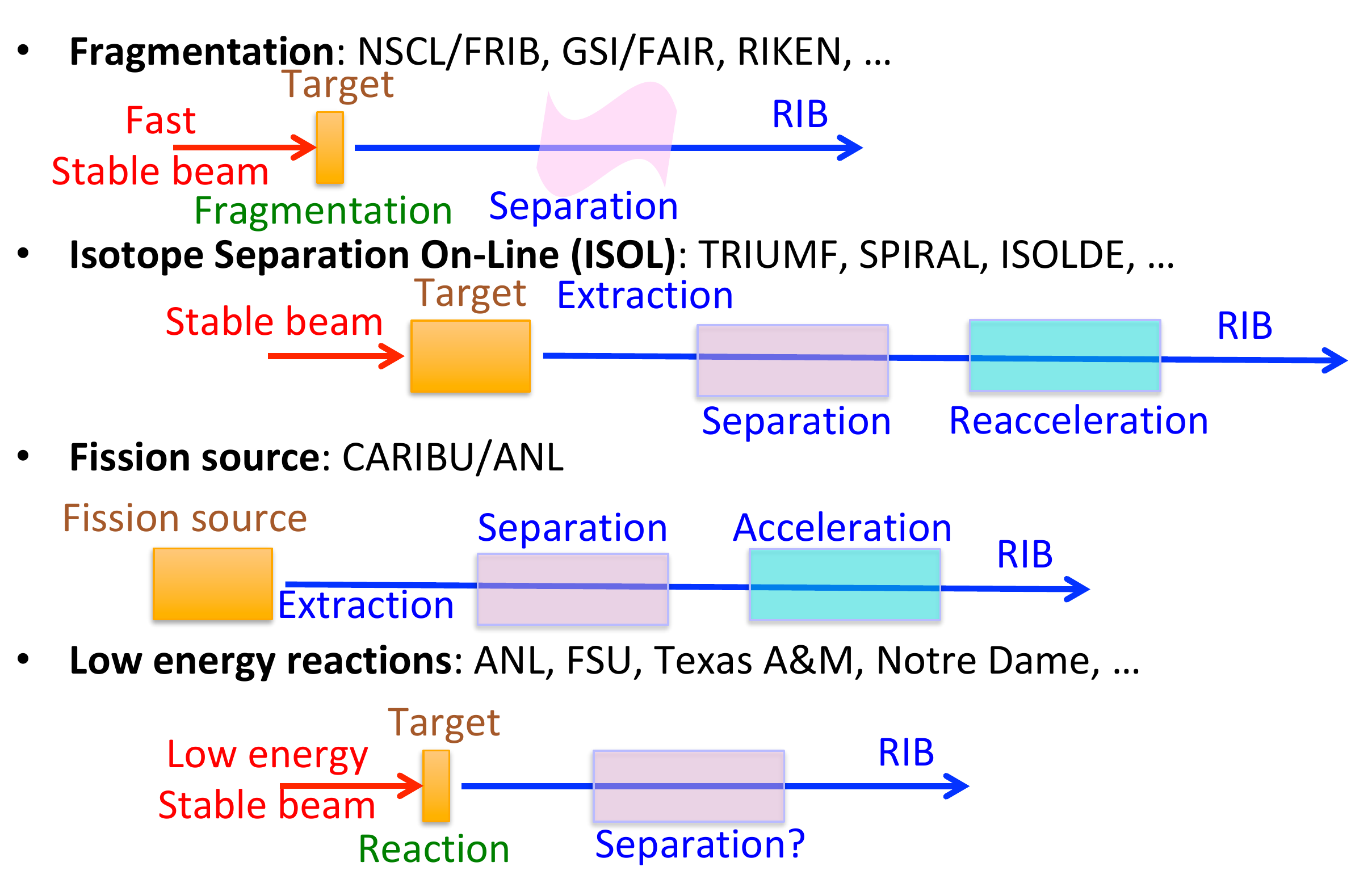}
\caption {Schematic illustration of the principles behind  fragmentation and ISOL facilities.}
\label{fig:production}
\end{center}
\end{figure}

The current overview of present and planned facilities and their associated experimental equipment to infer neutron capture rates is presented in this section. Although the focus of this review is indirect neutron capture rates, all of the facilities presented herein have broad science programs that variously include nuclear structure, nuclear applications, fundamental symmetries and the application of nuclear science to the betterment of society.  A range of different techqniues are employed to produce the isotopes of interest for study and include projectile fragmentation, isotope separation on-line, fission and charged particle reactions, see Fig. \ref{fig:production}.

Those facilities focused on direct neutron capture cross section measurements are mentioned here for completeness and to provide the reader additional reference material and are not discussed further.  Facilities dedicated to direct neutron capture cross section measurements include the DANCE~\cite{CoR07} 
facility at Los Alamos National Laboratory, n\_TOF~\cite{cue2013, bar2013a} at CERN, the GELINA facility at Geel~\cite{bensussan1978}, time-of-flight neutron measurements at iThemba LABS~\cite{mosconi2010}, and future neutron measurements at SARAF~\cite{SARAF2018} in Israel, the Neutron Science Facility at Spiral2 \cite{Ledoux2017}.

\subsection{Argonne National Laboratory}
The Argonne Tandem Linac Accelerator System (ATLAS) at Argonne National Laboratory provides over 50 MV of acceleration for beams of stable ions from protons through uranium using three superconducting linac sections.  Radioactive ion beams are delivered to experimental areas from the Californium Rare Ion Breeder Upgrade (CARIBU) project~\cite{sav2008}.  CARIBU provides beams of short-lived neutron-rich isotopes produced in the spontaneous-fission decay of $^{252}$Cf.  The rare isotopes can be reaccelerated or delivered at low-energy to various experiment stations.  Lighter radioactive ions can be produced in-flight using charge exchange or few nucleon transfer reactions \cite{har2000}.

GRETINA/GRETA: The GRETINA and GRETA devices have been described previously in the section on MSU.  At ANL, GRETINA has been placed in at ATLAS to perform Coulomb excitation studies coupling with the CHICO II particle detector and in front of the Fragment Mass Analyzer (FMA).

\textbf{HELIOS}: The HELical Orbit Spectrometer (HELIOS) \cite{wuo2007,lig2010} studies simple inverse kinematics reactions with radioactive ion beams.  The reaction target and detector are both located along the magnetic field axis of a large-bore superconducting solenoidal spectrometer.  The light mass reaction products emitted from the target are bent by the magnetic field back to the solenoid axis where they are detected suing an array of position sensitive silicon detectors.  The measurement of the reaction product’s target-to-detector distance and particle energy are translated into excitation energy and center of mass angle.  The heavy reaction products can be characterized by a silicon ${\Delta}E - E$ telescope positioned perpendicular to the beam axis.  Indirect capture measurements can be performed by coupling HELIOS with a photon detection system such as APOLLO.

\textbf{APOLLO}: The APOLLO array was developed by Los Alamos National Laboratory and coupled with the HELIOS spectrometer.  APOLLO consists of six LaBr$_3$:Ce and 15 CsI:Tl 2 inch x 3 inch cylindrical crystals.  Since the detectors must operate in the magnetic field of HELIOS they are read out using silicon photomultiplier.  The array allows for a close-packed geometry and has an efficiency of approximately 12\% at 1 MeV.

\textbf{ORRUBA}: The Oak Ridge Rutgers University Barrel Array (ORRUBA) \cite{pai2007} is an array of silicon detectors designed to perform transfer reaction measurements in inverse kinematics with an emphasis on $(d,p)$ reactions.  Two rings of silicon detectors are placed forward and backward of $\theta_{lab}$ = 90$^{\circ}$.  The silicon detectors used are a combination of non-position sensitive and position-sensitive resistive silicon strip detectors.  The array covers approximately 80\% of the azimuthal angles over the range of polar angles from 43$^{\circ}$ to 137$^{\circ}$.  The array was upgraded to SuperORRUBA \cite{bar2013b} building on the original ORRUBA design with increased sensitivity and resolution using double-sided silicon strip detectors.  The double-sided detectors contained 64 1.2mm x 4cm strips on the front face and 4 7.5 cm x 1 cm strips on the back face.

\subsection{Cyclotron Institute}
The Cyclotron Institute operates the K500 at Texas A\&M University.  The K500 can supply a variety of stable beams from protons through uranium at energies ranging from a low of 8 MeV protons to 3.5 GeV uranium. In-flight rare isotope beams can be provided by the Momentum Achromat Recoil Separator (MARS) facility \cite{tri1991,tri2002}.
The Cyclotron Institute is upgrading the capabilities of the facility to provide reaccelerated rare isotopes beams to all experimental areas.  As part of the upgrade the K150 cyclotron is being refurbished to serve as a driver for the production of rare isotopes using the IGISOL method in combination with a heavy ion guide and light ion guide.  The light ion guide will provide rare isotopes predominately from $(p,n)$ reactions.  Rare isotopes from heavy-ion induced reactions will be provided by the heavy ion guide after stopping in and extraction from a gas catcher.  The rare isotopes will then be injected into the K500 cyclotron for further acceleration.

\textbf{Hyperion}: Hyperion~\cite{hug2017} is a particle and $\gamma$-ray detector which replaces the previous array STARLiTeR~\cite{les2010}.  Hyperion contains  a highly segmented silicon telescope for charged particle identification surrounded by an array of up to fourteen Compton-suppressed HPGe detectors for coincidence particle-$\gamma$ measurements.  Each clover detector has an efficiency of 150\% in add-back mode (summing energy deposition from all four crystals within a single detector).  The Compton suppressor is a set of 16 isolated BGO crystals that surround the clover detector on all sides.  The array can split into two hemispheres.  The silicon telescope is based on annular double-sided detectors (Micron S2 design) segmented into 16 sectors and 48 rings.  Different thicknesses are chosen for the $\Delta E$ and $E$ detectors.  An aluminum shield can be mounted in front of the silicon detectors to reduce the impact from $\delta$ electrons.

\subsection{GSI}
Rare isotopes are produced from ion beams of stable ions from helium up to uranium accelerated up to a maximum rigidity of 18 Tm ( $1-4.5$ GeV/u) in the SIS18 synchrotron and impinged on a thick production target.  Ions are separated in-flight with the FRS~\cite{gei1992} fragment separator and delivered the experimental systems.  
The Facility for Antiproton and Ion Research (FAIR)~\cite{gut2006} currently being built in Darmstadt, Germany adjacent to the existing GSI facility, within a broad European collaboration. The central part of the FAIR facility adds a large circumference (1,100 meter) superconducting double-synchrotron fed by the GSI accelerators UNILAC and SIS18 following an upgrade to handle the expected high beam intensities.  A number of experiment programs will be pursued at FAIR one of which will include the investigation of nuclear structure and astrophysics using short-lived nuclei.

\textbf{LAND}: The Large Area Neutron Array (LAND) detector was designed and built to study neutrons emitted from near-relativistic heavy ion collisions ~\cite{bla1992} at the SIS facility of GSI, Darmstadt.  Emitted neutrons are forward focused and detected with a high efficiency in the LAND neutron detector consisting of a series of layers of two meter paddles with 10 cm x 10 cm cross sections where each layer is oriented orthogonally to the previous layer.  Each paddle consists of an alternating arrangement of scintillator and iron convertor.  Neutron momentum is determined from a combination of time-of-flight and position information.  
The Reactions with Relativistic Radioactive Beams (R$^{3}$B) experiment is a component of the Nuclear Structure, Astrophysics, and Reaction (NuSTAR) program at FAIR.  R$^{3}$B builds on the expertise developed using the LAND/ALADIN setup but with a number of improvements to resolution, size, and rate capability. The R$^{3}$B setup consists of~\cite{r3bgsi} a large-acceptance superconducting dipole magnet (GLAD), the New Large Area Neutron Detector (NeuLAND)~\cite{neuland}, the silicon tracker, the photon calorimeter and spectrometer (CALIFA)~\cite{cor2014} and in-beam tracking detectors for heavy fragments and evaporated protons.  

\textbf{HECTOR+} The HECTOR+ array~\cite{camera2014} consists of ten large volume LaBr$_3$:Ce detectors developed by the Milano group.  The detectors are 3.5 inches in diameter and 8 inches in length.  The detectors have been well characterized at high $\gamma$-ray energies and coupled to a number of spectrometers around the world.

\subsection{ISOLDE} 
ISOLDE is the radioactive ion beam facility located at CERN.  High-energy pulsed protons are delivered from the Proton-Synchrotron Booster onto a thick production target (\textit{e.g.} 50 g/cm$^{3}$ Uranium carbide).  Rare isotopes are produced through spallation, fragmentation, and fission reactions in the target.  The target is held at high temperatures (1000 $^{\circ}$C – 2000 $^{\circ}$C) and the radioactive ions are transported to an ion source through diffusion and effusion.  Ionization can occur in a hot plasma, on a hot surface, or through laser excitation.  The ionized isotopes are accelerated to a few tens of keV, mass separated, and delivered to different experimental stations.  Ions are reaccelerated by the Radioactive beam Experiment at ISOLDE (REX-ISOLDE).  In REX-ISOLDE, the ions from ISOLDE are cooled and bunched in a buffer gas of a Penning trap (REX-TRAP), charge bred to a higher charge state in an electron beam ion source (REX-EBIS), and post accelerated to a few MeV/nucleon by a room-temperature linear accelerator (REX-LINAC).  

The HIE-ISOLDE (High Intensity and Energy) boosts the energies of the reaccelerated rare isotopes beams from approximately 3 MeV/nucleon to ultimately 10 MeV/nucleon while also increasing the production rates due to an increase in the intensity and beam energy of the protons.

\textbf{Miniball}:  Miniball is an array of 24 six-fold segmented high-purity germanium detectors \cite{war2013}.  The outer contact of the crystal is six-fold segmented and there is no segmentation along the depth of the detector.  The detectors are tapered to enable close packing and were designed for low-multiplicity experiments with radioactive ion beams.  The germanium crystals have a diameter of 78 mm (before shaping) and a length of 70 mm. Each crystal is encapsulated in a permanently sealed Al can enabling maintenance on the cold electronics without exposing the germanium material.  Three or four crystals are contained in a single cryostat.  The array is used in Coulomb excitation and transfer experiments with radioactive ion beams at ISOLDE but has also been used at GSI and LISOL.

\textbf{T-REX}: The Transfer at REX (T-REX) setup~\cite{TREX} is optimized for the study of transfer reactions in inverse kinematics.  Particle detection is achieved through a number of ${\Delta}E - E$ telescopes constructed of silicon detectors.  A combination of position sensitive, resistive silicon strip detectors and double sided silicon strip detectors are used to cover backward and forward angles.  The resistive detectors are arranged in two boxes centered around 90 degrees while the DSSDs are circular in shape with space in the center for the beam to pass through.  Four PIN-diodes are arranged upstream of the device to serve as an active collimator and the beam profile can also be monitored using a thin CVD diamond detector.

\subsection{iThemba}

\textbf{K600 magnetic spectrometer}
The $K = 600$ magnetic spectrometer  at iThemba LABS is a high-resolution magnetic spectrometer for light ions. It has the capability to measure particles from scattering or direct reactions at extreme forward angles that include zero degrees, making it one of only two facilities worldwide with the capability (the other being at RCNP, Japan) of combining high-energy resolution with zero degree measurements at medium beam energies. The advantage of such measurements is the selectivity it provides to excitations with low-angular momentum transfer.
The K600 consists of five active elements i) a quadrupole, ii) two dipoles and iii) two trim coils.  The average flight path for a particle from target to detector is approximately 8m.   Three different focal-planes enable measurements at a momentum dispersion of either 6.2, 8.4 or 9.8 cm/\%.
The focal-plane detectors are positioned behind the second dipole, and consists of position sensitive multi-wire drift chambers and a pair of plastic scintillation detectors. By employing dispersion matching techniques an energy resolution of 30 keV FWHM can be achieved at 200 MeV for $(p,p’)$ reactions, exploiting the excellent resolving power (1/28000) of the spectrometer. For more details on the K600 refer to Ref. \cite{nev2011}.

\textbf{AFRODITE}:
AFRODITE (AFRican Omnipurpose Detector for Innovative Techniques and Experiments) is a gamma-ray detector array with the capability of detecting both high- and low-energy photons with a reasonably high efficiency by combining Compton Suppressed HPGe detectors (CLOVERS) with Low Energy Photon Spectrometer (LEPS) detectors. AFRODITE comprises of up to 14 Clover detectors which consist of four 50 mm x 70 mm HPGe crystals. These can be augmented with up to eight LEPS detectors which consist of four segmented planar Ge detectors of 2800 mm$^2$ x 10 mm.

The peak-to-total ratio for the Clover detectors operated in add-back mode are ~45-50\%, and the total efficiency at 1.3 MeV is approximately 2.5\%.  
A range of ancillary detectors can be combined with AFRDOITE, such as silicon or CsI charged-particle detectors, recoil detectors, CsI detectors, and fast neutron detectors for time-of-flight discrimination for neutron reaction channels. The fast-timing array of 8 2 inch x 2 inch LaBr$_3$:Ce detectors can also be added to AFRODITE.
During 2019 AFRODITE will be upgraded to 17 Clover detectors through funding from the GAMKA consortium. 

\textbf{ALBA}:
The African LaBr$_3$:Ce array (ALBA) consists of 23 large-volume (3.5 inch x 8 inch) LaBr$_3$:Ce detectors. ALBA can be used as a stand-alone $\gamma$-ray spectrometer or coupled to the K600 spectrometer or other particle detectors (silicon, CSI, neutron detectors). 
The first six detectors of ALBA are already available for measurements at iThemba LABS. A further 17 detectors will be delivered in 2019 through funding from the GAMKA consortium. 

\subsection{Jyv{\"a}skyl{\"a} Accelerator Laboratory}
At the University of Jyv{\"a}skyl{\"a} in Finland, rare isotopes are produced through the Ion Guide Separator Online (IGISOL) facility \cite{Moore2013}.  Neutron-rich rare isotopes are produced through proton-induced fission on either a $^{238}$U or $^{232}$Th target using the K130 cyclotron, and in the future also the MCC30 cyclotron. 
The resulting fission products are thermalized in a gas cell, extracted from the gas cell, accelerated to 30 keV, mass separated, and delivered to the experimental areas.  
See Ref.~\cite{igisolreview} for a review of the IGISOL facility. 

The JYFLTRAP double Penning trap is an essential part of the IGISOL facility and has been used for measuring many nuclear masses of neutron-rich isotopes~\cite{eronen2016}. 
This is particularly important for constraining neutron separation energies, a crucial ingredient in ($n,\gamma$) reaction-rate calculations as shown \textit{e.g.} by Vilen \textit{et al.}~\cite{vilen2018}. 

The Decay Total Absorption $\gamma$-ray Spectrometer (DTAS)~\cite{guadilla2016} has been successfully commissioned at IGISOL and has already provided \textit{e.g.} $\beta$-decay intensities of importance for the $r$-process nuclear reaction network calculations~\cite{guadilla2017} and could in principle be used for the $\beta$-Oslo method. 

\subsection{Legnaro National Laboratory}
The Legnaro National Laboratories (LNL) is on of the four national labs in the Italian Institute of Nuclear Physics (INFN).  The laboratory operates a 16 MV tandem accelerator coupled with a superconducting linac, ALPI, and a superconducting linac injector, PIAVE. The accelerated ions at intensities up to 100 pnA for most ions are then delivered to the experimental areas \cite{Puglierin2010}.

The Selective Production of Exotic Species (SPES) facility is the upgrade path envisioned for LNL \cite{SPESTDR}.  Rare isotope beams will be produced through the ISOL technique. Protons produced from a high current ($\sim$0.7 mA) and high energy (70 MeV) cyclotron will be impinged on a uranium carbide target to induce fissions ($\sim$ 10$^{13}$ fissions/s).  The rare ions produced are surface ionized, mass selected and charge bred.  The ions will then be inserted into the existing  PIAVE-ALPI accelerator facilities at LNL.

\subsection{Michigan State University}
\begin{figure}[tb]
\begin{center}
\includegraphics[clip,width=0.7\columnwidth]{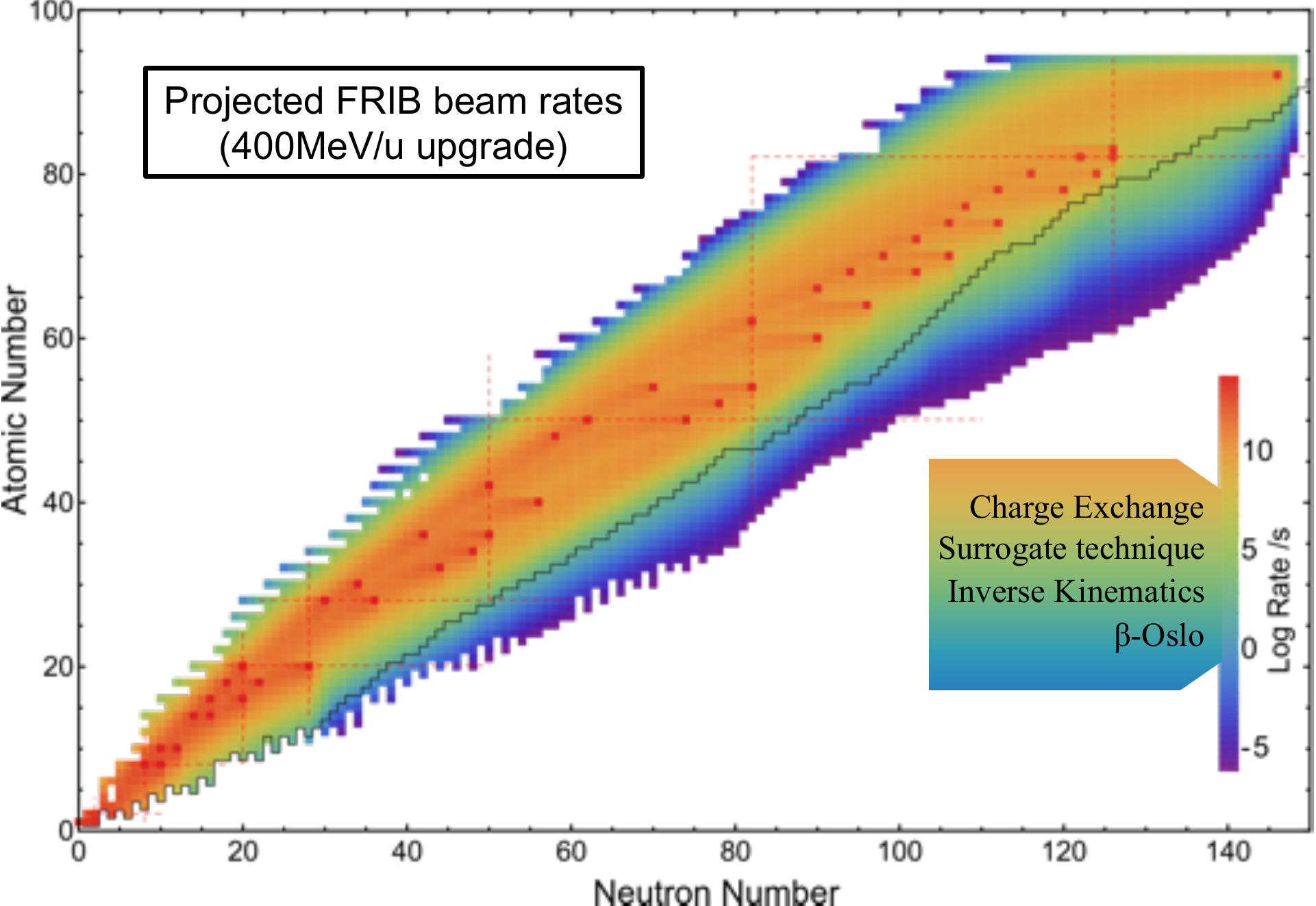}
\caption {Expected FRIB rates for the 400 MeV/nucleon upgrade. Indicated in the rate legend are the necessary intensities for the various indirect methods that could be used to constrain ($n,\gamma$) reaction rates of relevance for the $r$ process.}
\label{fig:FRIB_rates}
\end{center}
\end{figure}

At the National Superconducting Cyclotron Laboratory (NSCL), located on the campus of Michigan State University (MSU), rare isotopes are produced from beams of stable ions from helium up to uranium accelerated up to approximately 160 MeV/nucleon and impinged on a thick production target.  Rare isotopes are produced through the fragmentation process and desired species are separated in-flight with the A1900 \cite{mor2003} fragment separator and delivered to a suite of experimental systems.  The production and delivery of rare isotopes is chemically independent. In addition to the fast beam program, the facility has the capability to stop the high-energy ions in a gas cell \cite{cooper2014} and reaccelerate the ions to modest energy providing rare isotopes at thermal and MeV/nucleon energies.  
In 2009, Michigan State University was selected as the site for the construction of the Facility for Rare Isotope Beams (FRIB).  FRIB will be the world's highest power heavy-ion accelerator for the production of rare isotopes with the capability to produce approximately 80\% of all isotopes predicted to exist up to uranium.  
FRIB will maintain the capability present at NSCL to deliver fast, thermal, and reaccelerated rare isotopes beams to a variety of experimental stations.  Expected beam rates at FRIB are shown in Fig.~\ref{fig:FRIB_rates} and many nuclei along the expected $r$-process path can be studied.

\textbf{SuN}: The Summing NaI detector (SuN) is a total absorption spectrometer constructed from a cylinder of NaI:Tl 16 inches in diameter and 16 inches in length \cite{sim2013}.  A 1.7 inch borehole is located along the central axis of the cylinder.  The cylinder is segmented into a top and bottom half with four slices along its length for a total of eight distinct scintillator crystals.  The top and bottom of SuN are in separate aluminum housings and the four smaller crystals are optically, but not structurally, isolated.  Radioactive ions are deposited into the center of the SuN detector on either an active or inactive material.  A small double-sided silicon strip detector (DSSD) is used as the active stopping material to stop the incoming ion and detect the subsequent $\beta$-decay electron in fast beam experiments.  A moving tape collector is used for thermal ion beams and the $\beta$-decay electron is detected with a surrounding plastic scintillator.  In coincidence with the detection of the $\beta$-decay electron, SuN records the total energy deposited from delayed photons and the energy deposited into individual segments which provides the starting point to the $\beta$-Oslo method.  The photon detection efficiency of SuN is 85\% for $^{137}$Cs.  The $\beta$-Oslo technique using SuN is generally applicable to isotopes with a rate of $\sim$ 1pps at the experimental station.

\textbf{GRETINA/GRETA}: GRETINA is a novel $\gamma$-ray tracking array using HPGe to study the structure and properties of atomic nuclei~\cite{pas2013}.  GRETINA consists of 28 highly segmented coaxial HPGe crystals in groups of four placed into a single cryostat.  Each crystal is subdivided into 36 separate segments.  Sub-segment position resolution leads to superior Doppler energy reconstruction for fast moving beams and excellent in-beam gamma-ray energy resolution.  The detectors are designed to fit in a close packed spherical geometry encompassing one quarter of a sphere. GRETINA was completed in 2011 and has undertaken science campaigns at Lawrence Berkeley National Laboratory, the National Superconducting Cyclotron Laboratory, and Argonne National Laboratory.  At NSCL, GRETINA has been placed at the target position in front of the S800 spectrometer and measures photon emission following population of excited nuclear states through various reactions.  GRETINA is the first portion of the Gamma-Ray Energy Tracking Array (GRETA) detector to scale up to cover a full 4$\pi$ \cite{Deleplanque1999,Lee2008}.  

\textbf{S800}: The S800 is a high-resolution, high-acceptance superconducting spectrograph well suited for reaction measurements in inverse kinematics with radioactive ion beams \cite{baz2003} produced at high energies from the NSCL. It was designed for high precision measurements of both momentum and scattering angle of the reaction products.  It is preceded by an analysis line allowing for operation in either a focused or dispersion matched mode.  Coupled with a gamma-ray spectrometer (such as GRETINA) enables indirect determinations of reaction rates.

\textbf{HRS}: The High Rigidity Spectrometer (HRS)  \cite{HRS2018} is a future experimental device expected to be operational at FRIB for the analysis of charged and uncharged reaction products resulting from the collision of a rare-isotope beam with a target.  Two stages are currently envisioned with the first stage consisting of a large-gap dipole magnet to deflect charged particles followed by a second stage spectrometer section.  The HRS is designed to have a maximum rigidity of 7 Tm suitable to study many rare isotopes produced at FRIB at their maximum rates.  Further the spectrometer will have a momentum resolution of 1:1000, a large acceptance, and a large opening angle for accepting neutrons downstream from the reaction target.

\subsection{Notre Dame University}
The Nuclear Structure Laboratory (NSL) is located on the campus of the University of Notre Dame.  The facility hosts a number of accelerators. An FN Tandem Van de Graff Accelerator capable of reaching acceleration voltages above 10 MV provides stable beams from a few MeV to 100 MeV.  A separate KN Van de Graff accelerator has a maximum operating voltage of 5 MV producing high-intensity low mass beams.  A third 3 MV tandem accelerator focus on delivering beams of light ions (H and He).  Low-energy radioactive ion beams can be produced using the dual in-line solenoid system, TWINSOL.

\textbf{HECTOR}: The High Efficiency TOtal absorption spectrometeR (HECTOR) is a segmented NaI:Tl detector consisting of sixteen, 4 x 8 x 8 inches crystals.  Each crystal is individually encapsulated in an aluminum housing viewed by two photomultipliers.  The individual detectors are assembled into a 16-inch cube with a central borehole 2.55 inches in diameter.

\subsection{Oslo Cyclotron Laboratory}
The facility is built around a Scanditronix MC35 Cyclotron which can deliver high intensity light ions.  Protons, deuterons, $^{3}$He, and $^{4}$He with varying intensities between 50 and 100 $\mu$A can be delivered between 2 - 47 MeV depending on species. The accelerated beams can be delivered to six separate experimental stations.

\textbf{SiRi}: Light ion prodcuts from transfer reactions are detected using the Silicon Ring (SiRi) detector array~\cite{siri}.  The array consists of eight ${\Delta}E - E$ silicon telescopes. The silicon detectors are trapezoidal in shape and the $\Delta E$ detector is segmented into eight curved pieces with each segment having a constant scattering angle, $\theta$. The silicon detectors are arranged 5 cm from target at angles of $45^{\circ}$ or $135^{\circ}$ with respect to the incoming beam.

\textbf{OSCAR}: The Oslo Scintillator Array consists of 30 LaBr$_3$:Ce large volume 3.5 inches x 8 inches detectors. Each detector can be placed in three standard distances of 16, 22 or 25 cm from the target. The total efficiency is $\approx$ 15\% for $E_{\gamma} =1.3$~MeV with the detectors placed at 22 cm from target.

\subsection{RIKEN}
The Radioactive Isotope Beam Factory (RIBF) facility~\cite{yan2007} at the RIKEN Nishina Center produces rare isotope beams through the projectile fragmentation of a stable nucleus or the projectile in-flight fission process with uranium isotopes interacting with a stable target.  The high energy is made available from the combination of three ring cyclotrons with K-values of 570, 980, and 2500 MeV.  Beams of light ions are available up to 440 MeV/nucleon and heavy ions (such as $^{238}$U) with an energy up to 350 MeV/nucleon. Desired nuclei from the fragmentation reaction are selected and transported to experimental stations using the BigRIPS magnetic spectrometer \cite{kub2003}.  

\textbf{DALI2}: The Detector Array for Low Intensity radiation 2 (DALI2)~\cite{tak2014} is the successor to DALI for in-beam gamma-ray measurements with fast rare isotope beams.  The array consists of up to 186 individual detectors.  There are three different types of NaI:Tl detectors in DALI2.  The three detectors have dimensions of 45 x 80 x 160 mm$^3$, 40 x 80 x 160 mm$^3$, and 60 x 60 x 120 mm$^3$.  Each crystal is encapsulated in an aluminum housing 1-mm thick and has a typical energy resolution of approximately 9\% for $^{137}$Cs.  The detectors are arranged around the beam line twelve different layers with 6-14 detectors in each layer.

\subsection{SPIRAL}
At the Syst{\'e}me de Production d’Ions Radioactifs et d’Acc{\'e}l\'eration en Ligne (SPIRAL) facility rare isotope beams are produced through the ISOL technique using a light or heavy ion accelerated through the Grand Acc{\'e}l{\'e}rateur National d’Ions Lourds (GANIL) cyclotrons bombarding a target.  The rare isotopes diffuse out of the target at high temperatures are passed though a transfer line, ionized, and delivered to either a low-energy beam line or injected into the compact cyclotron Cyclotron pour Ions de Moyenne Energie (CIME).  Extraction from CIME delivers a beam ranging from 1.7-25 MeV/A for experimental study.

\textbf{AGATA}
The Advanced Gamma Tracking Array (AGATA) is a 4$\pi$ tracking array to study the structure and properties of atomic nuclei~\cite{Akkoyun2012}.  AGATA will consist of 180 encapsulated germanium crystals (9.0 cm in length and 8.0 cm in diameter).  The crystals are tapered to enable close packing, hexaconical in shape, and segemented in 36 independent segments (6 longitudinal and 6 transverse).Sub-segment position resolution leads to superior Doppler energy reconstruction for fast moving beams and excellent in-beam gamma-ray energy resolution. AGATA has been previously stationed at LNL (2010-2011) and GSI (2012-2014).  AGATA is currently deployed at GANIL starting in 2015 and is expected to remain until 2020.

\subsection{TRIUMF}
TRIUMF is Canada's particle accelerator center and provides a range of rare isotope beams following the bombardment of various target materials with 480 MeV proton beams delivered from the main cyclotron with intensities up to 100 $\mu$A.  The ions diffuse out of the target, are ionized, and delivered to the experimental stations.  The thermal beams can be used directly or reacclerated up to 1.8 AMeV in the ISAC-I facility \cite{dil2014} and up to 16.5 AMeV using the ISAC-II facility.

The Advanced Rare IsotopE Laboratory (ARIEL) is the next rare isotope facility being developed at TRIUMF \cite{dil2014a}.  ARIEL will use an electron linac to accelerate an electron beam up to 35 MeV with a power of 100 kW in addition to another proton beam from the cyclotron.  The electron beam will impinge on a uranium target and produce rare isotopes through the photofission process. The rare isotopes will be extracted using the new CANREB facilities and directed to ISAC for reacceleration.

\textbf{TIGRESS}: The TRIUMF-ISAC Gamma-Ray Escape Suppressed Spectrometer (TIGRESS) \cite{sve2005} is a gamma-ray spectrometer used for a broad range of nuclear physics at TRIUMF.  TIGRESS consists of an array of 32-fold segmented HPGe detectors. Each detector consists of four cylindrical HPGe crystals 60 mm in diameter and 90 mm in length (before shaping) that are electrically segmented into eight segments on their outer contacts for a total of 32 outer contacts.  The inner core contacts are also extracted for high-resolution energy measurements.  Two sides of each crystal are tapered at 22.5$^{\circ}$ to allow for close packing of the detectors.  Using waveform analysis, an average position resolution for the detectors was determined to be 0.44 mm allowing for the accurate correction of Doppler shifts. Compton suppression shields constructed of bismuth germanate (BGO) and cesium iodide (CsI) surround the TIGRESS HPGe detectors.  The shields suppress scattered background and improve the signal to background ratio.  The HPGe detectors can be operated with or without shields at a radius of 11.0 or 14.5 cm, respectively.  Auxillary detectors are placed inside TIGRESS for charged particle detection such as the BAMBINO or SHARC silicon detectors.

\section{Challenges and future prospects}
For all new methods highlighted here, there are still challenges that need to be addressed.  
For the Oslo method in inverse kinematics, the main challenge is 
achieving high enough intensities for the radioactive beam of interest,
limiting the range of nuclei that can be reached at present facilities. 
Also, obtaining sufficient excitation-energy resolution (better than  
1 MeV FWHM) might be difficult depending on the reaction kinematics and the available detector systems. 
Also, as both the level density and $\gamma$-ray strength function need to be normalized, uncertainties in absolute normalization can be quite large away from stability where no neutron-resonance data exist, introducing extrapolations and model assumptions into the analysis. 

For the $\beta$-Oslo method, one needs to take into account the rather narrow
spin range that is populated through $\beta$ decay in the daughter nucleus. 
This is rather straight-forward as long as the spin of the mother nucleus
is known, but this is not always  the case for more exotic nuclei. 
Again, as for the Oslo method in inverse kinematics, absolute normalization becomes an issue when measuring nuclei well away from the stability line.
Also, the determination of the initial excitation energy might 
suffer from incomplete summing, making a low-energy tail. This effect is typically small for the highly efficient SuN
detector~\cite{SuN}. Further improvement would include 
correcting for incomplete summing on the excitation energy in addition to the $\gamma$ energy;
the latter is already done and the former is work in progress. 
Finally, in cases where the $\beta$-decay $Q$-value is very high
(several MeV above $S_n$), the $\beta-n$ channel could be a 
significant decay mode. At present, it is not possible to discriminate
against neutrons in SuN; it would be highly desirable 
to implement that opportunity in future TAS detector designs. 

Regarding the surrogate method, it requires rather extensive modelling to determine
the transfer reaction cross section, optical modelling estimates for
the neutron emission and realistic level densities and $\gamma$ strength functions.
Applied to unknown neutron-rich nuclei, all these assumptions
could lead to rather large  uncertainties.
In the future, one could hope to pin down reliable level densities and $\gamma$ strength functions
and constrain the modelling of the nuclear reactions of interest. In this way,
the surrogate method will be complementary to other techniques in
the study of cross sections relevant for the $r$-process nucleosynthesis.

As for future prospects, and considering the huge achievements
both in nuclear-physics experiments and nuclear theory, as well 
as in astronomy observations and models, we are now at a very exciting
point in history. New radioactive-beam facilities will greatly extend the experimental reach of 
exotic nuclei, providing the ultimate testing ground for nuclear theory
and enabling many more ($n,\gamma$) reaction rates 
to be experimentally constrained, amongst other experimental information
such as $\beta$-decay rates and nuclear masses. This will in turn significantly 
improve the input for $r$-process nucleosynthesis 
simulations, limiting the parameter space which is
at present very large. Also, with the coming increased statistics on neutron-star mergers and neutron-star--black-hole mergers from gravitational-wave 
detectors such as advanced LIGO~\cite{advLIGO}, combined with the advances of rare-isotope facilities, we foresee a bright future for the nuclear-astrophysics community paving the way to understand every aspect of the astrophysical $r$ process.

\section*{Acknowledgments} 
A.~C.~L. gratefully acknowledges funding 
through ERC-STG-2014 under grant agreement no. 637686.
A.~C.~L. acknowledges support from the “ChETEC” COST Action (CA16117), supported by COST (European Cooperation in Science and Technology).
This work was supported by the National Science Foundation under Grants No. PHY 1102511 (NSCL) 
and No. PHY 1430152 (Joint Institute for Nuclear Astrophysics), and PHY 1350234 (CAREER). 
The authors would like to thank F.L. Bello Garrote, A.~G\"{o}rgen, V.~W.~Ingeberg, J.~E.~Midtb{\o}, V.~Modamio, T.~Renstr{\o}m, E.~Sahin, S.~Siem, O.~Sorlin, G.~M.~Tveten, M.~Wiedeking and F.~Zeiser for stimulating discussions. 


\end{document}